\documentclass[fleqn,usenatbib,useAMS]{mnras}

\usepackage{graphicx}	
\usepackage{xcolor}
\usepackage{dblfloatfix}
\let\oldAA\AA
\renewcommand{\AA}{\text{\normalfont\oldAA}}
\usepackage{amssymb,amsmath,stackengine,scalerel}
\usepackage{multicol}        
\usepackage{bm}		
\usepackage{pdflscape}	
\usepackage[euler]{textgreek} 
\usepackage{epstopdf}
\usepackage[utf8]{inputenc}
\usepackage[english]{babel}
\bibpunct{(}{)}{;}{a}{}{,} 
\usepackage{fontawesome}
\usepackage{comment}
\usepackage[para,online,flushleft]{threeparttable}
\usepackage{array,multirow,booktabs}
\usepackage{bigstrut}
\usepackage{braket}

\usepackage{natbib,twoopt}
\bibpunct{(}{)}{;}{a}{}{,} 
\newcommandtwoopt{\citeads}[3][][]{\href{http://adsabs.harvard.edu/abs/#3}%
{\def\hyper@linkstart##1##2{}%
\let\hyper@linkend\@empty\citealp[#1][#2]{#3}}}
\newcommandtwoopt{\citepads}[3][][]{\href{http://adsabs.harvard.edu/abs/#3}%
{\def\hyper@linkstart##1##2{}%
\let\hyper@linkend\@empty\citep[#1][#2]{#3}}}
\newcommandtwoopt{\citetads}[3][][]{\href{http://adsabs.harvard.edu/abs/#3}%
{\def\hyper@linkstart##1##2{}%
\let\hyper@linkend\@empty\citet[#1][#2]{#3}}}
\newcommandtwoopt{\citeyearads}[3][][]%
{\href{http://adsabs.harvard.edu/abs/#3}
{\def\hyper@linkstart##1##2{}%
\let\hyper@linkend\@empty\citeyear[#1][#2]{#3}}}
\hypersetup{breaklinks}
\usepackage[T1]{fontenc}
\usepackage{ae,aecompl}
\usepackage{newtxtext,newtxmath}

\newcommand{\Shk}{$S_{\rm HK}$}
\newcommand{\ihor}{$\iota$~Hor}

\title[]{\textit{Far beyond the Sun: II. Probing the stellar magnetism of the young Sun \texorpdfstring{$\boldsymbol{\iota}$}{i} Horologii from the photosphere to its corona}}

\author[E. M. Amazo-G\'{o}mez]{E. M. Amazo-G\'{o}mez
    \thanks{Contact e-mail: \href{eamazogomez@aip.de}{eamazogomez@aip.de}}$^{1}$,
    J. D. Alvarado-G\'omez$^{1}$,
    K. Poppenhaeger$^{1,15}$,
    G. A. J. Hussain$^{2,3}$,
    \newauthor
    B. E. Wood$^{6}$,    
    J. J. Drake$^{7}$,
    J.-D.~do~Nascimento~Jr.$^{7,16}$,
    F. Anthony$^{16}$,
    J. Sanz-Forcada$^{8}$,
    \newauthor
    B. Stelzer$^{9}$,
    F. Del Sordo$^{10,11}$,
    M. Damasso$^{12}$,
    S. Redfield$^{13}$,
    J. F. Donati$^{14}$,
    \newauthor
    P. C. K\"{o}nig$^{3,4}$,
    G. H\'ebrard$^{4,5}$,
    P. A. Miles-P\'aez$^{3}$
    \vspace{0.5cm} \\ 
    $^{1}$ Leibniz-Institut f\"{u}r Astrophysik Potsdam, An der Sternwarte 16, 14482 Potsdam, Germany\\
    $^{2}$ European Space Agency, Keplerlaan 1, 2201 AZ Noordwijk, Netherlands\\
    $^{3}$ European Southern Observatory, Karl-Schwarzschild-Strasse 2, 85748 Garching bei München, Germany\\
    $^{4}$ Institut d'astrophysique de Paris, UMR7095 CNRS, Sorbonne Universit\'e, 
    98 bis boulevard Arago, 75014 Paris, France\\
    $^{5}$ Observatoire de Haute-Provence, CNRS, Universit\'e d'Aix-Marseille, 04870 Saint-Michel-l'Observatoire, France\\
    $^{6}$ Naval Research Laboratory, Space Science Division, Washington, DC 20375, USA\\
    $^{7}$ Center for Astrophysics | Harvard \& Smithsonian, 60 Garden Street, Cambridge MA 02138, USA\\
    $^{8}$ Centro de Astrobiolog\'ia (CSIC-INTA), ESAC Campus, Camino Bajo del Castillo, E-28692 Villanueva de la Ca\~nada, Madrid, Spain\\
    $^{9}$ Eberhard Karls Universit\"{a}t, Institut f\"{u}r Astronomie und Astrophysik, Sand 1, 72076 T\"{u}bingen, Germany\\
    $^{10}$INAF–Osservatorio Astrofisico di Catania, via Santa Sofia, 78 Catania, Italy\\
    $^{11}$Institute of Space Sciences (ICE-CSIC), Campus UAB, Carrer de Can Magrans s/n, 08193, Barcelona, Spain\\
    $^{12}$ INAF-Osservatorio di Torino, Via Osservatorio, 20, 10025 Pino Torinese TO, Italy\\
    $^{13}$ Astronomy Department and Van Vleck Observatory, Wesleyan University, Middletown, CT 06459-0123, USA\\
    $^{14}$ CNRS-IRAP, 14, avenue Edouard Belin, F-31400 Toulouse, France\\
    $^{15}$ Universit\"{a}t Potsdam, Institut f\"{u}r Physik und Astronomie, Karl-Liebknecht-Straße 24/25, 14476 Potsdam-Golm, Germany\\
    $^{16}${Universidade Federal do Rio Grande do Norte (UFRN), Dep. de F\'isica, 59078-970, Natal, RN, Brazil}
    }

\date{Last updated today; in original form yesterday}

\pubyear{2023}

\begin{document}
\label{firstpage}
\pagerange{\pageref{firstpage}--\pageref{lastpage}}
\maketitle

\begin{abstract}
A comprehensive multi-wavelength campaign has been carried out to probe stellar activity and variability in the young Sun-like star $\iota$-Horologii. We present the results from long-term spectropolarimetric monitoring of the system by using the ultra-stable spectropolarimeter/velocimeter HARPS at the ESO 3.6-m telescope. Additionally, we included high-precision photometry from the NASA Transiting Exoplanet Survey Satellite (TESS) and observations in the far- and near-ultraviolet spectral regions using the STIS instrument on the NASA/ESA Hubble Space Telescope (HST). The high-quality dataset allows a robust characterisation of the star's rotation period, as well as a probe of the variability using a range of spectroscopic and photometric activity proxies. By analyzing the gradient of the power spectra (GPS) in the TESS lightcurves we constrained the faculae-to-spot driver ratio ($\rm S_{fac}/S_{spot}$) to 0.510$\pm$0.023, which indicates that the stellar surface is spot dominated during the time of the observations. We compared the photospheric activity properties derived from the GPS method with a magnetic field map of the star derived using Zeeman-Doppler imaging (ZDI) from simultaneous spectropolarimetric data for the first time. Different stellar activity proxies enable a more complete interpretation of the observed variability. For example, we observed enhanced emission in the HST transition line diagnostics \ion{C}{iv} and \ion{C}{iii}, suggesting a flaring event. From the analysis of TESS data acquired simultaneously with the HST data, we investigate the photometric variability at the precise moment that the emission increased and derive correlations between different observables, probing the star from its photosphere to its corona.
\end{abstract}


\section{Introduction}

\textcolor{black}{An accepted understanding among solar and stellar scientists is that the 11-year variations in global activity observed in the Sun are coherent and connected with the 22-year magnetic cycle. These cycles have been extensively studied and correlated with a synchronized behavior of the solar atmospheric layers \cite[i.e., photosphere, chromosphere, and corona][]{2015LRSP...12....4H,2022NatSR..1215877L}.  In order to approach a better magnetic understanding of other stars, stellar researchers have conjugated a set of multiple techniques to overcome the difficulties generated by the lack of quality and quantity data, if compared with the solar counterpart \cite[see review by,][]{2012LRSP....9....1R}. Techniques such as the Zeeman Doppler Imaging \cite[ZDI,][]{1997MNRAS.291..658D,2000MNRAS.318..961H,2002A&A...381..736P}, allow us to investigate the stellar large-scale polarity. Moreover, dynamo simulation studies emphasize the significance of considering essential parameters, such as stellar inclination, rotation, and observation phase coverage, in order to improve the recovery of more accurate ZDI magnetic diagnostics \cite[see,][]{2023arXiv230607838H}.} 

\textcolor{black}{By implementing the ZDI technique large-scale polarity and activity modulations, including magnetic reversals, have been detected only in a few stars \cite[e.g.,][]{2009MNRAS.398.1383F,2017MNRAS.471L..96J,2018A&A...620L..11B,2022A&A...658A..16B}, These studies have also identified some of them that exhibit solar-like cyclic behavior, such as 61 Cyg\,A \cite[][]{2016A&A...594A..29B}. However, for most stars, the time scale for magnetic field reversals is much shorter. Furthermore, these rapid reversals do not show a resemblance to the observed variability in chromospheric activity, unlike the case of the Sun \cite[e.g.,][]{2008MNRAS.385.1179D,2009MNRAS.398.1383F,2012A&A...540A.138M}.  }

\textcolor{black}{This situation becomes particularly challenging when studying other indicators of stellar magnetism such as coronal activity, as monitoring X-ray emissions over the entire cycle duration is difficult \cite[see Table\,3 in][]{2017MNRAS.464.3281W}. It is not enough available stellar data, and further research is required to gain a better understanding of the relationship between magnetic field behavior and activity variations in stars as we have for the Sun. This could involve supporting more long-term observational campaigns, as presented in this work, to study the coronal activity and polarimetric behavior of stars over longer time scales. }

\textcolor{black}{As mentioned, combining multiple instruments and techniques to monitor the magnetic variability of stars different than the Sun is very challenging, but still proven to be an effective way to render a robust description of them. Apart from the Sun, there are just a few systems that have been observed over multiple wavelengths and followed up regularly, \cite[e.g., $\alpha$ Cen A, $\epsilon$ Eri, $\xi$ Bootis, etc.][]{Pagano2004,2011A&A...527A..82D,2012A&A...543A..84R,2012Natur.491..207D,2013ApJ...770..133H,1993ApJ...412..797D,1996ApJ...462..948L,2000ApJ...544L.145H,2000ApJ...545.1074D,2006AJ....132.2206B,2008A&A...488..771J,2008MNRAS.385.1691N,2012ApJS..200...15A,2014A&A...569A..79J,2017MNRAS.471L..96J,2020A&A...636A..49C,2016A&A...594A..29B,2018A&A...620L..11B,2022A&A...658A..16B}. }

$\iota$ Horologii ($\iota$\,Hor, HD\,17051, HR\,810) is one of the few systems with a robust set of multi-wavelength records covering a period of years \cite[see][]{1995ApJ...438..269B,1997MNRAS.284..803S,2005ApJS..159..141V,2007ApJ...669.1167B,2017ApJ...850...80R}. The star has an estimated age of about $\sim625$~Myr~\cite[see][]{2008A&A...482L...5V} and similar characteristics to the young Sun \cite[i.e., F8V-G0V, $\sim~6080$K, $R_{*}\sim1.16~R_{\odot}$][see Table~\ref{tab_1}]{2010MNRAS.405.1907B}. The star hosts a giant Jupiter-like planet at approximately 1\,au \cite[see][]{2000A&A...353L..33K,Zechmeister2013}. \ihor\ is more active than the Sun, e.g. $\log L_{\rm X}/L_{\rm bol}=-4.9$ on average \citep{2013A&A...553L...6S,AlvaradoGomez2018}, and has one of the shortest chromospheric and coronal activity cycles known to date \cite[$\sim$1.6~years:][]{2010ApJ...723L.213M,2011A&A...532A...6S,2013A&A...553L...6S,2019A&A...631A..45S}. Besides, a magnetic activity cycle of $\sim$2.1 years was described in \cite[][hereafter \textcolor{blue}{Paper\,I}]{AlvaradoGomez2018} and \cite{2019shin.confE.110A,AlvaradoIII}. The cycles are analysed in this paper after extended observations, evidencing the possibility of a double-cycle behavior.

\textcolor{black}{Given the resemblance between \ihor\, and} the young Sun, its short magnetic and activity cycles (close to 10 times shorter than the 11-year solar cycle), the star represents a particularly interesting target to monitor in order to get a better insight into stellar magnetism generation and enable us to test \textcolor{black}{the solar activity paradigm (i.e., whether it is or not correct to apply the solar activity analogy to the star)}. Long-term observations, as performed for \ihor\,, benefit the development of more accurate models of stellar dynamos, which are critical to our understanding of magnetism throughout the main sequence: the magnetism driving the star's angular momentum loss and causing its magnetic activity to decrease over time \cite[][]{2011ApJ...743...48W,2012ApJ...746...43R,2014ApJ...794..144R,2016Natur.535..526W,2018MNRAS.479.2351W,2022A&A...662A..41R}. The rate at which activity, and hence high-energy radiation and stellar wind decrease, is ultimately critical for rotational and possibly magnetic dynamo evolution \cite[][]{2018ApJ...862...90G}. 

Here we show how a multiwavelength analysis allows probing activity across different atmospheric layers. By compiling the information contained in a set of activity indicators it is possible to build a global picture of the star's magnetic activity state. We argue that, due to its proximity, brightness, and youth, \ihor\ is a keystone star to build on and test the solar-stellar analogy. In this paper, we continue the follow-up to the star as part of our intensive monitoring of \ihor\, under the \emph{Far beyond the Sun} campaign (P.I. J.D. Alvarado-G\'omez). In \textcolor{blue}{Paper\,I} we described the first half of the data acquired, 3 ESO HARPSpol observational periods. Here, we cluster together the entire set of ESO HARPSpol data (6 observational periods) and simultaneous observations of the star obtained by TESS and HST telescopes.

\begin{figure*} 
\includegraphics[trim=0.cm 0.cm .0cm 0.cm, clip=true, scale=0.385]{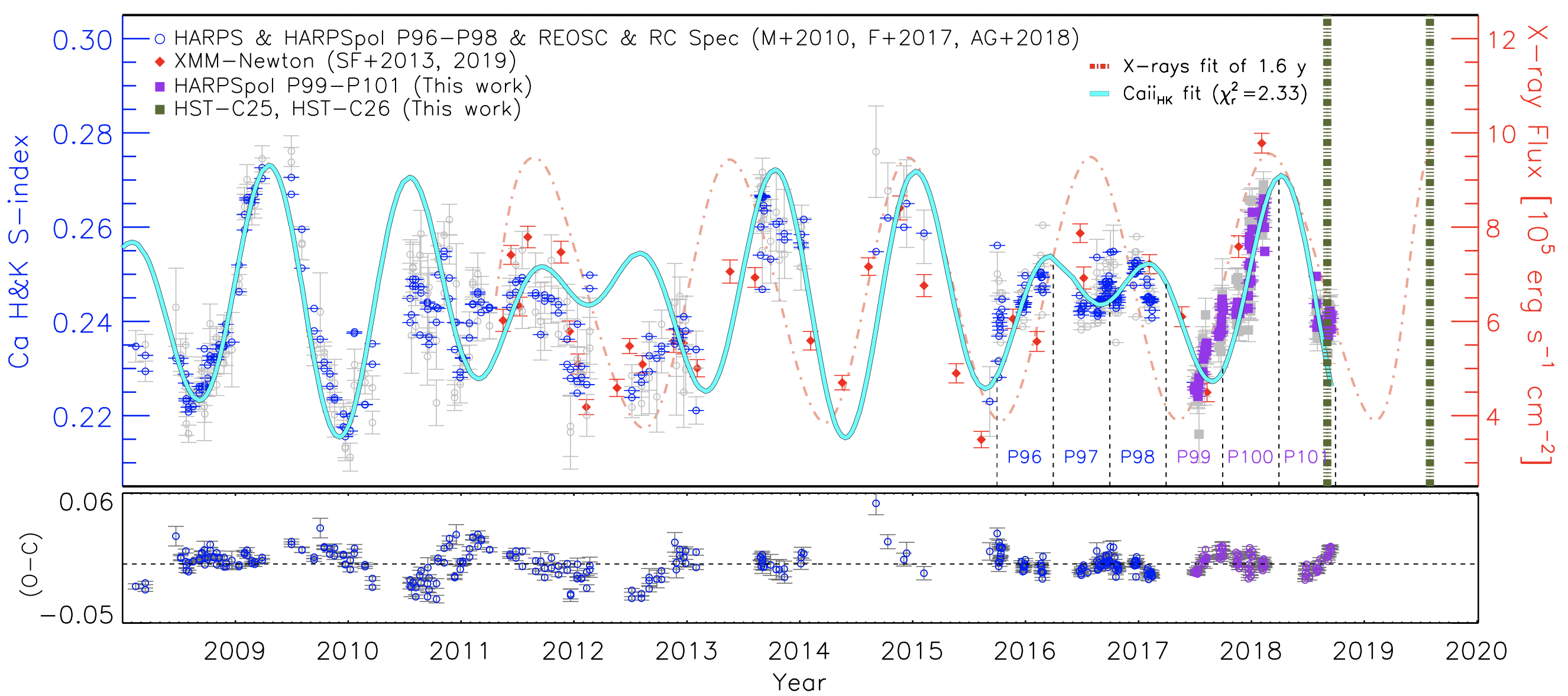}
\caption{Top panel: Long term follow up of \Shk\ value from the \ion{Ca}{ii}$~\rm{H\,\&\,K}$ core lines. Gray dots are the original scattered data before the smoothing averaging over the rotational period modulation. Blue points are in detail analysed in \citet[][M+2010]{2010ApJ...723L.213M}, \citet[][F+2017]{2017MNRAS.464.4299F}, and \citet[][AG+2018]{AlvaradoGomez2018}. X-rays observations in red-filled rhomboids for $\iota$-Horologii reported by \citealp[][SF+2013,2019]{2013A&A...553L...6S,2019A&A...631A..45S}. The purple-filled squares are new data presented in this work. As well as the Olive dashed line which corresponds to TESS and HST simultaneous observations. Bottom panel: Observed minus calculated residuals of the double period fitting.}\label{Fig1}

\end{figure*}

\begin{table}
\caption{Fundamental properties of $\iota$ Hor.}\label{tab_1}   
\begin{threeparttable}
\centering             
\begin{tabular}{l c c}    
\hline\hline        
Parameter & Value & Reference \\
\hline
Spectral Type & F8V\,--\,G0V & \protect{\cite{2010MNRAS.405.1907B}}\\ 
$T_{\rm eff}$ [K] & $6080 \pm 80$ & \protect{\cite{2010MNRAS.405.1907B}} \\
$\log(g)$ & $4.399 \pm 0.022$ & \protect{\cite{2010MNRAS.405.1907B}} \\
$R_{*}$ [R$_{\odot}$] & $1.16 \pm 0.04$ & \protect{\cite{2010MNRAS.405.1907B}} \\
$M_{*}$ [M$_{\odot}$] & $1.23 \pm 0.12$ & \protect{\cite{2010MNRAS.405.1907B}} \\
$v\sin i$ [km s$^{-1}$] & $6.0 \pm 0.5$ & \protect{\cite{AlvaradoGomez2018}} \\
$P_{\rm rot}$ [days]$^{\rm spec}$ & $7.70^{+0.18}_{-0.67}$ & \protect{\cite{AlvaradoGomez2018}} \\
$\overline{P_{\rm rot}}$ [days]$^{\rm photo}$ & $7.39^{+0.28}_{-0.16}$ & \protect{This work} \\
$\left<v_{\rm R}\right>$ [km s$^{-1}$]$^a$ & $16.927 \pm 0.001$ & \protect{This work} \\ 
$\left<\log(L_{\rm X})\right>$ & $28.78 \pm 0.08$ & \protect{\cite{2013A&A...553L...6S}}\\
Age [Myr]$^b$ & $\sim625$ & \protect{\cite{2008A&A...482L...5V}}\\
\hline                  
\end{tabular}
\begin{tablenotes}
{Summary of the main properties of $\iota$ Hor. The \small $(\rm spect)$ and \small $(\rm photo)$ refer to the spectroscopic and photometric estimations of the rotation period values, respectively. \small $^{(a)}$ Average value from the 6 multi-epoch HARPSpol observations reported in this work and \protect{\cite{AlvaradoGomez2018}}. \small $^{(b)}$~Age derived from HARPS asteroseismology. This value falls in between the estimates from gyrochronology \cite[$\sim 740$~Myr,][]{2007ApJ...669.1167B}, and from the level of X-ray \cite[$\sim 500$~Myr,][]{2011A&A...532A...6S}.}
\end{tablenotes}
\end{threeparttable}
\end{table}

\subsection{Data exploration}\label{dataset}

Under the \emph{Far beyond the Sun} campaign, the $\iota$-Hor system has been intensively monitored over six semesters, during ESO observing periods P$_{96}$ to P$_{101}$ (between October 2015 to September 2018). This allowed us to collect 199 data points using the spectropolarimetric mode of the HARPS instrument (HARPSpol) at the ESO 3.6-m telescope at La Silla Observatory \cite[see,][]{2003Msngr.114...20M,2011Msngr.143....7P}. 

In addition to the spectropolarimetric data, we analysed photometric time series obtained by the TESS telescope \cite{2015JATIS...1a4003R}. We describe in Section~\ref{HARPS} the complete spectropolarimetric data set gathered for this project. We outline the analysis of our 2.95-year HARPSpol follow-up, adding three additional semesters observing \ihor\, to those presented in \textcolor{blue}{Paper\,I}. We used Sector~2 and Sector~3, for a total data compilation of 54 days (23 Aug. -- 20 Sep. 2018, 20 Sep.-17 Oct. 2018 respectively). We also looked at sectors 29 and 30 (Aug. and Sep. 2020, respectively) obtaining similar results as described in Section~\ref{sec_TESS}. In addition, we retrieve the surface faculae to spot ratio, $\rm S_{fac}/S_{spot}$, by applying the GPS method over TESS photometric time series. We complemented our collection of data for $\iota$ Hor with acquired NUV and FUV spectra using the Space Telescope Imaging Spectrograph (STIS) instrument \cite[see,][]{1986SPIE..627..350W,1986ESASP.263..653G,1990BAAS...22S1283W} on board the Hubble Space Telescope observatory. We analyse two separate visits of 3 orbits each (one in the NUV and, two in the FUV regions), yielding 6 separate exposures for the target. Three observations were taken on 3 September 2018 over cycle 25 (P.I. J.D. Alvarado-G\'omez, HST proposal ID~15299) and three on 1 August 2019 during cycle 26 (P.I. J.D. Alvarado-G\'omez, HST proposal ID 15512). The STIS/HST observations were obtained using the Multi-Anode Microchannel Array (MAMA) detector by the E140M (FUV) and E230H (NUV) gratings. A more detailed description of the HST observations is presented in Section~\ref{HST}. Section~\ref{Simultaneous} contains a multi-technique (Zeeman-Doppler Imaging, spectropolarimetry, photometry), multi-wavelength (FUV, NUV, visible) comparative analysis carried out in one epoch consisting of simultaneous observations by HARPSpol, TESS, and HST. We present the main discussion and conclusions in Section~\ref{DandC}. The data availability is described in Section\,\ref{DA}. 

\section{HARPS spectropolarimetry}\label{HARPS}

The HARPS instrument on the ESO 3.6-m telescope at the La Silla Observatory is a high-precision, ultrastable \'echelle spectrograph, and velocimeter. It can attain a precision of $0.97\,\rm{m}\,\rm{s}^{-1}$ over a spectral range of $378~{\rm nm}$ to $691~{\rm nm}$, and has a spectroscopic resolution of $R=\,120\,000$ \cite[see][]{2003Msngr.114...20M,2005Msngr.120...22P}. In addition, the instrument offers a spectropolarimetric mode that enables observation of circularly and linearly polarised signatures across the entire spectral range \cite[see][]{2011Msngr.143....7P}. We used the spectropolarimetric mode of the instrument over six consecutive semesters, from ESO period P$_{96}$ to P$_{101}$ (as shown with vertical dashed lines in Fig.~\ref{Fig1}). The three first semesters (P$_{96}$-P$_{98}$) were analysed in \textcolor{blue}{Paper\,I}. In this work, we complete the analysis by adding 98 new spectropolarimetric data points collected over the final three semesters of the campaign (P$_{99}$-P$_{101}$).

Each observation sequence consisted of roughly one-hour exposures, divided into four sub-exposures between which the half-wave Fresnel rhombs were rotated to different angles to construct the circular and null polarization profiles. Homogeneously, as in \textcolor{blue}{Paper\,I}, the data reduction was carried out using LIBRE-ESPRIT \citep{1997MNRAS.291..658D}. The reduced data products consisted of unpolarized intensity profiles and circularly polarized Stokes-V signatures. Diagnostic null (N) spectra serve as checks for instrumental noise and artifacts where combining the individual sub-exposures helps to cancel out possible polarisation coming from the instrument. These spectra were created by applying the ratio method \citep[see][]{2009PASP..121..993B}.%

Making use of the instrument's high resolution and spectropolarimetric observational mode we analysed the radial velocities (RVs). We extracted information about the stellar activity from indicators such as the \ion{Ca}{ii} S-index \Shk, and the H${\upalpha}$ emission as index \rm I$_{\rm H\upalpha}$. We also probed the stellar magnetic behavior by measuring the effect of the longitudinal magnetic field $B_{\ell}$ over the spectra. We analyse the information obtained from the HARPSpol dataset in the following subsections.

\subsection{\ion{Ca}{ii}$~\rm{H\,\&\,K}$ index}\label{Ca}

\begin{figure} 
\includegraphics[trim=.1cm 0.cm 0.cm 0.cm, clip=true, scale=0.22]{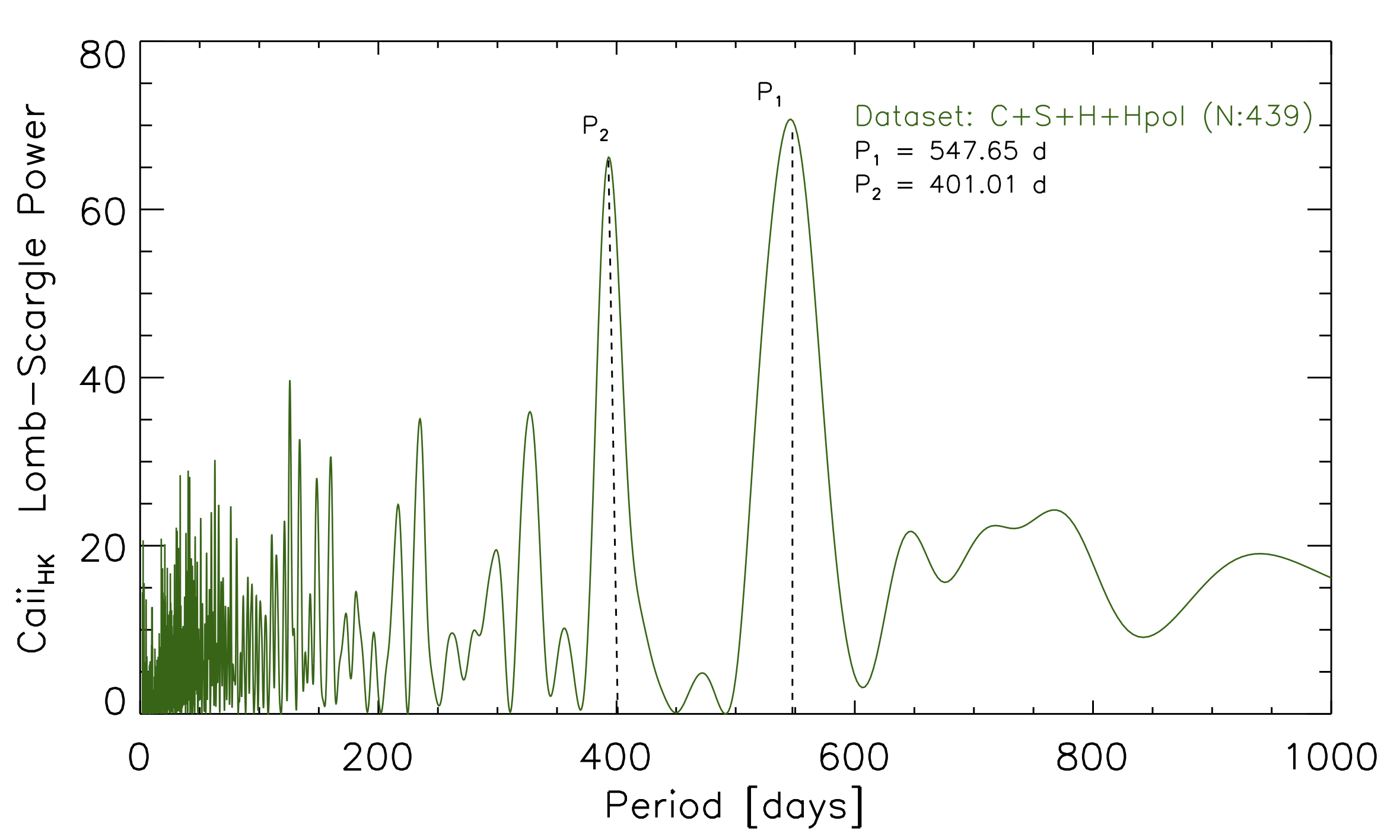}
\caption{Periodicity analysis on the long-term \Shk\ variations of $\iota$-Hor (Fig.\,\ref{Fig1}). Lomb-Scargle (LS) periodograms are recovering two main significant peaks above the noise level ($\rm LSP\sim40$) $\rm P_{1}=547.65\pm3$d (1.49$\pm$0.01~yr) and $\rm P_{2}=401.01\pm7$d (1.09$\pm$0.02~yr).}
\label{Periodogram}
\end{figure}

In the chromospheres of Sun-like stars, the source functions of ionized metals are generally dominated by collisional processes. In this context, the collisionally dominated \ion{Ca}{ii}$~\rm{H\,\&\,K}$ Fraunhofer lines and, the emission observed specifically in the line cores, turned out to be a good indicator of activity in this atmospheric layer (see~\citealt{1957ApJ...125..260T,2008LRSP....5....2H}). The enhancement of the line cores, which has been demonstrated to form in the high chromosphere, is a response to the presence of magnetically driven chromospheric heating and temperature rise \cite[see][]{1913ApJ....38..292E,1957ApJ...125..260T,1963ApJ...138..832W,1972ApJ...171..565S,1978PASP...90..267V,1978ApJ...226..379W,2008LRSP....5....2H}. The \ion{Ca}{ii} S-index, \Shk, was introduced by \citet{1978PASP...90..267V} as a consistent way to analyze the fluxes in the cores of \ion{Ca}{ii}$~\rm{H\,\&\,K}$ lines in the Sun, and was  quickly extended to Sun-like stars. The \Shk\ index is a dimensionless ratio that measures the enhanced emission of the $\ion{Ca}{ii}~\rm{H\,\&\,K}$ lines, centred at 396.85 and 393.37~nm, by comparing the emission flux in the line cores with that in the nearby continuum. We follow the approach and relations showed in ~\citet{2004ApJS..152..261W}:

\begin{figure*} 
\includegraphics[trim=0.2cm 0.cm 0.5cm 0.cm, clip=true, scale=0.7]{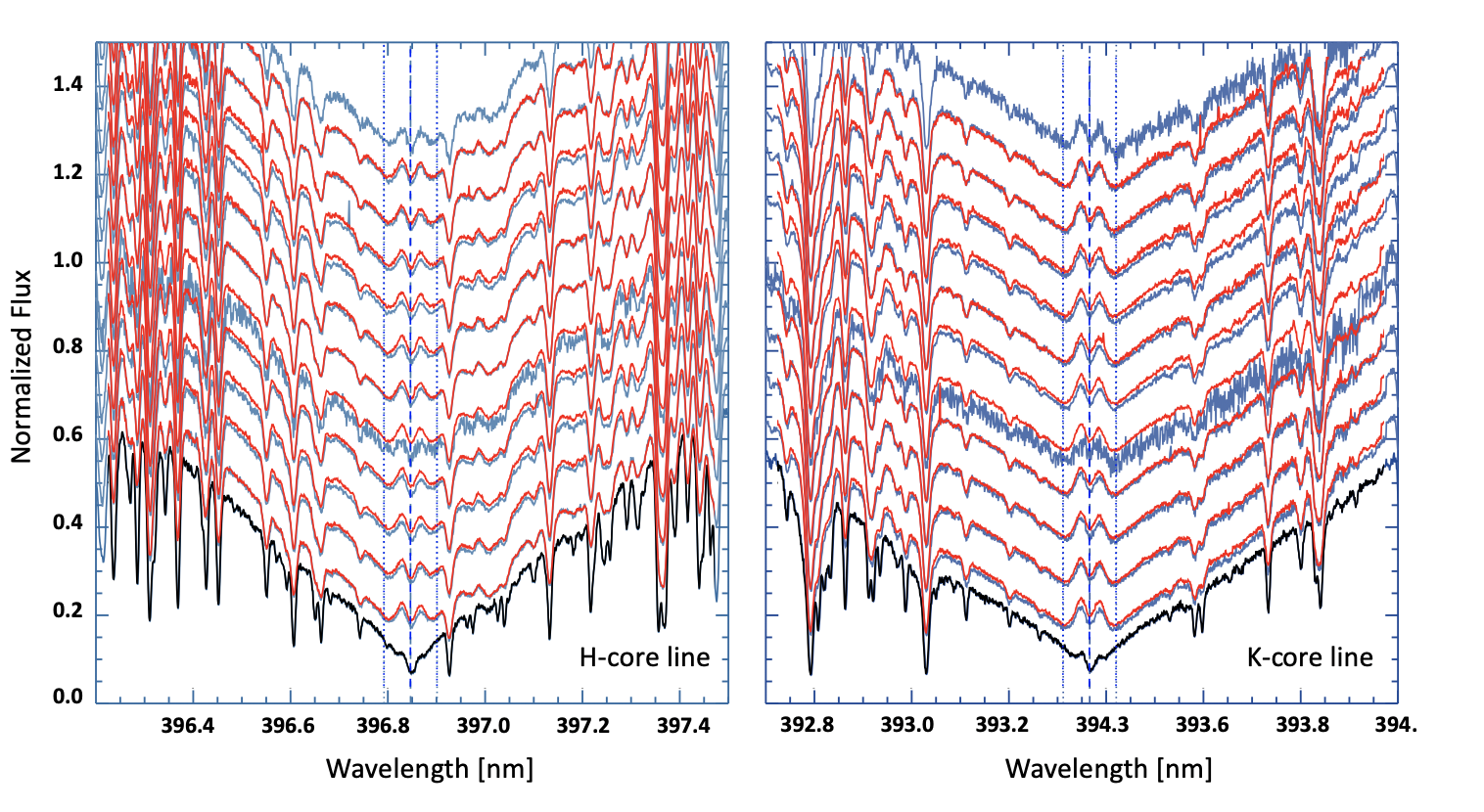}
\caption{\ion{Ca}{ii}$~\rm{H\,\&\,K}$ Fraunhofer core lines (left and right respectively) for the set of observations during the minimum period of activity in P$_{99- \rm A}$ (blue background plot) and comparison with the set of observations during maximum activity period in P$_{100- \rm C}$ (red over-plotted line), see comparative \ion{Ca}{ii} S-index \Shk\,  values in Figure~\ref{Fig1} for reference. The black spectrum at the bottom of both panels corresponds to the solar \ion{Ca}{ii}$~\rm{H\,\&\,K}$ at activity minimum, as obtained by the HARPS instrument on 12 April 2017.}\label{Fig2}
\end{figure*}

\begin{equation}
 S_{\rm HK} = \beta\left(\frac{{F_{\rm H} + F_{\rm K}}}{{F_{\rm R} + F_{\rm V}}}\right)\mbox{.}\label{Sindex}
\end{equation}

\noindent Here, ${F_{\rm H}}$ and $\,{F_{\rm K}}$ are the fluxes measured within $0.218~\rm{nm}$ wide triangular bandpasses centred on the line cores at $396.8469~\rm{nm}$ and $393.3663~\rm{nm}$, while 
$\,{F_{\rm R}}$ and ${F_{\rm V}}$ are the fluxes measured in two rectangular 2~nm wide bandpasses centred on the continuum on either side of the HK lines at  $400.107~\rm{nm}$ and $390.107~\rm{nm}$, respectively. The $\beta$ calibration factor depends on the implemented instrument. We used in this work the $\beta$ calibration coefficient for HARPS obtained by \cite{2015A&A...582A..38A}: \noindent $\beta_{\rm HARPS}\,=\,15.39\,\pm\,0.65$, with an offset of 0.058.

We estimated a percentile error of the \Shk\ (e\Shk) using the signal-to-noise of the spectral order corresponding to the $~\rm{H\,\&\,K}$ lines, as:

\begin{equation}
 eS_{\rm HK} = \frac{1}{\rm S/N}(S_{\rm HK}).\label{eSindex}
\end{equation}

In \textcolor{blue}{Paper\,I} we analysed the \Shk\ variability from the first half of the campaign (P${96}$ to P${98}$, blue points in Fig.~\ref{Fig1}), incorporating  into the analysis archive HARPS spectra, as well as, values obtained by SMARTS RC SPEC instrument reported in \citet{2010ApJ...723L.213M}, and, data by CASLEO/REOSC \cite[see][]{2017MNRAS.464.4299F}. The new \Shk\ values obtained during the ESO periods P$_{99}$ to P$_{101}$ are analysed here and listed in the journal of observations (see appendix section). 

The top panel of Figure~\ref{Fig1} shows the compilation of the four different datasets utilized for a multiparameter least-squares curve-fitting and the adjusted zero-level normalization (see, Table~\ref{Table:param}). 

We used the SYSTEMIC\,2 console package \cite[][]{2009PASP..121.1016M} to re-analyze the entire dataset, including the data from the second half of our HARPSpol campaign. The total data compilation comprises 439 \Shk\ values, which we used to perform a bootstrapping random periodicity analysis with $10^{4}$ trials, and allowing periods between p$_{\rm min}=$2.0~d and p$_{\rm max}=10^{4}$~d within a $10^{5}$ sampling grid. We removed the Earth's translational period of 365.25 days. We analysed the persistent periodicities in the long-term activity and found that a single activity cycle cannot explain the full dataset. Instead, the superposition of two sinusoidal functions provided a much better fit to the data (10 free parameters; see Table~\ref{Table:param}). The offsets in Table~\ref{Table:param} are the adjustment for de-trend and normalizing the different instrument sets of \Shk\, values. Additionally, we accounted for the scatter attributed to the stellar rotation (gray dots in\,Figure~\ref{Fig1}) by binning the instrument-specific data to the estimated rotation period of 7.8\,d obtained by applying the GPS method (see Sec.\,\ref{subsec_rotation}). We observed a reduced scatter in the amplitude of the time series, as well as a considerable improvement of the reduced $\chi^{2}_{r}$ of from 20.02 with the original data to 2.33 by fitting the averaged by-period time series (cyan line in\,Figure~\ref{Fig1}). However, we see that the phase and period of the fitted functions are not affected by the binning procedure, indicating the robustness of the solution. From the periodogram analysis (see Fig.\,\ref{Periodogram}), we found two superimposed periods: $P_{1}=547.65\pm3$d (1.49$\pm$0.01~yr) and $P_{2}=401.01\pm7$d (1.09$\pm$0.02~yr). 

\begin{table}
\caption{Two-period model for the \Shk\ activity evolution of $\iota$-Hor.}
\label{Table:param}   
\centering             
\begin{threeparttable}
\begin{tabular}{l c}    
\hline\hline         
Parameter & Value \\
\hline
& \\[-8pt]
\textit{1st Component} & \\
Period ($P_1$) & 547.653 $\pm$ 3 d (1.499 $\pm$ 0.012 yr)\\ 
Semi-amplitude ($A_1$) & 0.01414 $\pm$ 0.0010 \\
Phase ($\Phi_1$) & 4.73 $\pm$ 0.16 rad \\[3pt]
\textit{2nd Component} & \\ 
Period ($P_2$) & 401.012 $\pm$ 7 d (1.097 $\pm$ 0.023 yr)\\ 
Semi-amplitude ($A_2$) & 0.01556 $\pm$ 0.0008 \\
Phase ($\Phi_2$) & 3.56 $\pm$ 0.04 rad \\[3pt]
\textit{Offsets} $\left<S\right>_{\rm SET}$ & \\ 
CASLEO-REOSC (30)$^{a}$ & 0.2323 $\pm$ 0.0013 \\ 
SMARTS-RC Spec (143) & 0.2389 $\pm$ 0.0010 \\
HARPS PH3 (66) & 0.2438 $\pm$ 0.0011 \\ 
HARPSpol (199) & 0.2443 $\pm$ 0.0005 \\[3pt] 
\textit{Statistics} & \\ 
Reduced $\chi^2_{\rm r}$ & 2.33 \\ 
Data points & 439 \\
$\left<S\right>_{\rm ALL}$ & 0.2398 $\pm$ 0.011\\[1pt]
\hline
\end{tabular}
\begin{tablenotes}
{\small $^{(a)}$ Number of data points for a given instrument: REOSC at CASLEO, RC Spec at SMART and, HARPS at the 3.6m ESO telescope at La Silla.}
\end{tablenotes}
\end{threeparttable}
\end{table}

In contrast to the X-ray cycle of 588.5\,d (1.6~years) determined by \citet{2013A&A...553L...6S} and confirmed in \cite{2019A&A...631A..45S}, the total \Shk\ data showed a double periodicity; one of 1.49 years, closer to the periodic X-ray signal, and a secondary signal of 1.09~years. In the bottom panel of Figure~\ref{Fig1} we show the residuals from the optimized $\chi$ square fit, resulting in a double-period curve. We observe that the residuals are not totally flat, even with this optimal fit to the \Shk\ time series. Meanwhile, the section of \Shk\ data plot as purple points in Fig.~\ref{Fig1} are showing the modulation of the index over about one activity cycle synchronised with the X-ray observations with XMM-Newton telescope presented by \cite{2019A&A...631A..45S}. A possible reconciliation between having different cycles estimated by X-rays and chromospheric indexes could be due to a superposition of two predominant cycle modes (as has been suggested in \textcolor{blue}{Paper\,I} and also more recently in the \Shk\, cycle of $\kappa$~Ceti by \citealt{2022A&A...658A..16B}). In \textcolor{blue}{Paper\,I} was also proposed the possibility of a geometrical effect given by the misalignment of the observed features in the low chromosphere and the corona, where features located near the limbo edge can be traced in the corona but less clearly in the photosphere or chromosphere.

A comparison between the \ion{Ca}{ii}$~\rm{H\,\&\,K}$ line cores during minimum (P$_{99- \rm A}$, blue profile) and maximum activity periods (P$_{100- \rm C}$, red profile) is shown in Figure~\ref{Fig2}. The \ihor\ \ion{Ca}{ii}$~\rm{H\,\&\,K}$ line profile cores have a clear strong emission at both activity minimum and maximum periods. At activity maximum \ihor\ has higher levels of emission in the $~\rm{H\,\&\,K}$ cores at each epoch of observation, likely indicating a higher degree of coverage by chromospheric active regions at a range of longitudes.

\begin{figure*} 
\includegraphics[trim=.0cm 0.cm 0.0cm 0.cm, clip=true, scale=0.47]{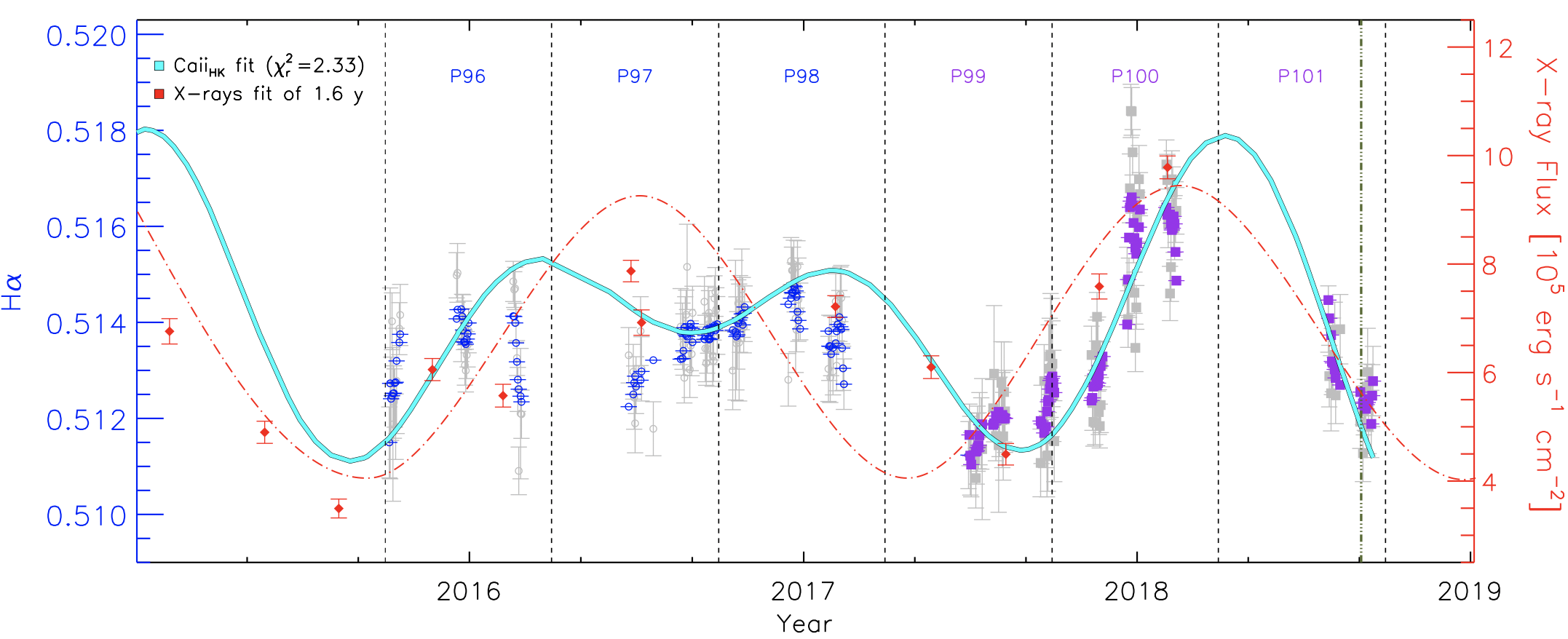}
\caption{Long term evolution of the H${\upalpha}$ index in \ihor. Gray dots are the original scattered data before the smoothing averaging over the rotational period. Blue open circles during observation periods P$_{96}$ to P$_{98}$ are reported in \textcolor{blue}{Paper\,I}. Purple-filled squares are the 98 values obtained during the observation periods P$_{99}$ to P$_{101}$. The cyan line corresponds to the reduced optimized $\chi$-square fit to the \Shk~dataset, resulting in a double-period curve. The red line is the 1.6-year fit from X-rays reported in \citealp[][]{2013A&A...553L...6S,2019A&A...631A..45S}.}
\label{Fig1_Ha}
\end{figure*}

In a similar way than performed for the \Shk, and in order to compare the magnetic behavior throughout other different stellar atmospheric layers, we analyzed the time series of H${\upalpha}$ emission as index I$_{\rm H\upalpha}$ and, the longitudinal component of the magnetic field $B_{\ell}$ in the following sections, see Figures~\ref{Fig1_Ha} and \ref{Fig1_Bl}, respectively. 

\subsection{H${\upalpha}$ emission as index I$_{\rm H\upalpha}$}\label{ha}

\begin{figure} 
\includegraphics[trim=.1cm 0.cm 0.5cm 0.cm, clip=true, scale=0.45]{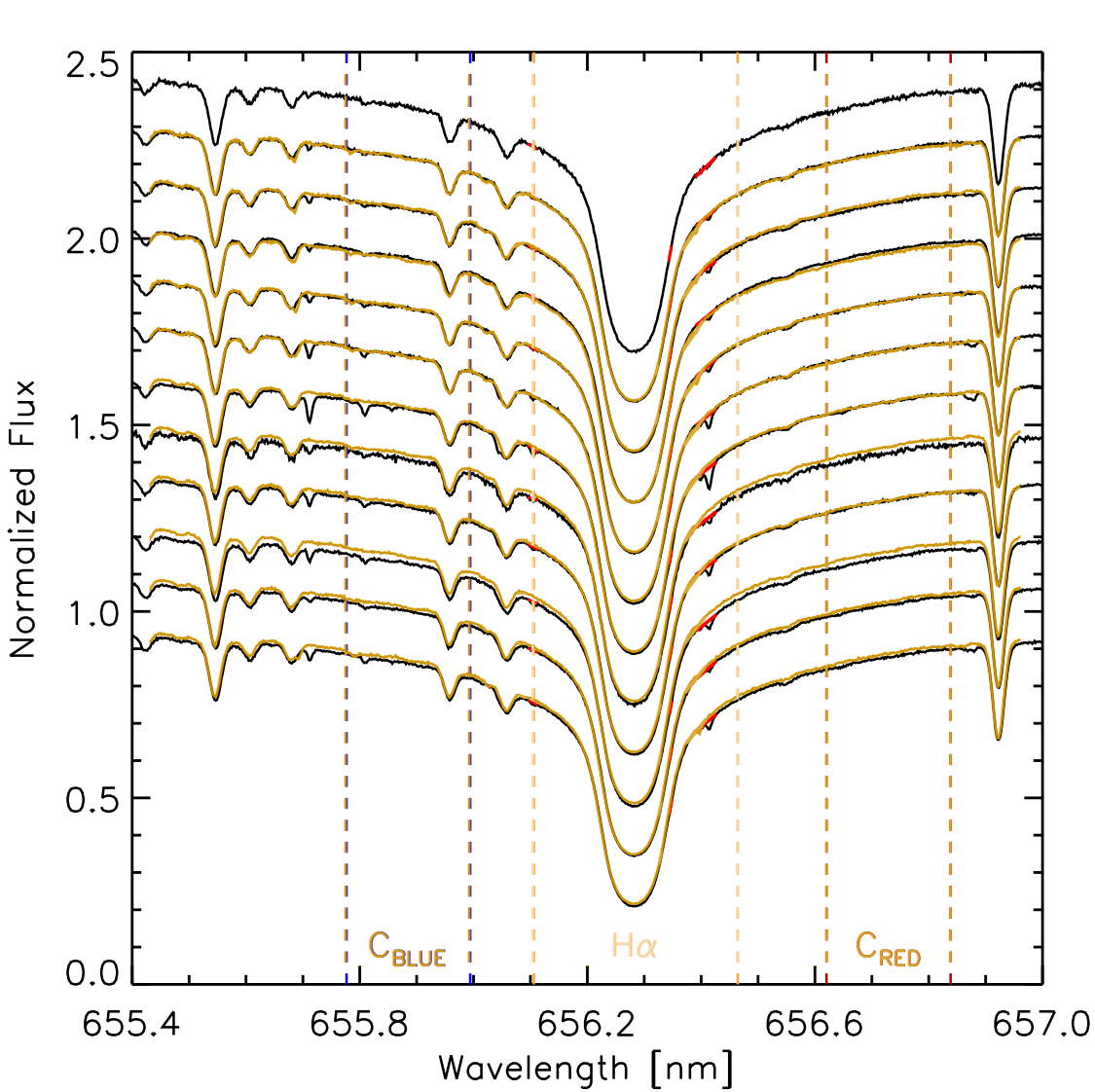}
\caption{H${\upalpha}$ lines for the set of observations during the period of minimum activity in P$_{99- \rm A}$ (black background plot) and comparison with the maximum activity period in P$_{100- \rm C}$ (golden over-plot). The corrections for telluric lines are shown in red; see Figure~\ref{Fig1} for reference.}
\label{Fig3}
\end{figure}

\begin{figure*} 
\includegraphics[trim=0.cm 0.cm 0.0cm 0.cm, clip=true, scale=0.46]{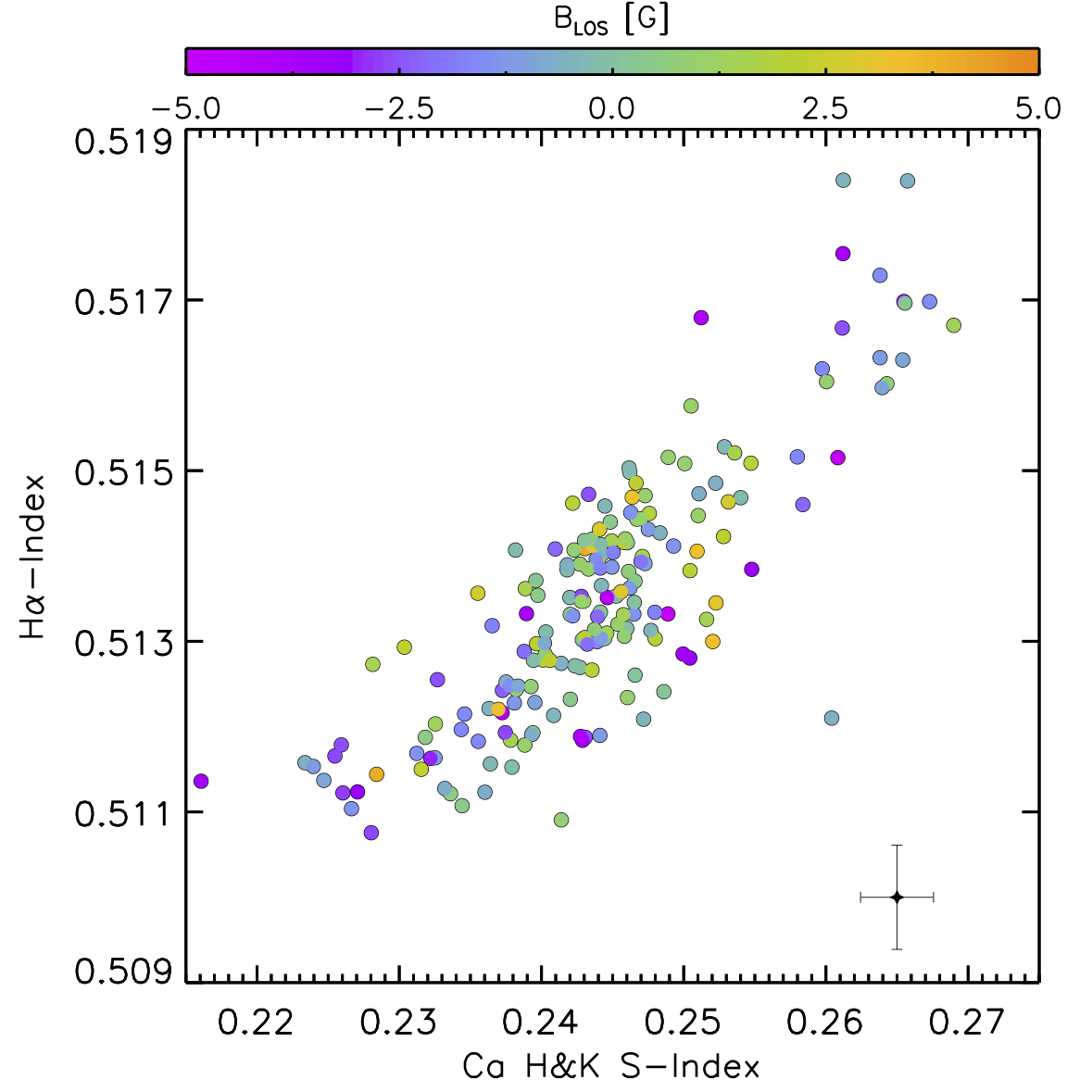}
\includegraphics[trim=3.15cm 0.cm 0.0cm 0.cm, clip=true, scale=0.46]{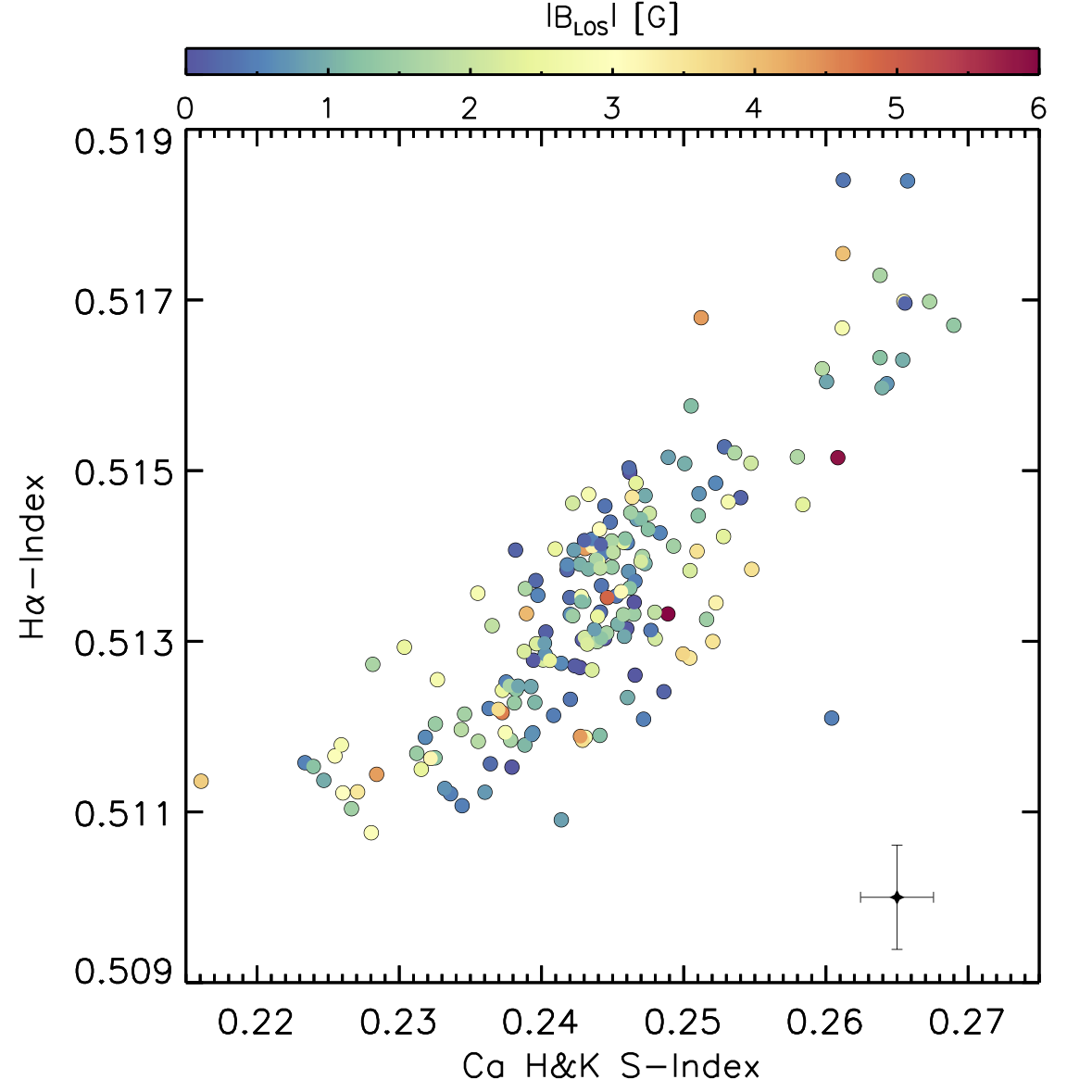}
\caption{Activity relation between I$_{\rm H\upalpha}$ and the \Shk\ index from the \ion{Ca}{ii}$~\rm{H\,\&\,K}$ cores. Left and right panels are coloured with B$_{\ell}$ and $|\rm B_{\ell}|$ respectively.}\label{Fig4}
\end{figure*}

\noindent The H${\upalpha}$ line centered at $656.2801~\rm{nm}$ provides another commonly used activity proxy of cool stars. The fluxes from the line wings and core trace emission from the upper photosphere and middle chromosphere, respectively. Multiple stellar features such as flares, prominences, spicules, filaments, etc., contribute to the emission of the H${\upalpha}$ line. For example, the core of the H${\upalpha}$ line, in particular its width and line shift, is an indicator of photospheric hot intergranular walls, known as the network or faculae if closer to a concentrated magnetic region. This line is also a tracer of the plages \cite[see,][]{1986ApJS...62..241L}. Similarly, the line has been used for tracing structures in rapidly rotating young convective stars, using "slingshot prominences" which appear as transient absorption features in H${\upalpha}$ \cite[see][]{2020MNRAS.491.4076J}. 

We calculated the H${\upalpha}$ activity indicator $I_{\rm H\upalpha}$ by integrating the flux over a $0.36$-nm bandpass around the H${\upalpha}$ line core at $656.2801$~nm (called $F_{\rm H\alpha}$), and dividing $F_{\rm H\alpha}$ by the flux integrated over two continuum regions of $0.22$-nm width on the red and the blue side of the H${\upalpha}$ line core: $C_{\rm B}$ at $655.885$ nm, and $C_{\rm R}$ at $656.730$~nm, as defined in \cite{2002AJ....123.3356G,2004ApJS..152..261W,2014MNRAS.444.3517M}, see Fig.~\ref{Fig3}:

\begin{equation}\label{eq2}
 I_{\rm H\upalpha} = \dfrac{F_{\rm H\alpha}}{C_{\rm B}+C_{\rm R}}\mbox{.}
\end{equation}

The relative positions and intensities of telluric to stellar lines, from one spectrum to another, can change due to different atmospheric conditions such as air mass and humidity during the observations. Those Earth atmospheric lines around the H${\rm \upalpha}$ region have been identified by using the Rowland table of “The solar spectrum” \cite{1966sst..book.....M}. The effects of the telluric lines inside the considered core and wing regions of H$\upalpha$ at about 655.8\,nm, 656.1\,nm, 656.4\,nm and 656.9\,nm (see Table~\ref{telha}1) are corrected by subtracting the area under the line profile and applying a cubic-spline fitting. We used a mask of width~W (in m\text{\normalfont\AA}) to correct the spectrum, as shown coloured in red in Fig.~\ref{Fig3}. 

We compare H${\rm \upalpha}$ line profiles from activity minimum and maximum (ESO P$_{99- \rm A}$ and P$_{100- \rm C}$) in Fig.~\ref{Fig3}. We do not observe a substantial difference in the line core for the different activity regimens, instead, we observe some slight excess towards the blue, $C_{\rm B}$ and, red $C_{\rm R}$ wing regions of the line that will make the biggest impact on the resultant indicator. 

Following Figure~\ref{Fig4} one can say that the H${\upalpha}$ index, I$_{\rm H\upalpha}$, is correlated with the chromospheric 
\ion{Ca}{ii}$~\rm{H\,\&\,K}$ index \Shk, which is quantified by a moderate positive Pearson correlation of $\rho_{S_{\rm HK}, I_{\rm H\upalpha}} = 0.82$. The correlation is very broad, given the high scatter on the H${\upalpha}$ index values. The values for $I_{\rm H\upalpha}$ obtained during ESO P$_{99}$ to P$_{101}$ are reported in Table~\ref{table_data_B3}2, and Table~\ref{table_data_B3}3.

\subsection{Longitudinal magnetic field}\label{sec_bl}

\noindent Circularly polarized light gives us a direct estimation of the magnetic field along the line-of-sight, $B_{\ell}$, unlike the indirect information obtained through the activity indicators discussed thus far (\Shk, $I_{\rm H\upalpha}$ and $L_X$). However, circularly polarised (Stokes V) signatures in Sun-like stars typically have relative amplitudes of 0.1\%. It is therefore not possible to obtain magnetic field measurements with a sufficiently high level of signal-to-background noise (S/N) in individual photospheric lines. We employ the multi-line technique of Least Squares Deconvolution, LSD, to produce a mean Stokes V signature with the co-added S/N of thousands of lines in order to robustly detect and analyse Zeeman signatures from stars like \ihor\ \cite[][]{1997MNRAS.291..658D,2010A&A...524A...5K}.

\begin{figure} 
\includegraphics[trim=0.2cm 0.2cm 0.2cm 0.2cm, clip=true, width=0.49\textwidth]{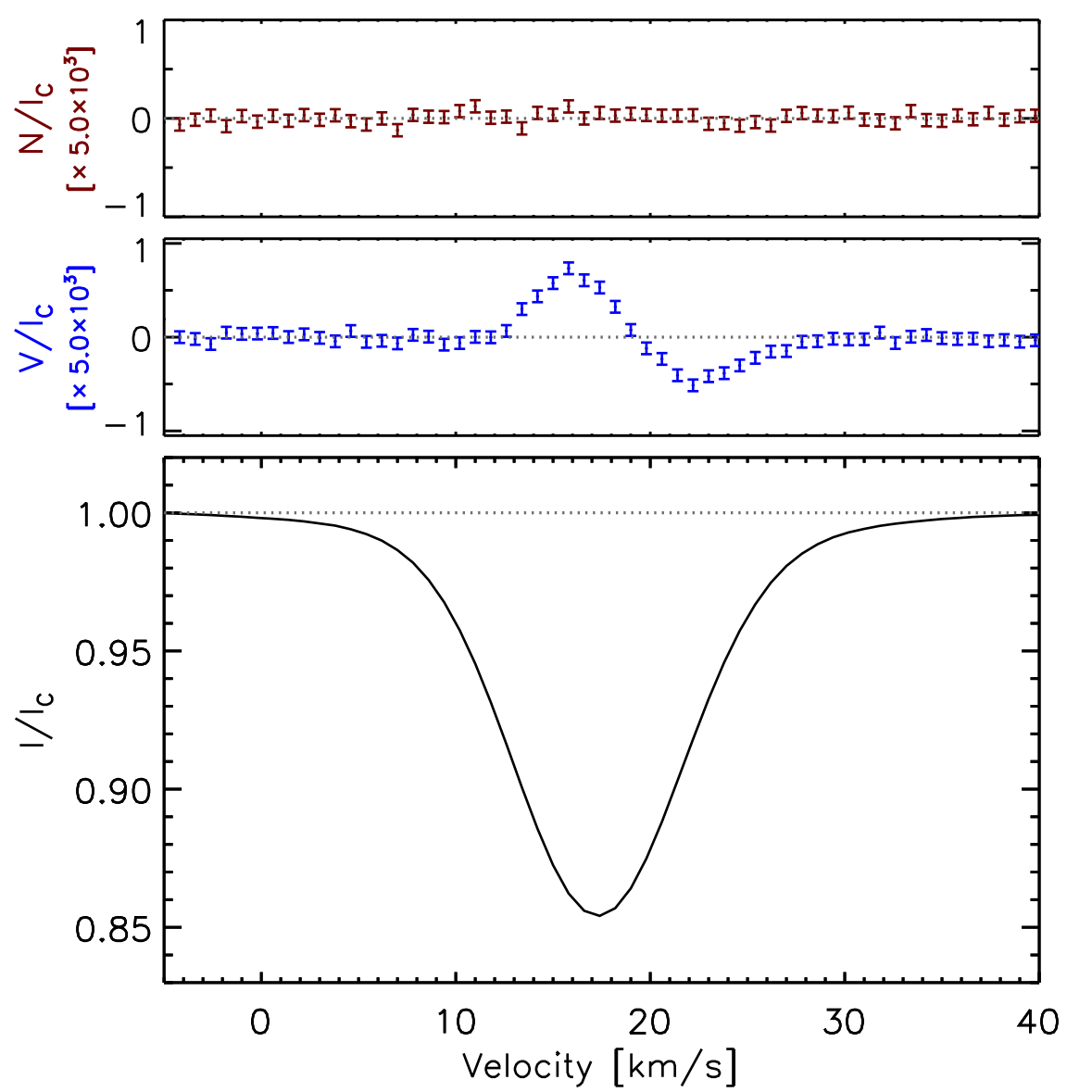}
\caption{Extracted LSD profile for \ihor~ from the HARPSpol observations acquired in 08.09.2018. The different panels show the intensity (I), circular polarization (Stokes V), and diagnostic null (N) profiles, normalized to the continuum intensity (I$_{\rm C}$). Both, LSD Stokes V and N, have been enhanced by a constant factor ($5\times10^{3}$) for visualization purposes.}
\label{Fig:LSD}
\end{figure}

\begin{figure*} 
\includegraphics[scale=0.47]{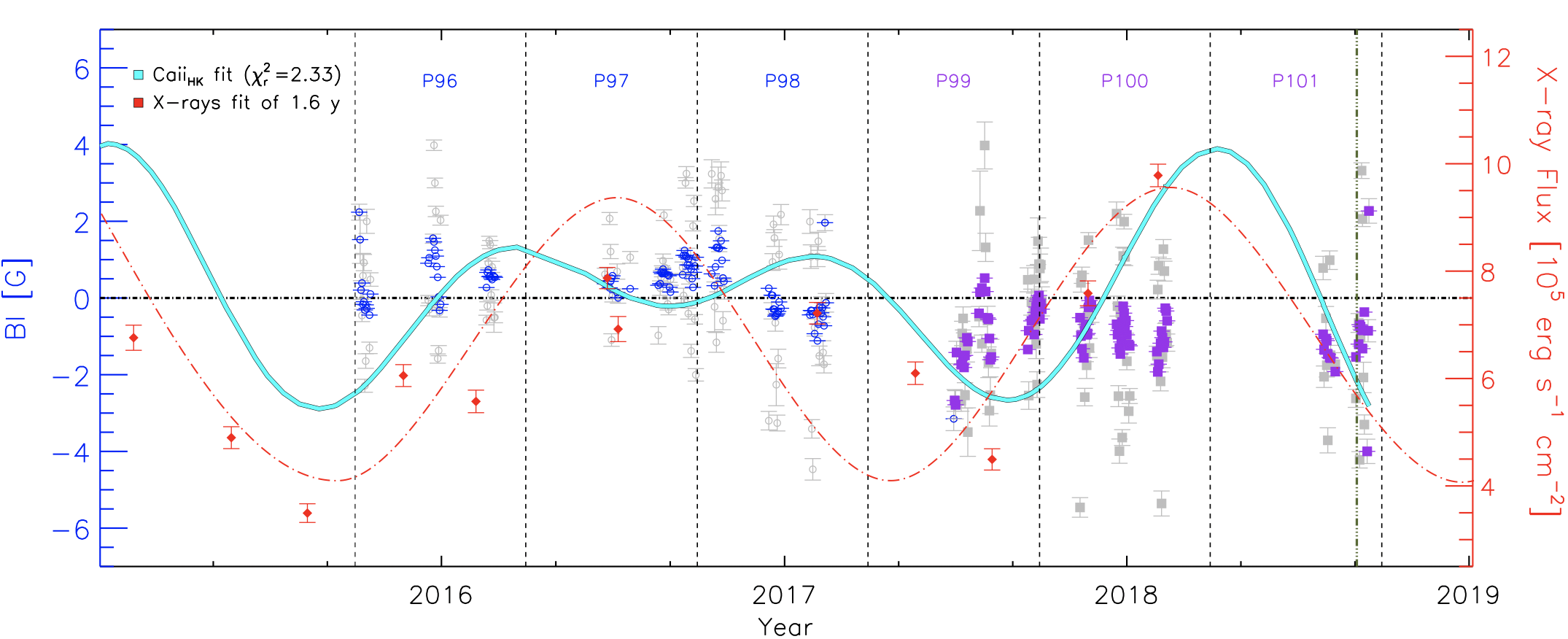}
\caption{Long term follow up of $B_{\ell}$ in \ihor. Gray dots are the original scattered data before the smoothing averaging over the rotational period. Blue open circles during observation periods P$_{96}$ to P$_{98}$ are reported in \textcolor{blue}{Paper\,I}. Purple-filled squares are the 98 values obtained during the observation periods P$_{99}$ to P$_{101}$. The cyan line shows the optimized reduced $\chi$-square fit to the \Shk~dataset, resulting in a double-period curve. The red curve is the 1.6-year cycle from X-rays reported in \citealp[][]{2013A&A...553L...6S,2019A&A...631A..45S}.}
\label{Fig1_Bl}
\end{figure*}

As described in the first paper of this study, a photospheric line list, tailored to the stellar properties of \ihor, is required to retrieve the LSD profiles on each night. This spectral line list, containing information on line depths, rest wavelengths, and Land\'e factors, is generated with the aid of \textcolor{black}{the Vienna Atomic Line Database \cite[VALD3][]{2015PhyS...90e4005R,2017ASPC..510..518P}. The mask computed corresponds to $T_{\rm eff}$= 6000 K, $\log(g)$ = 4.5 and microturbulent velocity of $1.0$~km s$^{-1}$)}. Literature values of $T_{\rm eff}$, $\log(g)$ are in Table \ref{tab_1}. The microturbulent velocity of $1.04$~km s$^{-1}$ is used as reference \cite[see][]{2010MNRAS.405.1907B}. A depth threshold of 0.05 in normalized units, was used to retrieve the line list from the database. The line mask was then refined following the method discussed in \citet{2015A&A...582A..38A}, matching the mask line depths to a high S/N HARPSpol spectrum of \ihor~(acquired during the first epoch of observations in Oct. 2015). The mask was then cleaned from very broad lines (e.g., Ca~H\&K, H$\alpha$) and line blends that would result in large deviations from the self-similarity assumption required by the LSD procedure \cite[see][]{1997MNRAS.291..658D,2010A&A...524A...5K}. The final optimized line mask for \ihor~had $8834$ entries within the HARPSpol spectral range. This optimized line mask was used to retrieve nightly LSD profiles within a range of velocities between $-5.0$~km~s$^{-1}$ and $45.4$~km~s$^{-1}$, with a velocity spacing appropriate for the HARPS spectrograph ($\sim0.8$~km~s$^{-1}$). The typical S/N in the retrieved LSD Stokes V profiles varied between $5\times10^4$ and $8\times10^{4}$, depending on the S/N of the original spectrum (see Tables~\ref{table_data_B3}1 and \ref{table_data_B3}2). An example LSD profile is presented in Figure\,\ref{Fig:LSD}.

Once we obtain the LSD profiles, we employ the \textcolor{black}{ center of gravity technique \cite[COG technique is a commonly used method to determine the center of a photospheric line profile. At sufficiently high spectral resolution, it can be used to determine the radial velocity of the star by measuring the center of the Stokes intensity profile in velocity space, see][]{0004-637X-592-2-1225} to measure the radial velocity and compute the disk-integrated longitudinal magnetic field. }

In the weak-field approximation regime, the longitudinal component of the magnetic field $B_{\ell}$, is linearly proportional to the amplitude of the Stokes $V$ parameter. In this regime, the Stokes V amplitude can be computed as follows:

\begin{equation}
\frac{V}{I} = -g_{\rm{eff}} C_{\rm{z}} \lambda ^2 \frac{{\rm d} I }{ I {\rm d} \lambda} B_{\ell},
 \label{weak_field1}
\end{equation}

\noindent where $V$ is the amplitude of the Stokes $V$ profile, $I$ is the intensity at a central wavelength $\lambda$, $C_{\rm z}\,=~467nm^{-1}~\rm G^{-1}$, and $g_{\rm eff}$ is the effective Land\'e factor (a dimensionless quantity describing the magnetic sensitivity, or response of the line, to the magnetic field).

The LSD Stokes $I(v)$ and $V(v)$ profiles represent the continuum intensity and circularly polarized profiles respectively. $B_{\ell}$ can be estimated (in Gauss) from the following expression:

\begin{equation}\label{eq3}
B_{\ell} = -714\dfrac{\int v V(v){\rm d}v}{\lambda_{0}\bar{g}\int\left[I_{\rm c} - I(v)\right]{\rm d}v},
\end{equation}

\noindent In our HARPSpol dataset the central wavelength is $\lambda_{0} \simeq 509\,$nm, the mean Land\'e factor for the integrated LSD Stokes I $\bar{g}\,\simeq\,1.198$, and $v$ is the velocity in km\,s$^{-1}$. 

The main source of uncertainty in $B_{\ell}$, $\sigma_{B_{\ell}}$, comes from the velocity limits over which the flux is integrated into Eq.\,\ref{eq3}, and the propagation of errors from the LSD technique (see \citealt{2014MNRAS.444.3517M}). We used a velocity range between $5.4$ and $30.2$ km~s$^{-1}$, which maximizes the $B_{\ell}/\sigma_{B_{\ell}}$ ratio and is sufficient to encompass the entire profile. We check for instrumental noise by applying LSD and COG techniques to the null spectra and compared them with the $B_{\ell}$ measurements. This enables us to evaluate whether our $B_{\ell}$ values have contributions from instrumental polarization. The resulting values of $B_{\ell}$, $\sigma_{B_{\ell}}$, and $N_{\ell}$, are listed in columns $10 - 12$ of Table~\ref{table_data_B3}2, and Table~\ref{table_data_B3}3.

We look for quantified correlations between the different activity indicators \Shk, I$_{\rm H\upalpha}$, $B_{\ell}$ and the absolute value of the longitudinal magnetic field $|B_{\ell}|$. \textcolor{black}{The last one, even though suffering from cancellation effects, allows us to compare the magnitude of the radial field with the activity indicators (see Fig.~\ref{Fig4}).} We observe a clear correlation between the \Shk\, and I$_{\rm H\upalpha}$ indicators, expressed with a Pearson coefficient of $\rho_{(S_{\rm HK},I_{\rm H\upalpha})} = 0.82$, as described in Sec\,\ref{ha}. Such a correlation is not evident between $B_{\ell}$ and the activity indexes. For example, the coefficient of correlation between $B_{\ell}$ and \Shk\,  is quantified by a very weak positive Pearson correlation of $\rho_{(B_{\ell},S_{\rm HK})} = 0.025$. In the case of $B_{\ell}$ and I$_{\rm H\upalpha}$ the coefficient is $\rho_{(B_{\ell},I_{\rm H\upalpha})} = 0.046$. For the values obtained between $B_{\ell}$ and $|B_{\ell}|$ we do observe a slightly higher anti-correlation ($\rho_{(B_{\ell},|B_{\ell}|)} = -0.26$). The compilation of correlation coefficients between the different indicators of activity, and the RVs calculated in the following section are organized in Table~\ref{PiersonCoeff}.

\subsection{Radial velocity evolution}
\label{sec:RV_evolution}

For the analysis of the radial velocity (RV) evolution presented in this section, we included pipeline-processed spectroscopic HARPS~PH3 archival data\footnote{\url{http://archive.eso.org/wdb/wdb/adp/phase3_main/form}} acquired between November 2003 and December 2016 ($60$ spectra), together with the HARPSpol dataset acquired between October 2015 and September 2018 ($199$ spectra), resulting in $259$ data points for the radial velocity time series.

\subsubsection{Least-Squares deconvolution and bisector profiles} 

The RVs are measured from the LSD profiles. As outlined in Paper\,I, the RV of each star was measured by extracting the LSD Stokes\,I profile and measuring the centroid from a least-squares Gaussian fit to the data. We found that these measurements were very consistent with RVs derived from ISPEC, using the fitting procedure of the cross-correlation function from the HARPS/SOPHIE G2 line mask with typical RV errors of 2\,ms$^{-1}$. To measure the asymmetry of the line profile, which could be due to physical mechanisms related to the stellar activity (e.g. starspots), we also computed the bisector as defined in \citet{2015MNRAS.451.2337S}, and the value of $BIS$, i.e. the velocity span between the average of the top and the bottom parts:

\begin{equation}
\label{eq_BIS}
  BIS=BIS_{\rm top}-BIS_{\rm bottom},
\end{equation}

where $BIS_{\rm top}$ is the average of the bisector between 60 and 90\% of the total contrast in the flux of the line profile and $BIS_{\rm bottom}$ is the average of the bisector between 10 and 40\% of the total contrast, where 100\% corresponds to the continuum and 0\% to the minimum of the line profile. 

Across the whole time series, the value of the $BIS$ does not exceed $143~\mathrm{m\,s^{-1}}$, with an average value of $\overline{BIS}\approx92.1~\mathrm{m\,s^{-1}}$, which represents about $82$\% of the overall measured RV variation. \textcolor{black}{The fact that the BIS value has the same order of magnitude than the RV variation indicates that the asymmetry of the LSD profile induced by the stellar activity cannot be neglected.} By applying the technique to three different values of the spectral binning of the LSD profile in velocity-space, $\updelta v\:=0.8$, $0.4$ and $0.2~\mathrm{km/s}$, we confirmed that the choice of $\updelta v$ does not affect the calculation of the $BIS$ value. The results shown here are based on the $BIS$ values derived for $\updelta V\:=0.8~\mathrm{km/s}$.

\begin{figure*}
\includegraphics[width=\textwidth, clip=true]{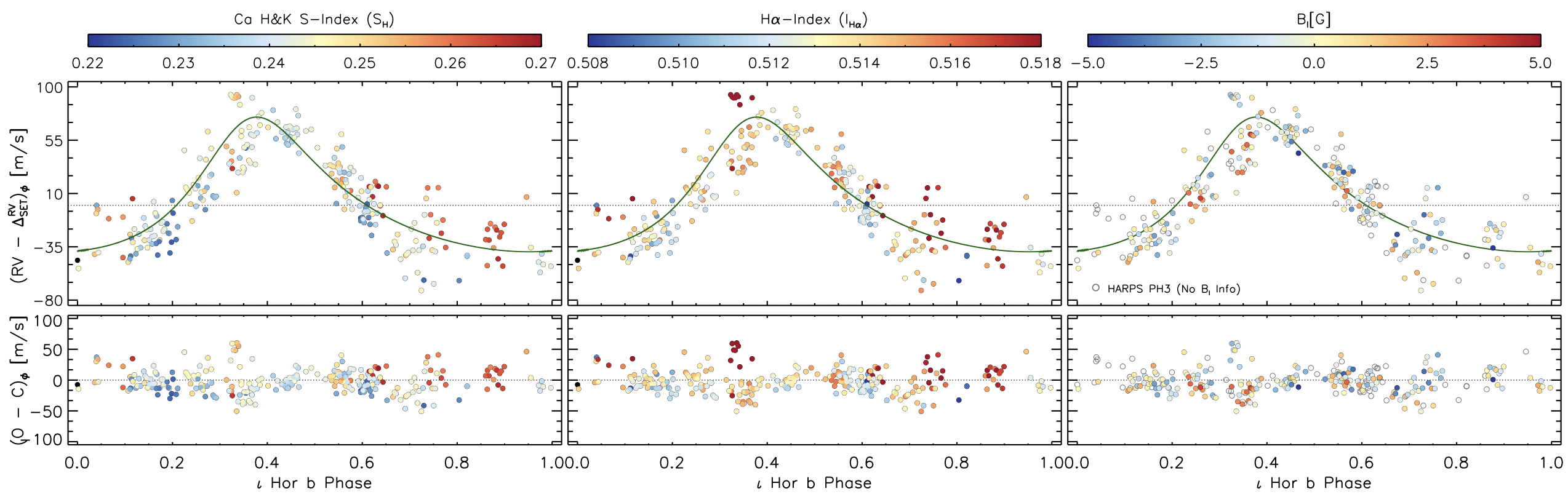}
\vspace{-10pt}\caption{RV variations of $\upiota$~Hor determined from the combined HARPS~PH3 and HARPSpol datasets. The phase and orbital solution (solid line) have been fitted using the parameters reported by \citet{2013A&A...552A..78Z}. The residuals are shown in the bottom panel. The color codings denote the corresponding values of \Shk\ (\emph{left}), I$_{\rm H\upalpha}$ (\emph{middle}), and $B_{\ell}$ (\emph{right}) for each observation (when available).}\label{Fig6}
\end{figure*}

In \textcolor{blue}{Paper\,I}, \cite{AlvaradoGomez2018} first noted a hint of an \emph{activity gradient} in the RV-subtracted residuals, which we can further demonstrate with this more complete dataset (see \emph{left} and \emph{middle} panels of Figure~\ref{Fig6}). The stellar activity levels, as traced by the chromospheric spectral indices, follow the distribution of RV residuals, with the (O-C) values tending to be positive during higher levels of \ion{Ca}{ii}$~\rm{H\,\&\,K}$ \Shk\ and the I$_{\rm H\upalpha}$, and negative (O-C) values corresponding to lower activity levels. On the other hand, the disk-integrated longitudinal magnetic field $B_{\ell}$ (Fig.~\ref{Fig6}, \emph{right}), tracing the larger scale activity \emph{jitter}, shows no apparent coherence with the RV variations, confirming that $B_{\ell}$ does not trace the global activity level in the same way that the chromospheric activity indicators do. While this dependence found in the residuals confirms the presence of activity-related {\em jitter}, the upper panels in Fig.~\ref{Fig6} show no such dependence on activity, confirming that the RV variability in $\upiota$~Hor is caused by the presence of a planet. 

\section{TESS photometry}\label{sec_TESS}

Building on the legacy of the Kepler mission, the Transiting Exoplanet Survey Satellite TESS \citep{2015JATIS...1a4003R} -- also a planet-hunting mission -- has enhanced our ability to explore the photometric variability of about 200 000 bright nearby stars across the whole sky in an area 400 times larger than that covered by the Kepler mission \citep{2010Sci...327..977B}. The TESS mission has allowed us in this work to analyse simultaneously high-resolution photometric, spectropolarimetric and UV observations for \ihor. In this section, we analysed the \ihor\ light curves obtained by the TESS mission for two consecutive sectors, S2 and S3, from 23 August 2018 to 17 October 2018. We also pre-analysed sectors S29 and S30, but given the high level of noise present in those sectors, we focus here on data from S2 and S3, which were also acquired closer to the dates of our simultaneous HST NUV/FUV and spectropolarimetric observations.

TESS light curves were obtained from the High-Level Science Products (HLSP) on the Mikulski Archive for Space Telescopes (\href{http://archive.stsci.edu/}{MAST}). In order to improve the analysis for the TESS dataset we stitched, normalized, de-trended, and reduced the noise on sectors S2 and S3 following the procedure as described in \cite{Almeida,2022AJ....163..257A}. In this process, we derived the TESS long-cadence (30-min and 10-min cadence) light curves for $\iota$~Hor by performing the photometric extraction from the full-frame images (FFI) data. First, we downloaded a $35 \times 35$ pixel cutout image centered on our target using the \texttt{TESScut} tool on MAST servers \citep{Brasseur2019}. Next, we applied a constructed custom mask aperture that includes each pixel from the target star and extracted the data from the observed fields \texttt{TIME} [BJD], \texttt{FLUX} [e$^-$/s], and \texttt{FLUX\_ERR} [e$^-$/s]. Then, we constructed vectors with observing times, measured fluxes, and flux uncertainties, respectively. Following the same steps, we also estimate the background signal considering the surrounding pixels of the object where the point-spread function appears to be zeroed. Before de-trending the light curve we performed a background subtraction. The resulting light curve is divided by the median and converted into normalized fluxes.

\subsection{Faculae to spot ratio}

Starspots and faculae imprint characteristic patterns on photometric time series, diminishing or enhancing the observed stellar brightness, while transiting the stellar disk. In this section, we focus on recovering a quantitative description of the stellar surface through the facular-to-spot ratio ($\rm S_{fac}/S_{spot}$), by implementing the gradient of the wavelet power spectra method \cite[GPS, see][]{PaperI,Eliana1,Eliana2}. Precise information on the stellar rotation period is required in order to properly analyse the stellar surface and the $\rm S_{fac}/S_{spot}$. Consequently, we implement independent methods to retrieve the stellar rotation period and compare our findings with previous analysis in the literature.

\subsubsection{Rotation Period}\label{subsec_rotation}

The stellar rotation period is key for determining a number of stellar properties and important for recovering an estimate of the distribution of features over the stellar surface. In order to obtain a better value of this parameter, we applied seven different methods to recover photometric periodicities related to the stellar rotation period. \textcolor{black}{We used quasi-periodic Gaussian Processes \cite[QP-GP, see][]{2017A&A...599A.126D,celerite1,2018RNAAS...2...31F,2020A&A...634A..75B}, the Generalized Lomb-Scargle periodogram \cite[GLS, see][]{2009A&A...496..577Z}, the Autocorrelation Function \cite[ACF, see][]{2013MNRAS.432.1203M,2014ApJS..211...24M,Santos_2019}, Wavelet Power Spectra \cite[PS, see][]{1998BAMS...79...61T,2014A&A...572A..34G,Santos_2019}, and the Gradient of Power Spectra \cite[GPS e.g., see][]{PaperI,Eliana1,Eliana2,ElianaThesis}. We applied those different methods (QP-GP, GLS, ACF, PS, and GPS) }to analyze the combined, normalized, and stitched sectors 2 and 3 on TESS photometry (from 23 August 2018 to 17 October 2018, see Table\,\ref{tab_Periods}. Additionally, we applied the GP method independently for sectors, S2 and S3. We do not observe strong deviations from any of the methods when applied to S2 and/or S3. More detailed descriptions of the different methods that we used can be found in Appendix\,\ref{App:rotation}. 

In Table~\ref{tab_Periods} we compile the results from the methods applied and show the results from the GLS, ACF, PS, and GPS methods in Figure~\ref{fig:IotHor_GPS}. The rotational modulation from the TESS lightcurves is consistently recovered by all seven of the implemented methods. The five methods applied to the normalized and stitched S2\,+\,S3 LCs gave us an estimate of the rotation period ranging from 7.23 to 7.94 days. Those values obtained are in agreement with the 7.718\,$\pm$\,0.007\,d obtained for the combined S2\,+\,S3 LCs analysed in \citet[][]{2019A&A...631A..45S}. The two additional methods applied independently to S2 and S3 retrieved values between 6.64\,d to 7.48\,d for S2, and 7.01\,d to 7.24\,d for S3. We noted that this range of periods is lower than the average period of 8.43\,$\pm$\,0.02\,d for S2 and 7.95\,$\pm$\,0.02\,d for S3 previously recovered from the GLS analysis of the pipeline-reduced TESS lightcurves in \citet[][]{2019A&A...631A..45S}. We considered that this discrepancy in the analysis of the individual sectors might be due to the different lightcurve extraction procedures. In \citet[][]{2019A&A...631A..45S}, lightcurves were extracted after testing different apertures constructed from a summed image of all cadences from a particular sector and choosing the aperture for which the lightcurve showed the lowest standard deviation. As described earlier, the lightcurves we use here were extracted after subtracting the background signal measured from surrounding pixels where the point spread function from the target should be zero.

\begin{table} 
\centering
\begin{tabular}{l c c r}
 \hline\hline
 Method & Prot & Error \\
     & [days] & [days]\\
 \hline\hline
  & & \\[-2pt]
 GP (S2) & 6.64 & $^{+0.28}_{-0.09}$ \\ 
 GP (S3) & 7.01 & $^{+0.16}_{-0.09}$ \\   
 QP-GP (S2 + S3) & 7.23 & 0.20\\
 GLS (S2 + S3) & 7.73 & 0.21\\
 ACF (S2 + S3) & 7.69 & 0.23\\
 PS  (S2 + S3) & 7.94 & 0.32\\
 GPS (S2 + S3) & 7.78 & 0.18\\ 
 \hline
 \end{tabular}
 \caption{Rotation period analysis from different independent methods.} \label{tab_Periods} 
 \end{table}

\subsubsection{Gradient of Power Spectra, GPS method}

Recent studies have suggested that our Sun exhibits particular characteristics that differentiate it from even its closest stellar analogues. For example,  \citet{2016ApJ...826L...2M} suggest that Sun-like stars in the middle of their main sequence life could start showing hints of a dynamo shutdown, and transition towards a different dynamo regime. \citet{Reinhold518} found that the solar variability seems to be an outlier if compared against stars with similar temperature, age, and rotation periods (when available for those slowly rotating stars).

Further analysis by \citet{PaperI,ElianaThesis} showed that the rotation period of stars with complex light curves, as for the solar case, can be reliably determined by implementing a novel technique based on the profile of the gradient of power spectra, GPS. In \citet{Eliana1} the rotation modulation is detected at all solar activity levels, where other methods have failed. By characterizing the particular shape generated by facular (M-like shape) or spot (V-like shape) transits, recorded in the total solar irradiance (TSI) and compared simultaneously with MDI observations, it was possible to infer whether the stellar surface was dominated by facular or spot regions. The new method developed in \citet[][]{PaperI,Eliana1,Eliana2,ElianaThesis} allows us not just to infer, but to quantify the degree of spot- or faculae-dominance on the stellar surface based on solar and stellar light curves. The quantification is made through the ratio between the bright and dark features, $\rm S_{fac}/S_{spot}$. Through the application of the GPS method to both solar-like LC simulations and TSI observations, it has been found that Sun-like stars exhibit distribution across three distinct regimes. Those regimes characterize either the dominance by the presence of spots or faculae on the stellar light curves and, stars that seem to be in a transition between the two branches. Interestingly, a detailed characterization of the solar brightness variations found that the Sun lies in the middle of the branch between the spot- and faculae-dominant regimes. The entire methodology is described, tested and applied in \citet[][]{PaperI,ElianaThesis}. 

\begin{figure*}
\centering 
\includegraphics[trim={0 0 0 0cm},clip,width=0.74\textwidth]{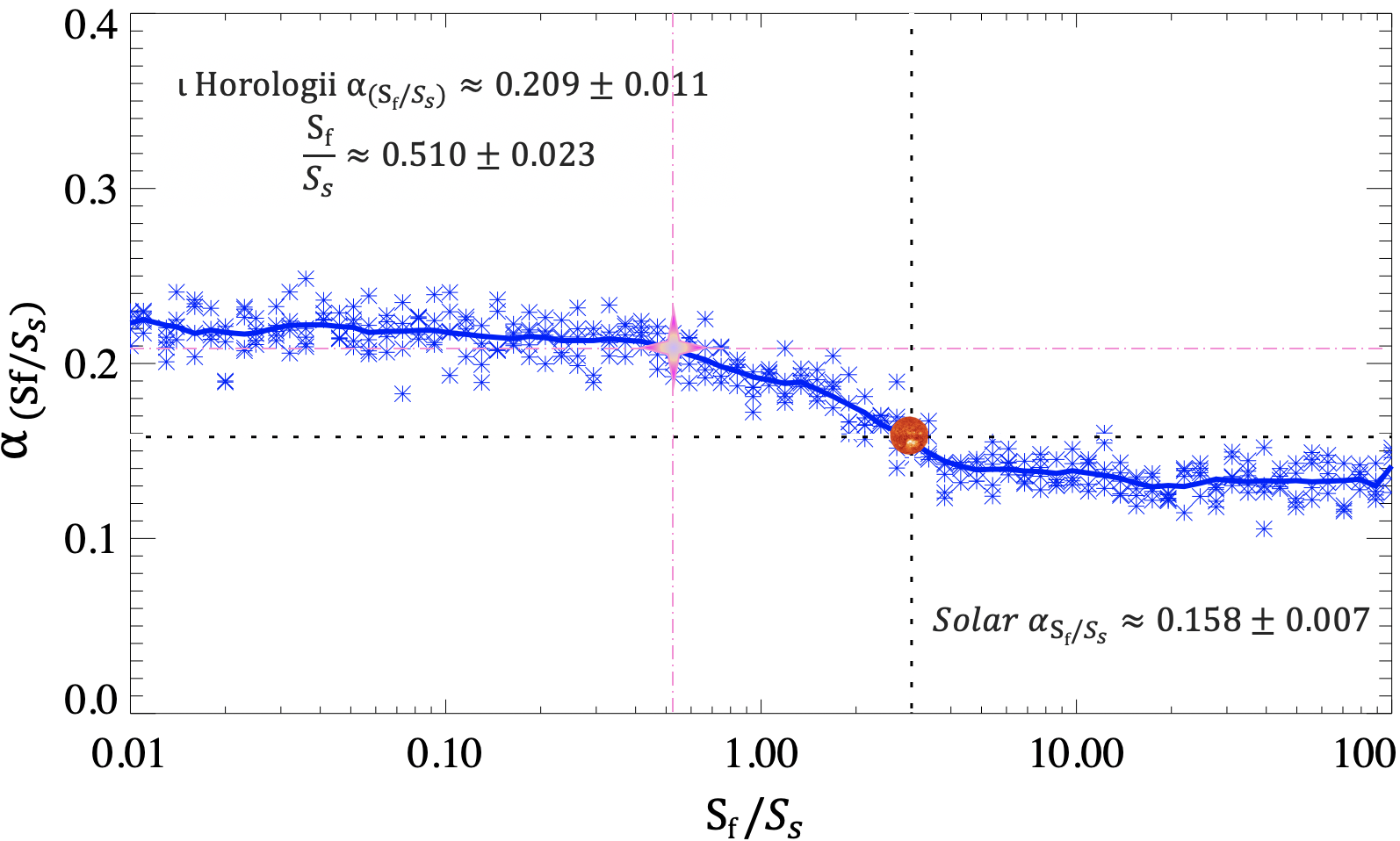}\hspace{0.0cm}
\caption{In blue the GPS $\alpha$-Factor of 400 modeled faculae to spot ratio ($\rm S_{fac}/S_{spot}$) adapted from \citet{PaperI}. The $\alpha$-Factor is proportional to the high-frequency inflection point found in the gradient of the power spectra, GPS, and inverse to the rotation period, as $\alpha$-Factor $= {\rm HFIP}/P_{\rm rot}$. The Sun is located in the transition region between the branches of spot-dominated (left) and faculae-dominated stellar surfaces (right). In \citet{Eliana1}, we found an $\alpha_{\odot}$=0.158 which corresponds to a $\rm S_{fac}/S_{ spot}\sim3$. The pink cross shows the position of \ihor\ and demonstrates that it is located in the spot-dominated branch of the diagram with a $\alpha_{\iota}=0.209 \pm 0.011$ and a $\rm S_{fac}/S_{spot}$=$0.510 \pm 0.023$.}
\label{Fig8}
\end{figure*}

From the stellar photometric brightness variations of \ihor\ we derived the GPS $\alpha$-factor and compared this against the faculae-to-spot driver ratio ($\rm S_{fac}/S_{spot}$) for 400 modeled light curves analysed in \citet[][see Fig.\,\ref{Fig8}]{PaperI}. The $\alpha$-factor is proportional to the high-frequency inflection point (HFIP\,=\,1.63\,d) found in the GPS, and inversely proportional to the rotation period. It is expressed as: $\alpha$-F$\,=\,{\rm HFIP}/P_{\rm rot}$. We found that \ihor\ has a high alpha-factor $\alpha$-F\,$=\,0.209\,\pm\,0.011$ (c.f. 0.158\,$\pm$\,0.007 for the Sun). As shown in Fig.~\ref{Fig8}, \ihor\ (pink cross) is located in the spot-dominated branch of the diagram of $\alpha$-Factor vs. $\rm  S_{fac}/S_{spot}$. This indicates that the stellar surface is spot dominated, with a facular-to-spot ratio ($S_{\rm fac}/S_{\rm spot}$) of $0.510\,\pm\,0.023$.

\section{HST/STIS data}\label{HST}

\begin{figure*}
\centering 
\includegraphics[trim={0 0 0 1cm},clip,width=0.45\textwidth]{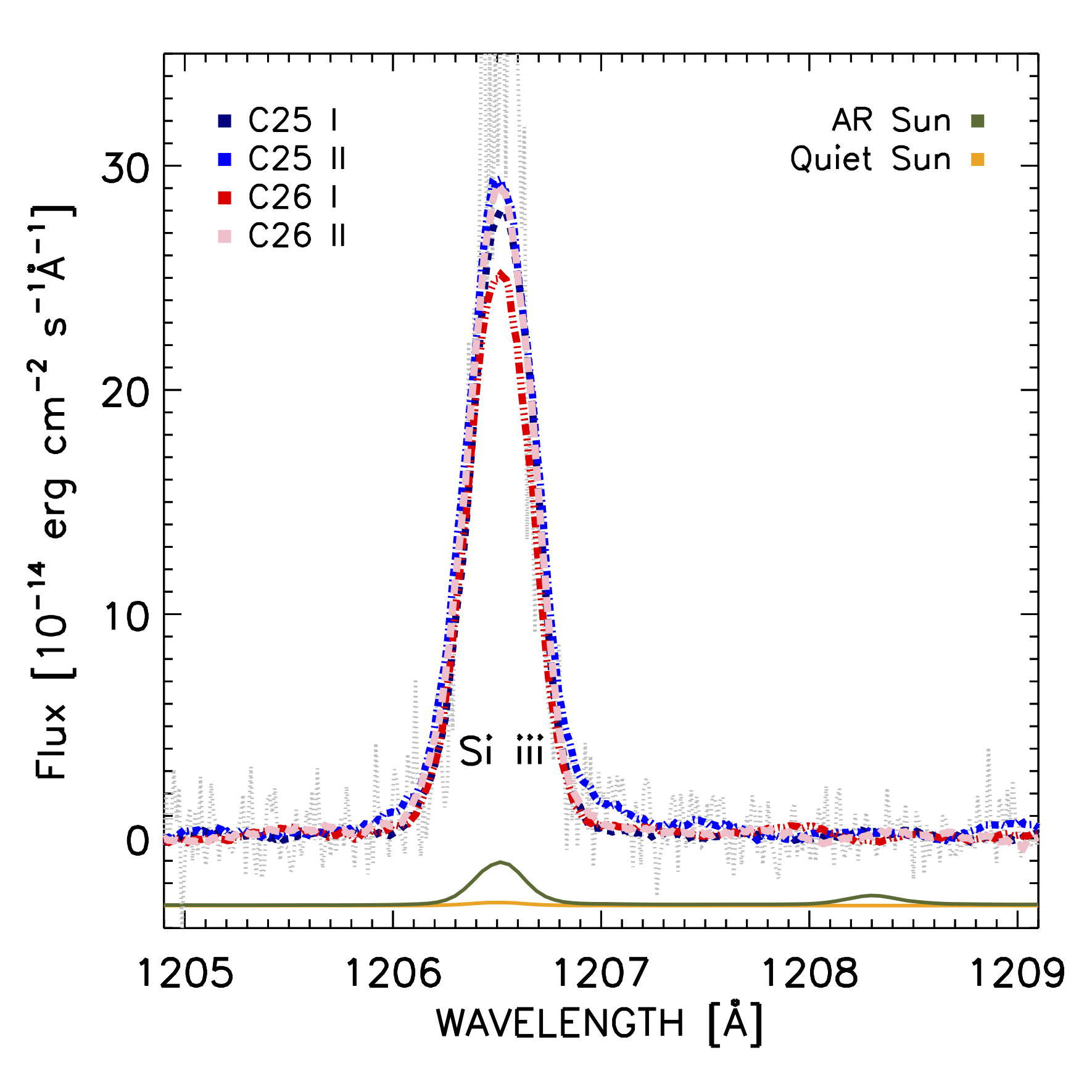}\hspace{0.0cm}
\includegraphics[trim={0 0 0 1cm},clip,width=0.45\textwidth]{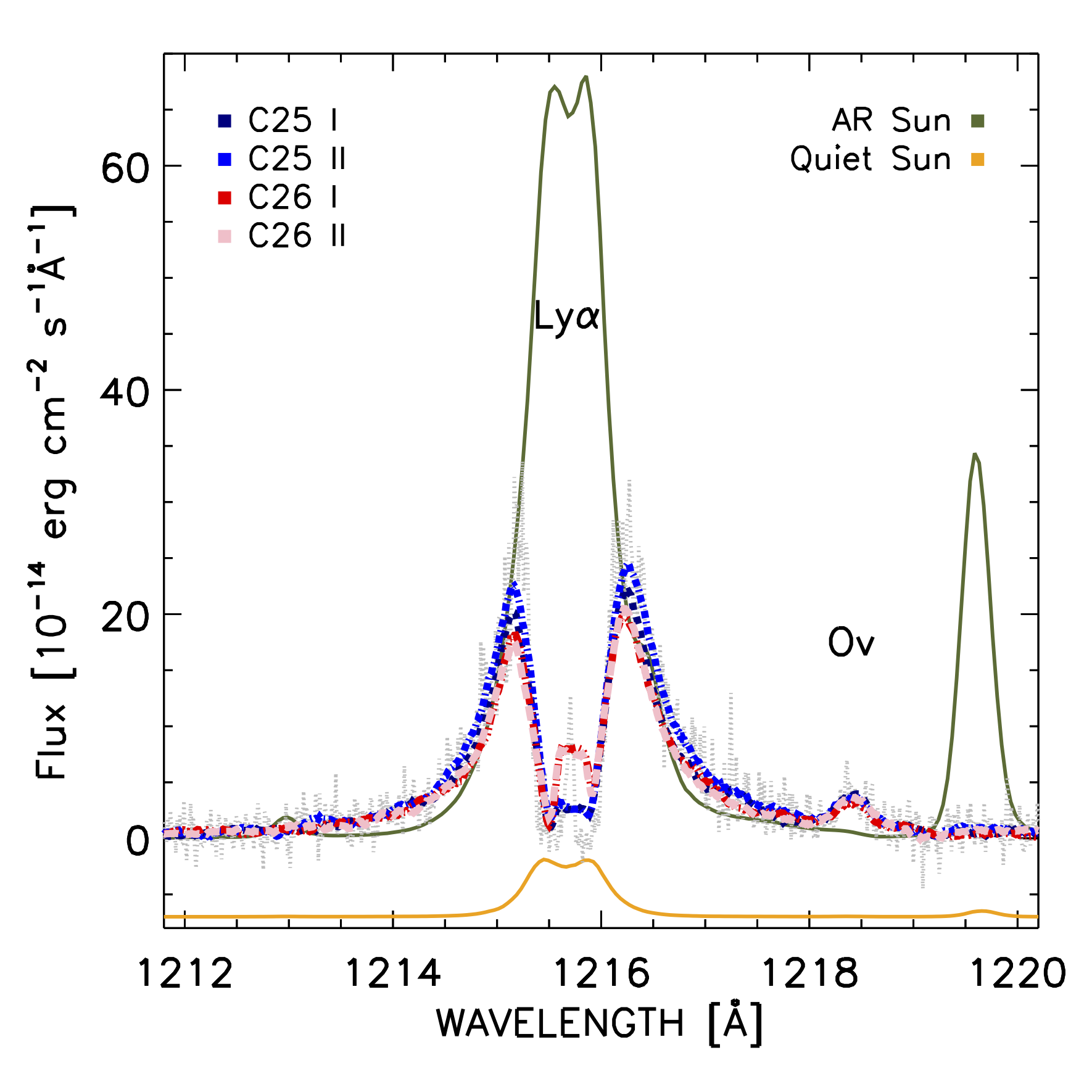}\hspace{0.0cm}
\includegraphics[trim={0 0 0 1cm},clip,width=0.45\textwidth]{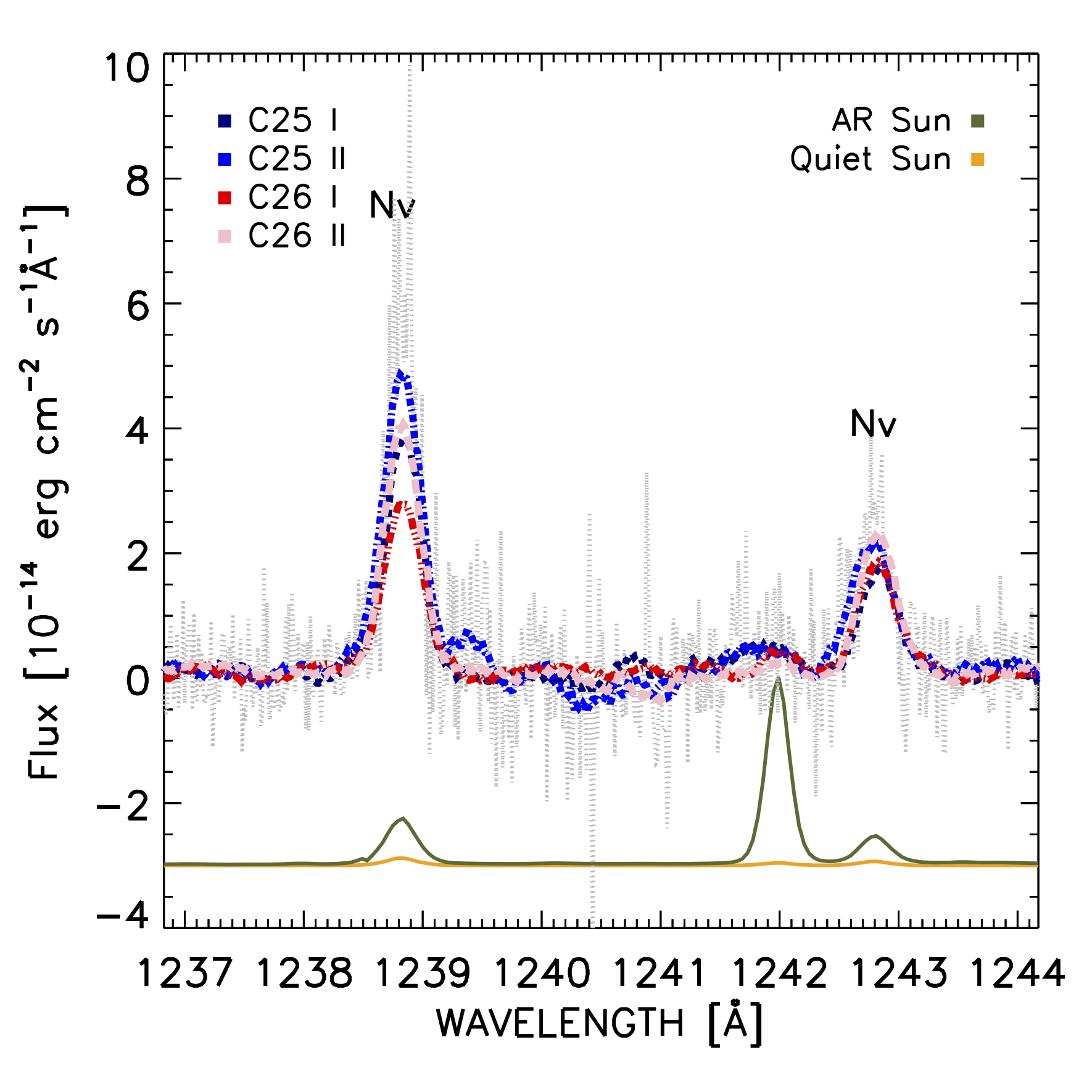}\hspace{0.0cm}
\includegraphics[trim={0 0 0 1cm},clip,width=0.45\textwidth]{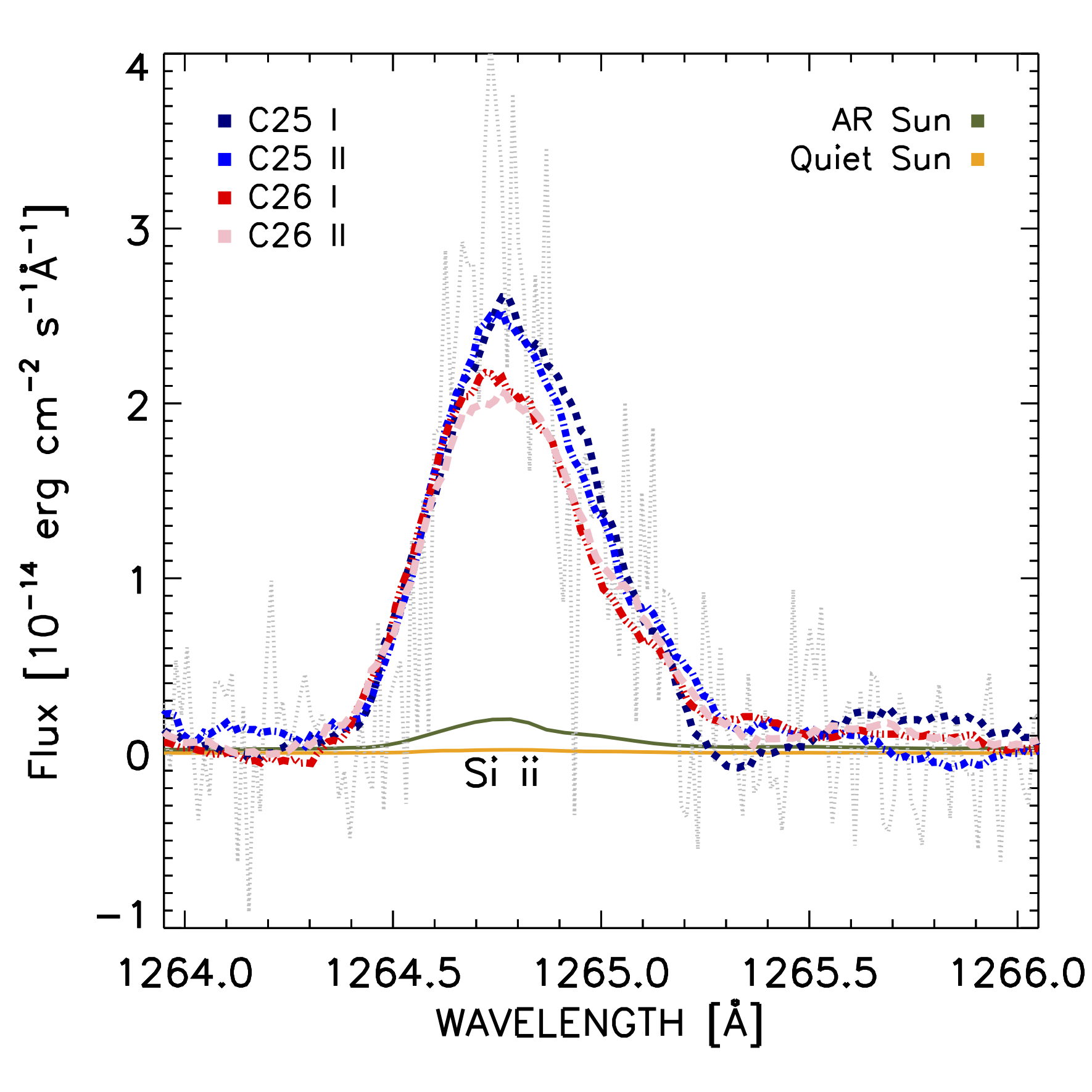}\hspace{0.0cm}
\includegraphics[trim={0 0 0 0cm},clip,width=0.45\textwidth]{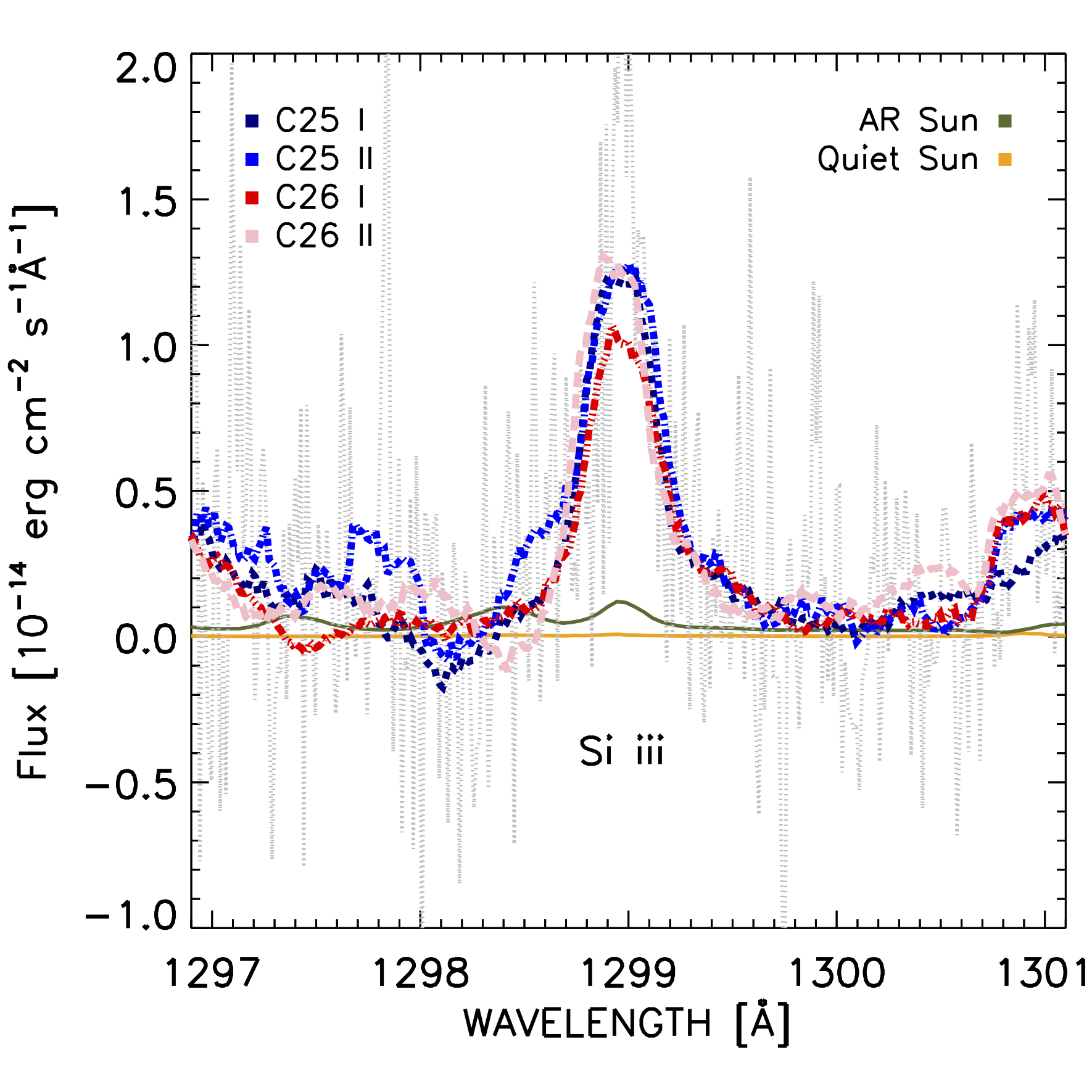}\hspace{0.0cm}
\includegraphics[trim={0 0 0 0cm},clip,width=0.45\textwidth]{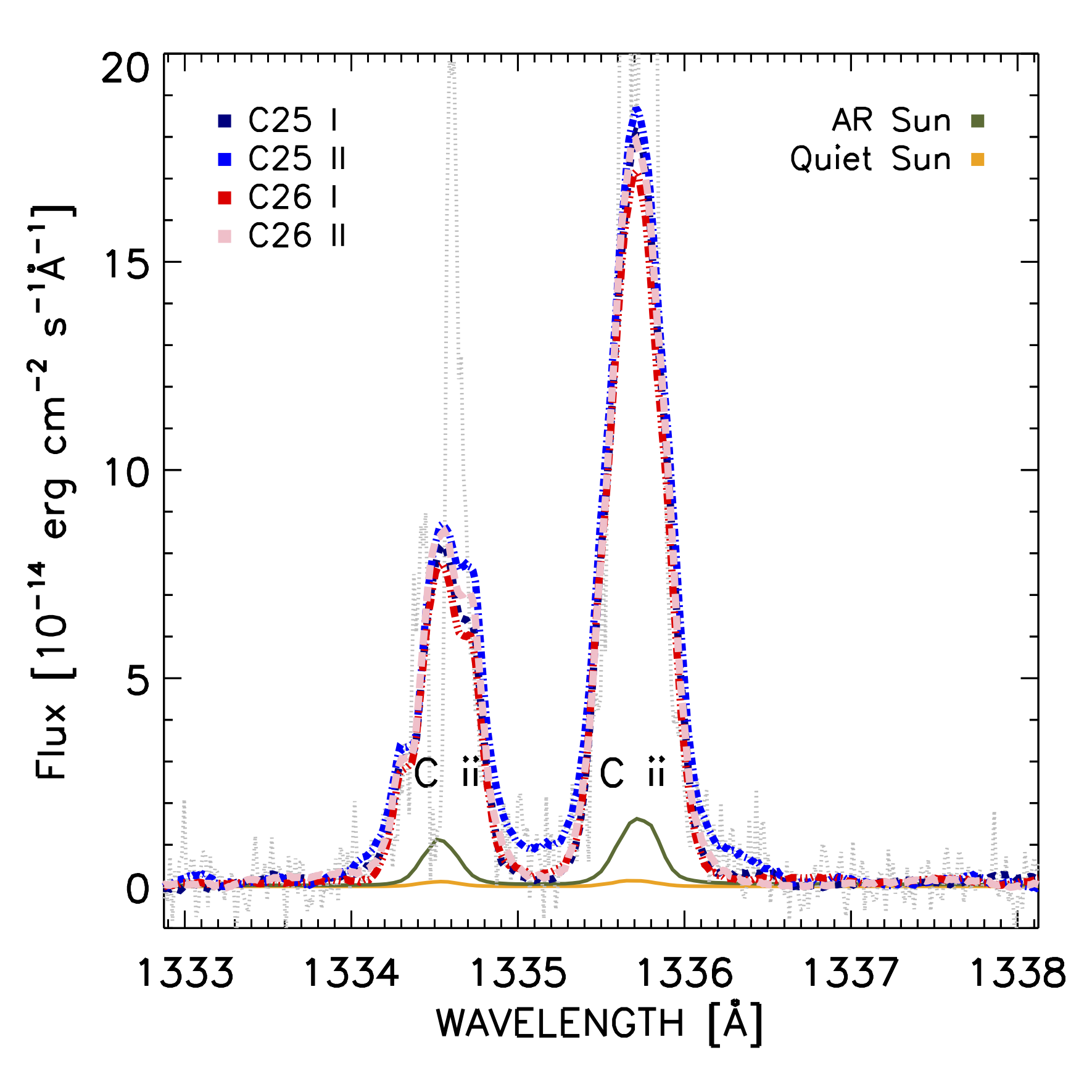}\hspace{0.0cm}
\caption{\ion{Si}{III}, Ly\,$\alpha$, \ion{O}{V}, \ion{N}{V}, \ion{Si}{II} and \ion{C}{II} lines of \ihor\ at four HST visits compared with solar active (green) and quiet regions (yellow). Lines observed during visits I and II for cycle 25 are colored in purple and blue, respectively. Lines observed during visits I and II for cycle 26 are colored in red and rose, respectively. See additional UV lines in Sec.~\ref{App:HST} of the appendix.}
\label{Fig9}
\end{figure*}

\begin{figure*}
\centering 
\includegraphics[trim={0 0 0 0cm},clip,width=1\textwidth]{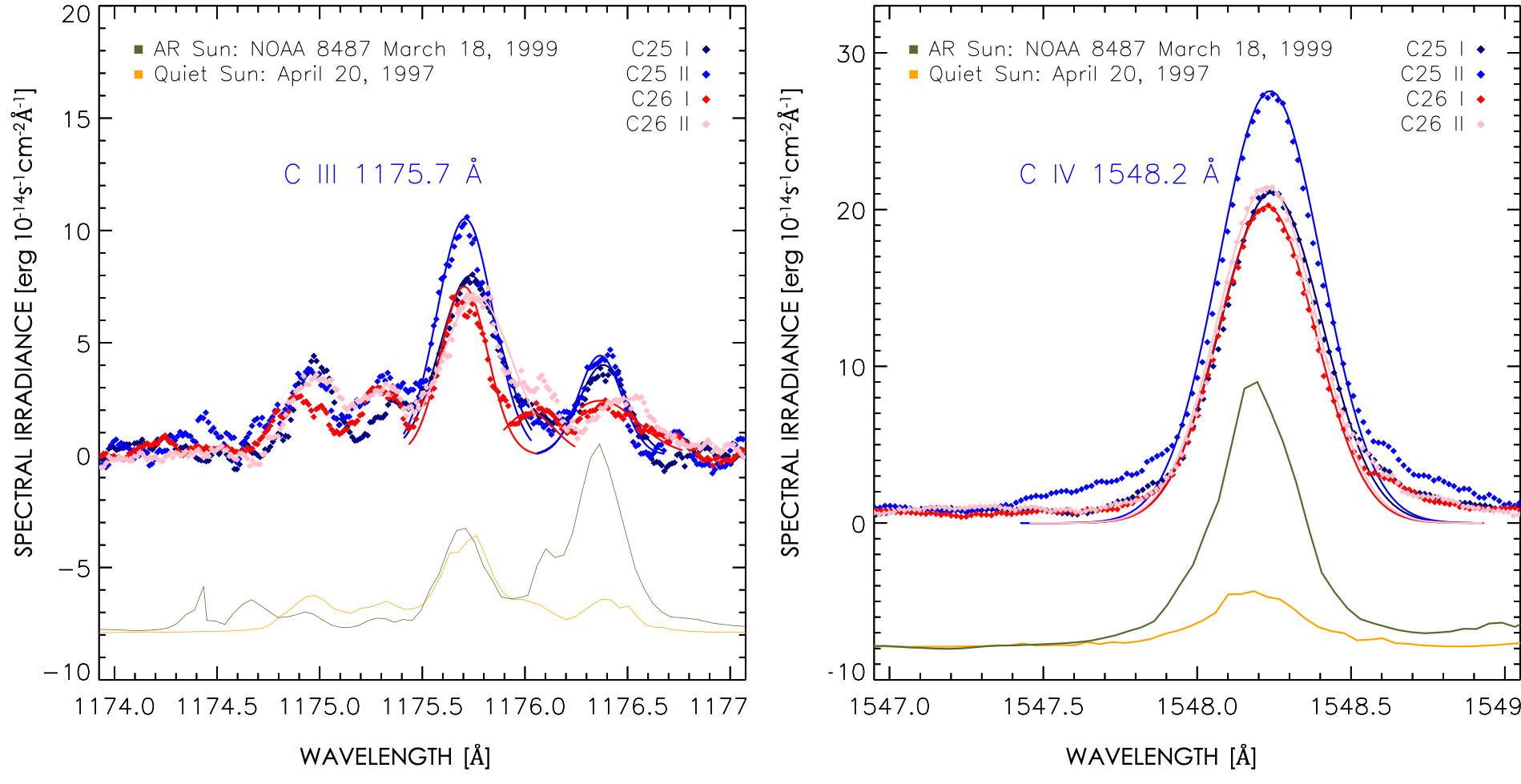}\hspace{0.0cm}
\caption{A comparison of \ion{C}{III} and \ion{C}{IV} smoothed lines obtained for \ihor\ at four different epochs in HST Cycles 25 and 26 and active (green) and quiet (yellow) regions on the Sun. Lines observed during visits I and II for cycle 25 are colored in purple and blue, respectively. Lines observed during visits I and II for cycle 26 are colored in red and rose, respectively. A clear enhancement is seen in \ihor\ during the second visit in Cycle 25, an indication of a possible flaring event.}
\label{Fig10}
\end{figure*}

\begin{table}
\centering
\begin{tabular}{l c c c c c r}
 \hline
 \hline 
 Flux density  & C25-I  & C25-II & C26-I & C26-II & Sun & Sun    \\
 \tiny{10$^{-14}$erg s$^{-1}$cm$^{-2}$\AA$^{-1}$}   &  & & & &  AR  &   QR    \\
 \hline
 
F$_{\rm \ion{C}{IV}}$ 1548.2~\AA  &   9.47  &   13.09 &   8.79   &  9.48  & 4.31 & 1.46  \\
F$_{\rm \ion{C}{III}}$ 1175.7~\AA &   4.53  &   6.55  &   3.98   &  4.43  & 1.83 & 1.74  \\
F$_{\rm \ion{Si}{IV}}$ 1402.7~\AA &   4.17  &   4.25  &   3.81   &  4.19  & 3.15 & ---  \\
F$_{\rm \ion{O}{IV}}$ 1401.1~\AA  &   0.43  &   0.48  &   0.37   &  0.44  & ---  & ---   \\
 \hline
 \hline
\end{tabular}
\caption{Measured flux densities for \ion{C}{III}, C{\sc iv}, \ion{Si}{IV} and \ion{O}{IV} lines for visits I and II during Cycles 25 and 26 of STIS/HST observations compared with solar active and quiet regions (called AR and QR respectively) observed by SUMMER/SOHO.}\label{fluxes_HST}
\end{table}

In the following paragraphs, we discuss spectroscopic observations of \ihor\ acquired during HST cycles 25 and 26 in the far- and near-ultraviolet (FUV \& NUV) regions. We used the STIS/HST instrument with the E140M grating and FUV-MAMA detector, and E230H grating with NUV-MAMA detector, covering 1150$-$1700~\AA\, and 1620$-$3100~\AA, respectively\footnote[2]{We shift from nm to \AA~in this section in order to facilitate comparisons with other studies analysing STIS/HST spectra.}. We acquired data over six orbits during cycles 25 and 26, three orbits in each cycle. The orbits during cycle 25 were acquired in Sep.~3 2018, and on Aug.~1 2019 for cycle 26. The E230H NUV-MAMA observations setup had an exposure of 2019~s and, the E140M FUV-MAMA involved two exposures of 3141~s. The spectra were reduced by the pipeline calibration code CALSTIS from the STIS science software package.

\subsection{Lines in the FUV}\label{sec:FUV}

We performed a comparison of the relative amplitudes of selected spectral lines between our STIS/HST stellar observations and solar UV spectra of a quiet and active region acquired by the Solar Ultraviolet Measurements of Emitted Radiation (SUMER) on board the Solar and Heliospheric Observatory (SOHO). The solar comparison data were of a quiet region on 20 April 1997, and the active region AR-NOAA8487 on 18 March 1999 \citep[see][]{Curdt2001,Curdt2004} and are available at \href{https://www2.mps.mpg.de/homes/curdt/}{Werner Curdt's homepage}.

We used two Gaussians functions, one thin and one thick, to fit the flux lines, such as \ion{S}{II}, \ion{S}{III}, \ion{N}{V} \ion{Si}{IV}, \ion{C}{IV}, \ion{C}{III}, \ion{O}{IV}, which are emission lines with broad wings, see Figs.~\ref{Fig9} to\,\ref{Fig12}. The narrow line fits the core of the line profiles and may correspond to a turbulent wave dissipation or possible Alfvén wave heating mechanism \cite[see][]{1997ApJ...478..745W}. The second broader Gaussian is used to fit the wings, in analogy with solar observations, this may be caused by flare heating effects or, as suggested by \citet{2001A&A...374.1108P}, due to magnetoacoustic waves traveling through the transition region to the corona. In particular \ion{C}{IV} $\lambda$1548.2~\AA\ line is characteristic of the transition region, with an approximate temperature of 10$^5$\,K. We find that the stellar spectra in the \ion{C}{IV} region are comparable with a typical observation of a solar active region (green line in Fig.~\ref{Fig10}). Comparing the four spectra acquired during Cycles~25 and 26 (see dotted lines, left panel of Fig.~\ref{Fig10}) we observe an enhancement of the emission of \ion{S}{III}, \ion{C}{IV}, \ion{C}{III} and, \ion{O}{IV} during the second FUV orbit of Cycle~25 (C25-II) which are detected as a higher amplitude in the line cores and broadening of the wings. The enhanced emission in the analysed lines suggests a possible flaring event. The total flux from the combined narrow and broad fitted Gaussian profile fits are shown in Table~\ref{fluxes_HST}. Although we noticed an increase in flux in the FUV lines during C25, we could not find any clear evidence of a flare in the TESS-LC data that we analyzed. However, there was a slight rise in flux at the start of C25-II, which can be seen in the zoomed-in bottom right panel of Fig.~\ref{Fig15}. 

\subsection{Searching for an Astrosphere around \ihor}\label{iHor:Astrosphere}

The original objective of our HST program was to measure the stellar mass loss rate of \ihor\ employing the astrospheric technique developed by Wood et al.~(\citeyear{2002ApJ...574..412W, 2005ApJS..159..118W}). This procedure is based on the detection of the Hydrogen wall emerging from the interaction between the stellar wind and the local interstellar medium (ISM). This absorption has been observed for a number of stars and is one of the few ways to detect mass loss from low-mass stars \citep{2005ApJ...628L.143W,2021ApJ...915...37W}. The astrospheric H\,{\sc i} signature (captured in the blue wing of the Lyman-$\alpha$ line) is highly blended with ISM absorption. Therefore, searching for astrospheric absorption involves analyzing the ISM properties observed toward the star as well.

\begin{figure*}
    \centering
    \includegraphics[trim={0 0 0 0cm}, clip, width= 1.\textwidth]{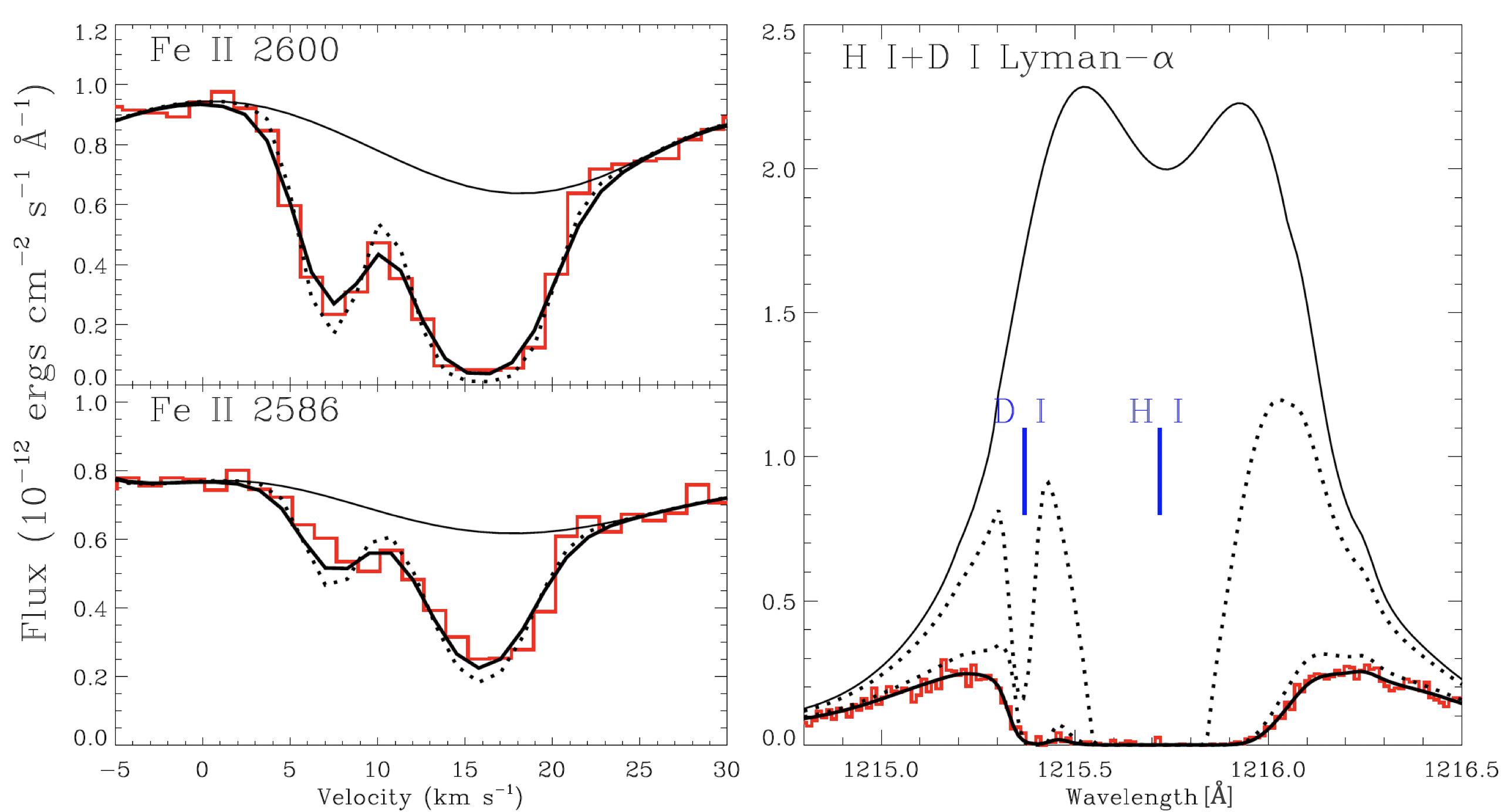}
    \caption{Fits to ISM absorption lines observed by HST toward \ihor\, with two \ion{Fe}{II} lines shown on the left, and the \ion{H}{I}+\ion{D}{I} Lyman-$\alpha$ line shown on the right. The \ion{Fe}{II} spectra are shown on a heliocentric velocity scale. The data are shown as red histograms. The thin black solid lines above the absorption are the assumed stellar emission background. Most of the stellar \ion{H}{I} Lyman-$\alpha$ emission is absorbed by the ISM. Two distinct \ion{Fe}{II} velocity components are observed, and the combined absorption of the fit is shown both before (dotted lines) and after (thick solid lines) instrumental smoothing. The two components are too blended to be resolved in the Lyman-$\alpha$ spectrum, but the Lyman-$\alpha$ fit includes both components, with parameters constrained by the \ion{Fe}{II} fit. The individual components are shown as dotted lines, and their combined absorption after instrumental smoothing is the thick solid line that fits the data.}
    \label{fig:ISM}
\end{figure*}

Figure~\ref{fig:ISM} shows \ion{Fe}{II} and Lyman-$\alpha$ absorption observed towards \ihor. The \ion{Fe}{II} lines are the best available lines for studying the ISM velocity structure. Two well-separated velocity components are seen toward \ihor. The two \ion{Fe}{II} lines with rest wavelengths of 2586.650~\AA\ and 2600.173~\AA\ are fitted simultaneously, using procedures extensively used in the past \cite[e.g.,][]{2008ApJ...673..283R}. Each absorption feature is defined by three parameters: a central wavelength (or velocity), a column density, and a Doppler parameter associated with the width of the line. The two \ion{Fe}{II} velocity components have velocities of $v=7.4$ and $v=15.9$ km~s$^{-1}$, logarithmic column densities (in cm$^{-2}$ units) of $\log N_{\rm \ion{Fe}{II}}=12.53$ and $\log N_{\rm \ion{Fe}{II}}=13.15$, and Doppler parameters of $b_{\rm \ion{Fe}{II}}=1.94$ and $b_{\rm \ion{Fe}{II}}=3.09$ km~s$^{-1}$, respectively.

For Lyman-$\alpha$, the \ion{H}{I} absorption is highly saturated and very broad (see Figure~\ref{fig:ISM}). This is true for even the nearest stars with the lowest ISM column densities, but \ihor\ seems a particularly extreme case for such a nearby star. Normally, narrow absorption from interstellar deuterium (\ion{D}{I}) is observed to the left of the broad \ion{H}{I} absorption, but for \ihor\ even the \ion{D}{I} absorption is saturated and the \ion{H}{I} absorption is so broad that it is almost completely blended with \ion{D}{I}. We fit the \ion{D}{I} and \ion{H}{I} absorption simultaneously, once again using procedures more extensively described elsewhere \cite[e.g.,][]{2005ApJS..159..118W,2021ApJ...915...37W}. Although the two ISM components seen in \ion{Fe}{II} are not separable in the much broader \ion{D}{I} and \ion{H}{I} absorption, we consider the two components in the Lyman-$\alpha$ fit, constrained by the \ion{Fe}{II} fit. In particular, we force the velocity separation and column density ratio to be the same, and for simplicity, we simply assume the two components have identical Doppler parameters. With these constraints, we measure an \ion{H}{I} Doppler parameter of $b_H=11.1$ km~s$^{-1}$, and for the two components we measure \ion{H}{I} column densities of $\log N_H=18.09$ and $\log N_H=18.72$, respectively.

Our ISM-only analysis fits the Lyman-$\alpha$ data quite well, as shown in Figure~\ref{fig:ISM}. We, therefore, see no evidence of astrospheric absorption toward \ihor\ that would allow us to say something about the strength of its stellar wind. The ISM \ion{H}{I} column density is much too high, and the resulting Lyman-$\alpha$ absorption is too broad to detect the astrospheric absorption signature. The total ISM \ion{H}{I} column density of $\log N_H=18.81$ is in fact the third highest known within 25~pc \cite[][]{2005ApJS..159..118W,2021ApJ...915...37W}. The only higher values are toward HD~82558 ($d=18.3$ pc, $\log N_H=19.05$) and HD 203244 ($d=20.8$ pc, $\log N_H=18.82$).

\subsection{\ion{Mg}{II}$~\textrm{H\,\&\,K}$ line}\label{sec:MgII}

\begin{figure*}
    \centering
    \includegraphics[trim={0 0 0 0}, clip, width= 1.\textwidth]{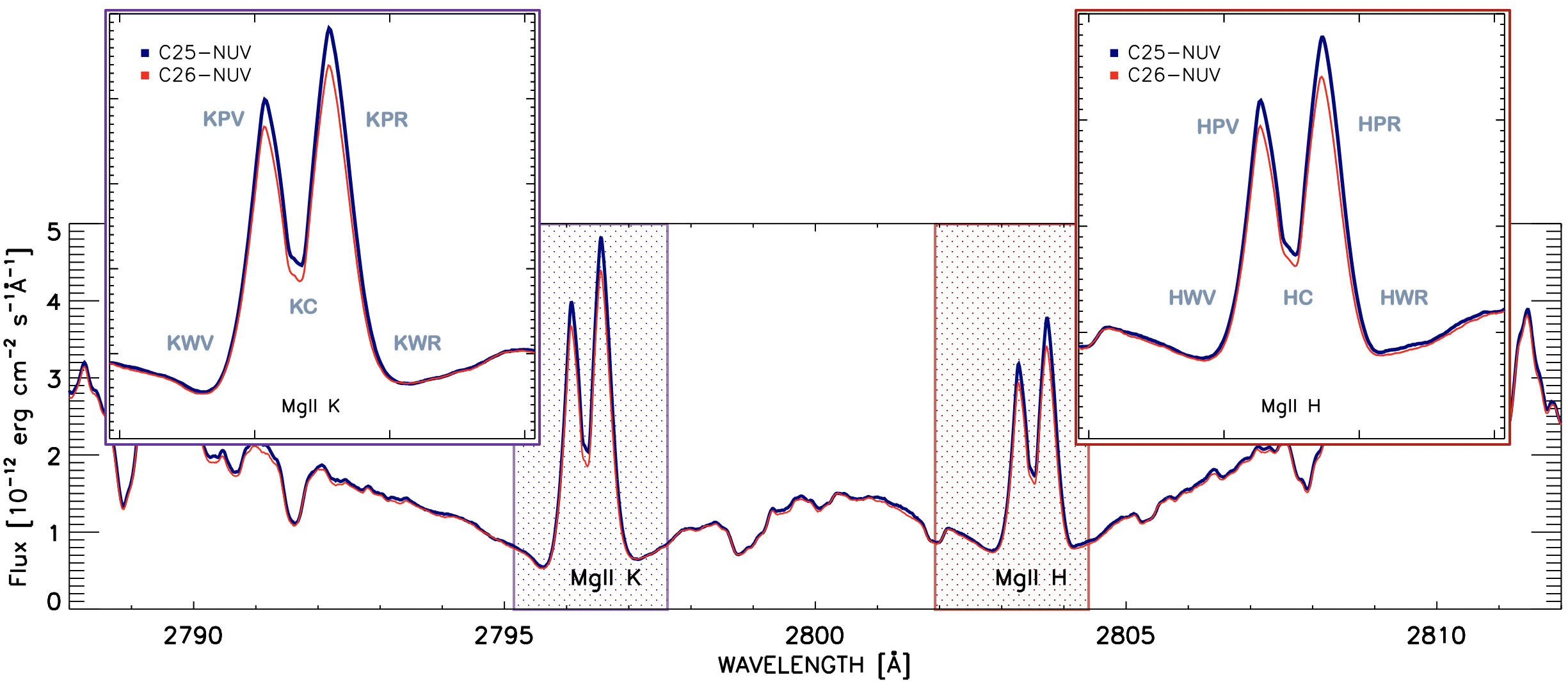}
    \caption{\ion{Mg}{II}$~\textrm{H\,\&\,K}$ profiles observed by HST/STIS for the NUV single orbit during cycles 25 and 26. We describe the different features of the \ion{Mg}{II}$~\textrm{H\,\&\,K}$ line as from the wavelengths on the violet side of the $~\rm{H\,\&\,K}$ lines, we denote the minimum of the base wing as HWV and KWV, the first peak HPV and KPV. The core of the lines HC and KC. The second peak is located on the red side with respect to the core, or central depression of the line, as HPR and KPR. Finally, the minimum of the wing base is located on the red side with respect to the line core as HKR and KWR.}
    \label{fig:MgII}
\end{figure*}

The wings, peaks, and cores of the spectral line \ion{Mg}{II}$~\textrm{H\,\&\,K}$ are considered reliable indicators for recovering activity information from the stellar photosphere, chromosphere, and near transition region. These lines are located in the near-ultraviolet (NUV) spectral range [k=2795.528\AA~and h=2802.704\AA]. It is known that the line is sensitive to the strength of the magnetic field and the hydrodynamics conditions, helping to determine atmosphere conditions at a certain region. For example, in the Sun, the profile features at the base of the core line (see in Fig.~\ref{fig:MgII}, KW and HW) are formed in the photosphere (high~600~km, mid photosphere), while the core peaks are formed in the middle chromosphere (KPV and HPV at about high~1200~km and the KPR and HPR at about 1550~km, mid chromosphere) being very useful for calculating the velocities of inflows or outflows in the specific region of the atmosphere. The deep profiles between the core peaks (KC, HC) are known to be very sensitive and saturate during flaring events, they are formed in the upper chromosphere around 2200~km, about 200~km below the transition region \cite[see][]{2013ApJ...778..143P,2013ApJ...772...89L,2013ApJ...772...90L,2015ApJ...811..127S}. We put in the stellar context these solar features to analyse the photospheric, chromospheric, and near transition region layers of \ihor.

The relative difference between \ion{Mg}{II} fluxes for the NUV orbits of cycles 25 and 26 suggests increased activity during cycle 25, as is also observed for the FUV lines. Even though higher activity was expected in cycle 26 than in cycle 25, following the predicted cycle trends from Ca and X-rays observed in Fig.~\ref{Fig1}, we observed a slightly enhanced flux in the NUV and FUV regions during cycle 25. We suggest the enhanced flux during cycle 25 could be associated with the presence of an active region (AR) and with elevated chromospheric activity (see characteristic AR shape during simultaneous HST-C25 observations in Fig.~\ref{Fig:Simult}). Unfortunately, the time of observation of the \ion{Mg} line, during the first orbit of HST, was not simultaneously performed with the observation of the enhanced \ion{C} lines, during the third orbit of HST, C25II, as shown in Fig~A\ref{Fig15}. Therefore, we do not observe saturation in the cores of \ion{Mg}{II}\,\rm{H\,\&\,K} lines which indicates that the flaring event, suggested by the enhanced \ion{C}{IV} and \ion{S}{IV} lines in the FUV during the orbit II in C25, occurred after the first NUV orbit in C25. The relative intensity between the core peaks of \ion{Mg}{II}\,\rm{H\,\&\,K} lines, are more prominent and easier to analyse than $~\textrm{H\,\&\,K}$ cores of \ion{Ca}{ii} in C25 and C26. Following \cite{Avrett2012SupportingCF} the relative difference between \ion{Mg}{II}\,\rm{H\,\&\,K} core lines indicates an averaged stellar outflow.

\section{Simultaneous observations}\label{Simultaneous}

\begin{figure*} 
\includegraphics[trim=0.cm .0cm .0cm 0.cm, clip=true, scale=0.4]{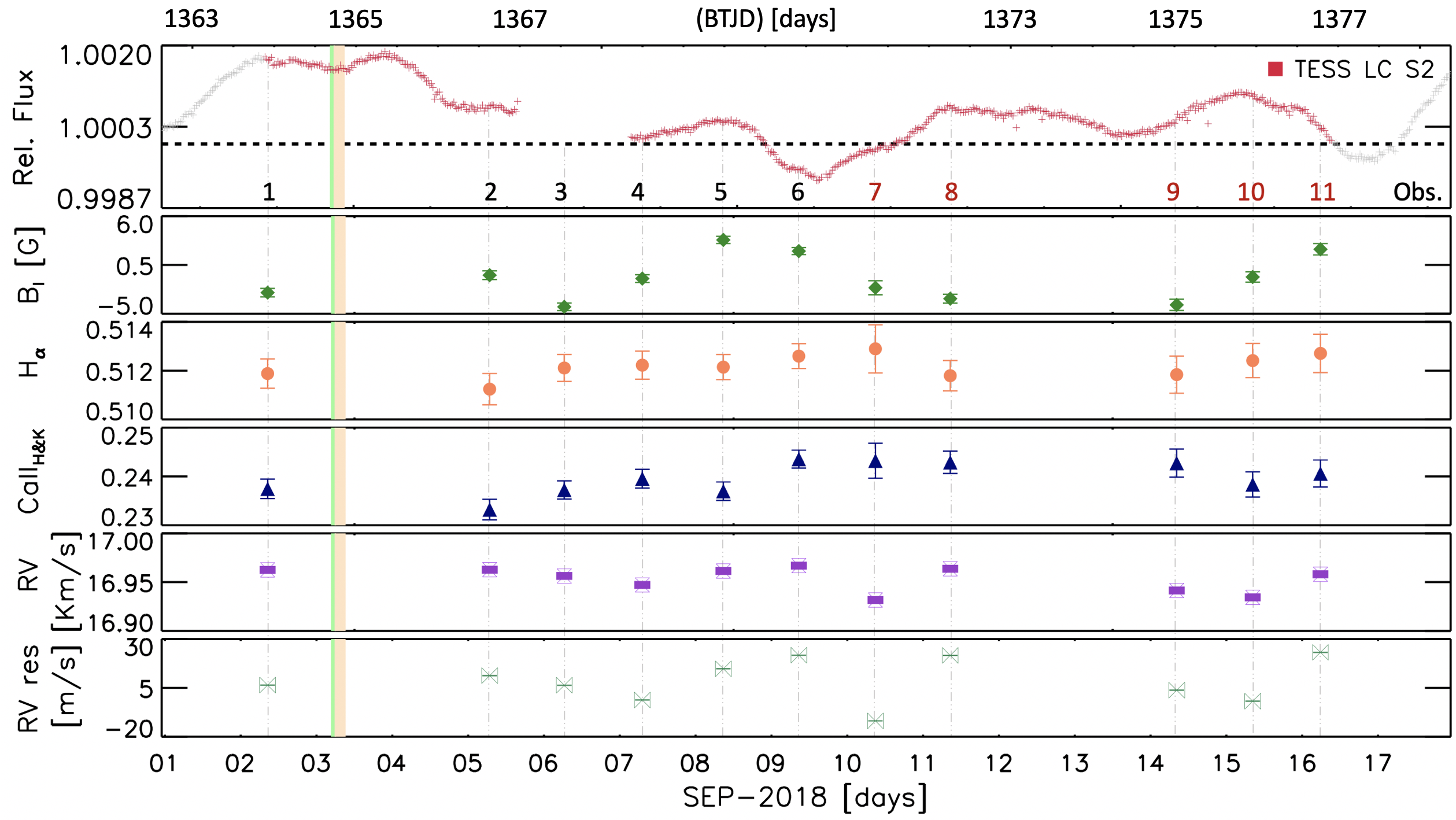}
\caption{Simultaneous observations of \ihor. The top panel shows the normalized light-curve section [1362.5 to 1379.5 BTJD days] obtained during S2 by the TESS satellite. The following panels show the simultaneous longitudinal magnetic field B$_{l}$, the H$\alpha_{\rm Index}$, the flux of calcium line cores, \ion{Ca}{ii}$~\textrm{H\,\&\,K}$, expressed in terms of the \ion{Ca}{ii}$~\rm{H\,\&\,K}$ \Shk, and radial velocities RVs. All values obtained from observations by the ESO/HARPSpol instrument for the EPOCH-P101C under the \emph{Far beyond the Sun} campaign. RV error values are in the order of $2 \rm m\,s^{-1}$, which makes it difficult to resolve in the scale of $ \rm Km\,s^{-1}$, go to Table.~\ref{table_data_B3}2 to see the RV error and S/N values per observation night. The green-bisque dash lines indicate 3 simultaneous observations of the NUV \& FUV regions obtained by the HST STIS/NUV-MAMA instrument, with E230H and E140M gratings, respectively.}\label{Fig:Simult}
\end{figure*}

In this section, we focus on the data acquired simultaneously during the campaign. These correspond to the ESO P101$_{\rm C}$ HARPSpol spectropolarimetric run (Aug - Sep 2018), the Sector~2 observations from TESS (starting on 23 August until 20 September 2018 and overlapping from August 29 to 16 September), and the coordinated Cycle 25 HST/STIS observations (3 September 2018). This combined dataset allows us to compare the various manifestations of the magnetism of \ihor, at different layers of the stellar atmosphere, on a sub-rotational period timescale. 

\begin{figure*} 
\includegraphics[width=0.99\textwidth]{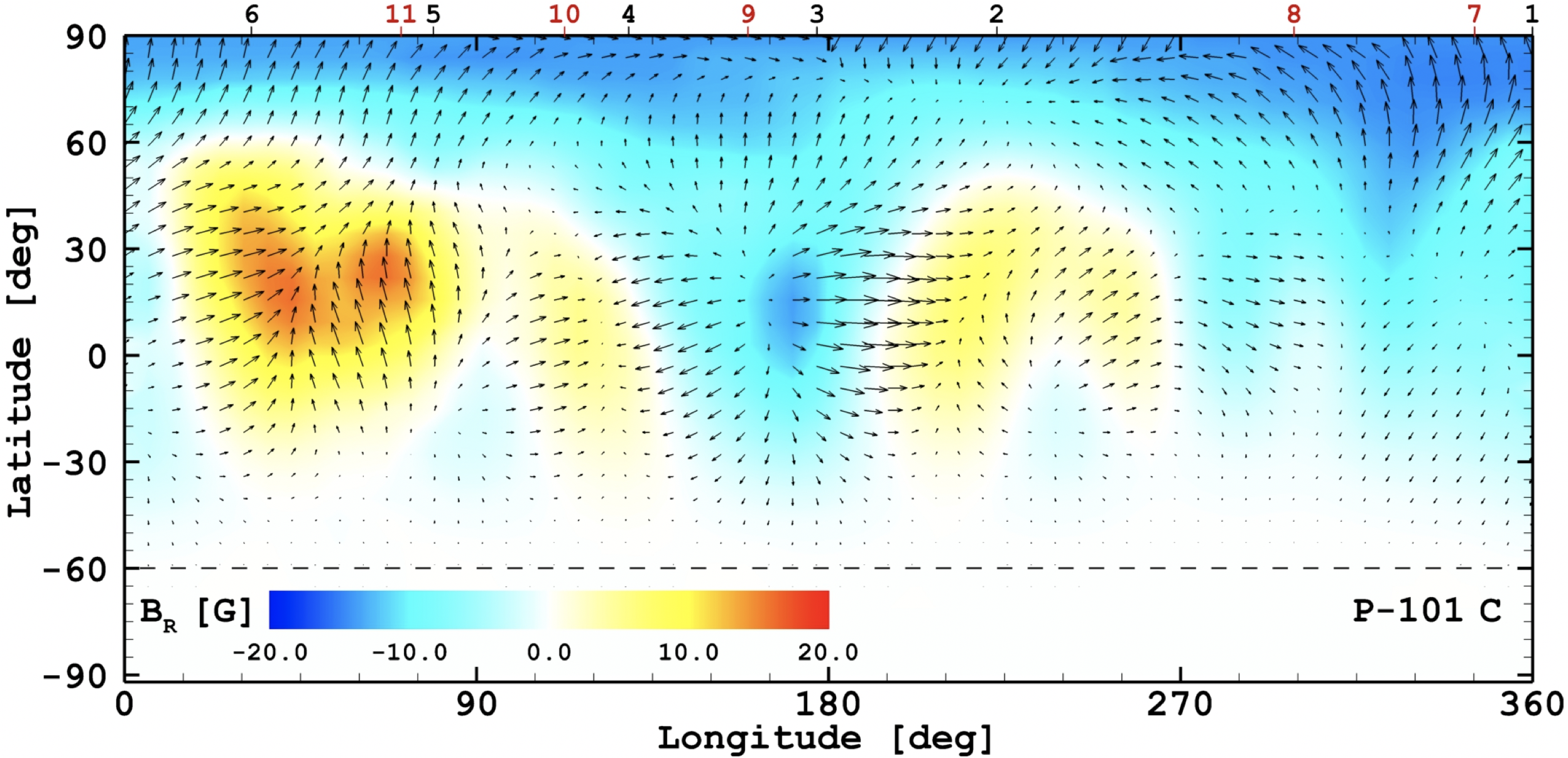}
\caption{Zeeman Doppler Imaging (ZDI) reconstruction of the large-scale magnetic field of \ihor~based on HARPSpol data acquired from 2 to 16 September 2018 (P101$_{\rm C}$). The color scale shows the radial component of the magnetic field ($B_{R}$) in Gauss, while vectors trace the azimuthal ($B_{A}$) and meridional ($B_{M}$) fields normalized in size to their magnitude $(B_{A}^2 + B_{M}^2)^{1/2} = 17.7$~G. Two consecutive rotations were employed to retrieve the ZDI map, whose phase coverage is indicated by the black and red tick marks in the upper x-axis, labeled by the night of acquisition during the run (see~Fig.~\ref{Fig:Simult}). The horizontal dashed line corresponds to the visibility limit imposed by the inclination of the star ($\sim$\,60$^{\circ}$).}\label{Fig16}
\end{figure*}

\begin{figure*} 
\includegraphics[trim=0.5cm 0.75cm 1.0cm 1.0cm, clip=true, width=0.49\textwidth]{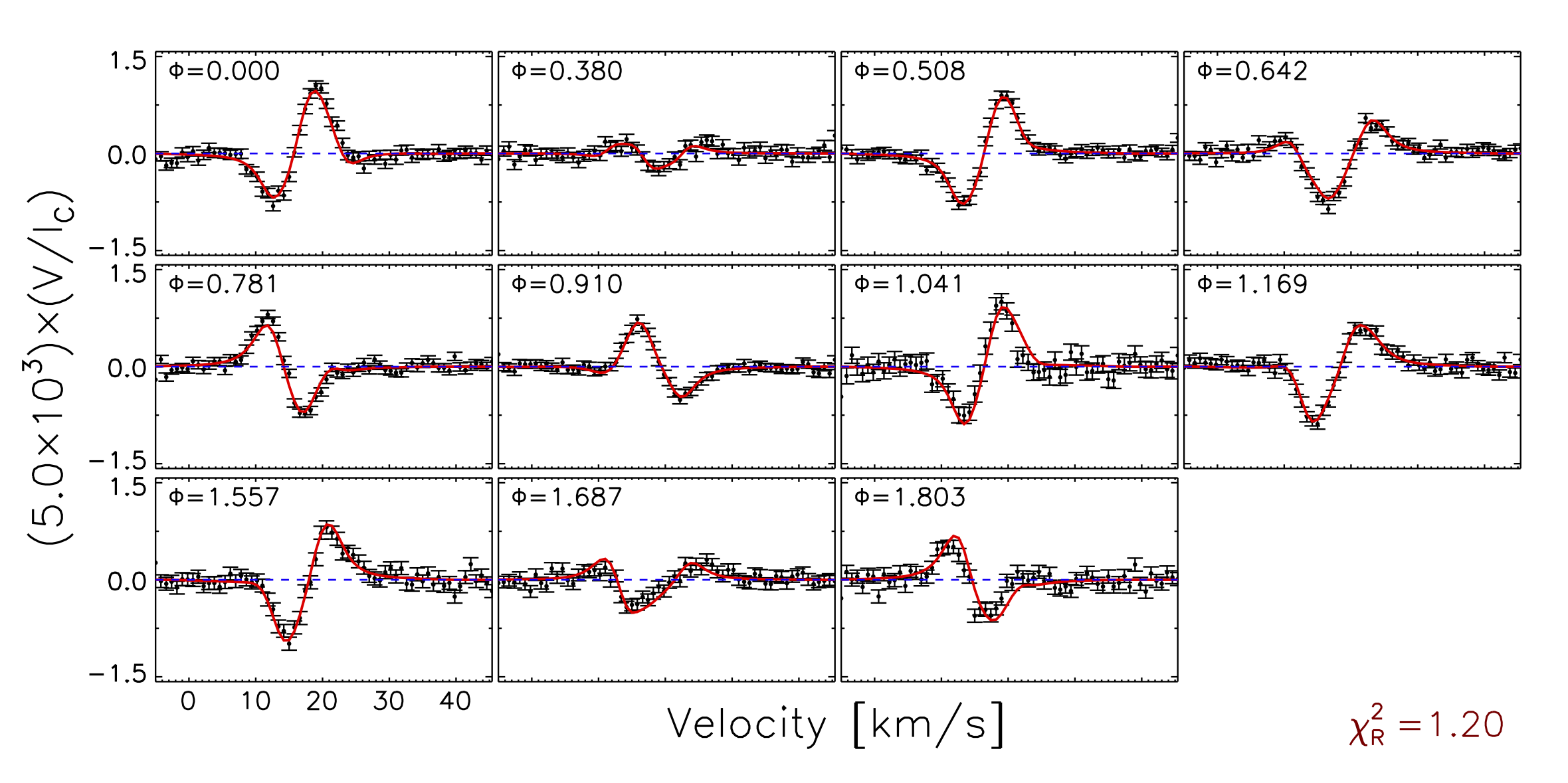}\,\includegraphics[trim=0.5cm 0.75cm 1.0cm 1.0cm, clip=true, width=0.49\textwidth]{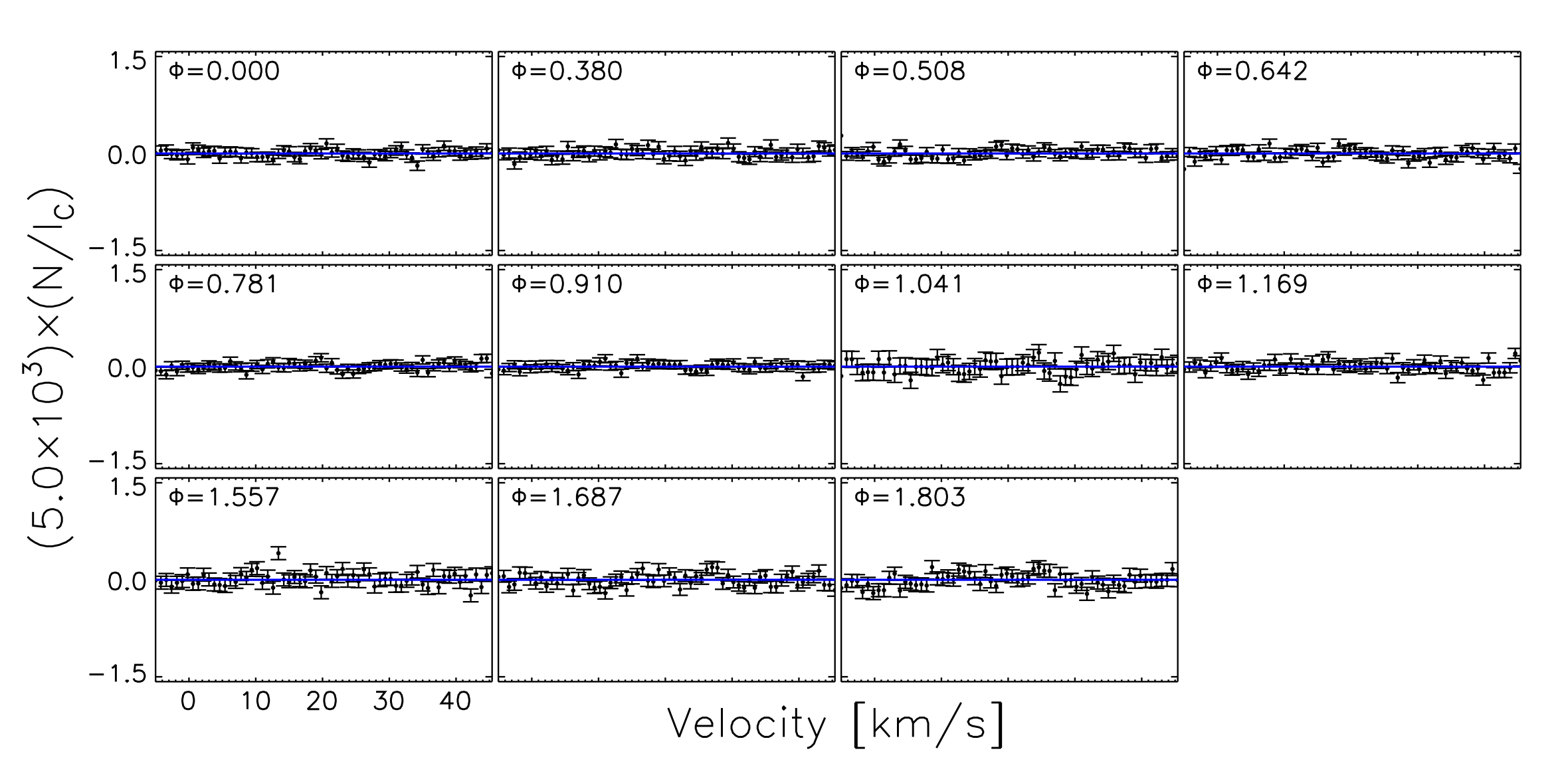}
\caption{Sequence of LSD Stokes V (left) and N (right) profiles used to reconstruct the ZDI map presented in Fig.~\ref{Fig16}. Observations (black symbols with errors), fits provided by the ZDI magnetic field distribution (red), and the achieved optimal reduced $\chi^2_{\rm ZDI}$ are shown. The data in both panels have been enhanced by a constant factor ($5\times10^{3}$) for visualization purposes. We include the observing phase ($\phi$) for each night, taking the first night of this epoch as $\phi=0.0$ (BJD = 2458363.85483) and using a $P_{\rm ROT} = 7.7$~d. The blue line in the right panel indicates the zero level.}
\label{Fig:ZDI-FIT}
\end{figure*}

The different panels of Fig.~\ref{Fig:Simult} show the temporal evolution of various observables during this particular run. This includes photometry from TESS, as well as chromospheric/photospheric activity indicators (\Shk, I$_{\rm H\upalpha}$), radial velocity (RV), and longitudinal magnetic field measurements ($B_{\ell}$) from the HARPSpol spectra. We also indicate the specific date of our HST/STIS visit probing the transition region and corona. While a detailed Zeeman-Doppler Imaging (ZDI) analysis of \ihor~will be presented in a forthcoming paper, we complement the aforementioned time series description with the corresponding ZDI magnetic field reconstruction associated with this epoch (Figs.~\ref{Fig16} and \ref{Fig:ZDI-FIT}). The ZDI reconstruction follows the implementation described by \citet{2000MNRAS.318..961H,2001MNRAS.322..681H,2002ApJ...575.1078H}, assuming a $P_{\rm rot} = 7.7$~d (Sect.~\ref{subsec_rotation}) and a inclination angle of $60^{\circ}$. Despite the relatively good phase coverage of this epoch, it was not sufficient for an estimate of the differential rotation profile of the star. Therefore, solid-body rotation was assumed for the ZDI map shown in Figs.~\ref{Fig16}. A Milne-Eddington local line profile was used which was tailored to a high S/N Stokes~I spectrum of the star, together with a linear limb-darkening law coefficient set to 0.74 \cite[appropriate for the spectral type of \ihor~and the bandpass of HARPSpol, see][]{2005PASP..117..711G}. The code uses a maximum-entropy regularization approach by \cite{1984MNRAS.211..111S} to guarantee the uniqueness of the solution. Following the methodology described in \citet{2015A&A...582A..38A}, a grid of maximum-entropy ZDI solutions, with their corresponding fits to the LSD Stokes V profiles,\textcolor{black}{ is constructed from which the optimal reduced $\chi^2_{\rm ZDI}$ value is obtained (Fig.~\ref{Fig:ZDI-FIT})}. The final map is presented in Fig~\ref{Fig16}, displaying the three components of the large-scale magnetic field (Radial: color-scale, Meridional, and Azimuthal: vectors) in a latitude-longitude Mercator projection. The phase coverage is represented by tick marks in the upper x-axis, labeled by the corresponding observing night of the run as indicated in the bottom panel of Fig.~\ref{Fig:Simult}.

As can be seen by comparing Figures~\ref{Fig:Simult} and~\ref{Fig16}, the first night of observations revealed a predominant negative polarity in the large-scale field (see panel for $B_{\ell}$) associated with the visible pole of the star. The next three nights were lost owing to bad weather conditions in La Silla, which unfortunately coincided with our coordinated HST/STIS visit. Still, the photometric variability displayed an increase of $\sim$\,0.2\% with respect to the normalized level, which could be associated with the presence of some flaring activity over those phases as registered in the FUV/NUV spectra (Sect.~\ref{HST}). That region of the ZDI map (retrieved mostly with the information provided one rotation later by nights 7 and~8) reveals relatively weak magnetic fields towards lower latitudes ($\lesssim$\,10~G) with the appearance of a positive polarity region registered during the second observed night. From there and for the following 5 nights both activity indicators, \Shk and I$_{\rm H\upalpha}$, slowly increase to reach a maximum on night 7 of the run. After a small gap, the TESS light curve shows a dip at the $0.1\%$ around night 6, indicating the presence of spots at that particular phase. This agrees with the distribution obtained on this region of the ZDI map, which shows the strongest low-latitude mixed-polarity regions of the reconstruction, reaching up to $20$~G in magnitude (nights 5 and 6). Such local changes in the field polarity are clearly visible in the sinusoidal behaviour of $B_{\ell}$ during those observed phases (Fig.~\ref{Fig:Simult}, second panel). 

From the beginning of the run, the RV remains relatively stable, except for night 7 [10 to 11 Sep. 2018] where it goes below the long-term average given in Table~\ref{tab_1}. As can be seen from Table~\ref{table_data_B3}2 and Fig.~\ref{Fig:Simult}, this could be mainly due to the relatively higher noise in the spectrum of that night, resulting also in larger uncertainties for all the derived quantities. However, the presence of a spot group around nights 5-6 [8 to 9 Sep. 2018], as can be contrasted with TESS photometry, could have also contributed to this relatively large drift in the RV. 

Bad weather prevented observations on 11-12 September, over which the photometry showed a relatively stable behaviour close to the nominal level. After this gap, the RV appears slightly lower during nights 9 to 11 than at the beginning of the run but still within the uncertainty limits for its long-term average. In contrast, $B_{\ell}$, \Shk\, and I$_{\rm H\upalpha}$ display very similar behaviour as during nights 3 to 5, indicating a relatively stable magnetic and activity configuration on the stellar surface over the course of the two observed rotations. One example of this is the polarity flip and relatively quick variation of $B_{\ell}$ during these three nights (by roughly $6$~G), which precede by approximately half a day a dip in the photometric light curve indicating the presence of a starspot group coming into view (see top panel of Fig.~\ref{Fig:Simult} around BTJD 1369.5 and 1378.5). This combined dataset illustrates that while the large-scale field remains stable over two rotation periods (as demonstrated both in the $B_{\ell}$ values and in our combined map) the changing lightcurve and RV values hint at substantial continuous flux emergence happening on smaller spatial scales but still sufficient to modify the brightness of the star and the shape of the intensity profile.

\section{Discussion and Conclusions}\label{DandC}

As presented in \textcolor{blue}{Paper\,I}, \ihor~constitutes the closest located star to the intersection between the cycle branches emerging in the $P_{\rm rot}-P_{\rm cyc}$ diagram \citep{1998ApJ...498L..51B, 2007ApJ...657..486B, 2016ApJ...826L...2M}. Given that the active (A) and inactive (I) cycle periods are similar, a beating pattern is expected in the temporal evolution of different activity indicators. This is indeed observed in \ihor, where the intense spectropolarimetric monitoring of the star, combined with archival observations, allowed us to refine and retrieve $P_{\rm cyc}^{A} = 1.499 \pm 0.012$~yr and $P_{\rm cyc}^{I} = 1.097 \pm 0.023$~yr. In addition, our analysis of the rotation period of \ihor~(Set.~\ref{sec_TESS}) moves the star even closer to the active-inactive branches intersection, with a $P_{\rm rot}$ close to $8$~days (refined from previous estimates at $P_{\rm rot}>8$~days, see e.g., \citealt{2010ApJ...723L.213M}). Note also that for sparsely sampled observations (or broader binning of the available data) the beat period of $P_{\rm beat} \simeq 4.49$~yr will dominate over the shortest periodicity (see e.g.~\citealt{2017MNRAS.464.4299F}). Similarly, the $\sim1.97$~yr period identified in \textcolor{blue}{Paper\,I} was not recovered in the analysis of the whole dataset, mainly due to the tighter constraints on shorter timescales placed by the 98 additional spectropolarimetric observations to the 2-component fit (see~Table~\ref{Table:param}).

The location of \ihor\ in the spot-dominated regime over rotation timescales ($\rm S_{fac}/S_{spot}\,\sim$\,0.51) is in line with its higher level of activity with respect to the Sun, reflecting its youth and relative rapid rotation. The Sun is closer to the faculae-dominated stars (towards the bottom right in Fig.~\ref{Fig8}) close to finalizing its transition from the spot-dominated (left) to the faculae-dominated stars (right). \citet{Eliana1} found that the Sun displays a roughly constant $\rm S_{fac}/S_{spot}\sim$3 along its activity cycle. It is therefore expected that \ihor\ would also have a stable $\rm S_{fac}/S_{spot}$ value across its activity cycles, although further observations would be needed to confirm this.

As discussed in Sect.~\ref{iHor:Astrosphere}, our search for an astrospheric signal from \ihor~was hampered by the extremely large interstellar absorption along that line of sight (resulting in the third largest known column density within 25~pc at $\log N_{H} = 18.81$). Given the coronal activity levels of the star ($F_{\rm X} \simeq 3.5 - 10.0~\times10^{5}$~erg~s$^{-1}$cm$^{-2}$), and following the mass loss-activity relation proposed for Sun-like stars ($\dot{M}_{\bigstar} \propto$ $F_{\rm X}^{1.34}$, \citealt{2005ApJ...628L.143W}), the stellar wind mass loss rate of \ihor~is expected to somewhere between the range $\dot{M}_{\bigstar} \sim 9 - 80~\dot{M}_{\odot}$ over the course of its activity cycle. Note however that in periods of high activity, the X-ray flux of \ihor~approaches the observed break in the mass loss-activity relation ($F_{\rm X} \sim 10^{6}$~erg~s$^{-1}$cm$^{-2}$), so that the actual $\dot{M}_{\bigstar}$ value could be much lower than the expectation. The results for the FUV/NUV line fluxes described in Sects.~\ref{sec:FUV} and \ref{sec:MgII}, the emission measure analysis presented by \citet{2019A&A...631A..45S}, and the different ZDI maps retrieved for the star, will be used in a forthcoming study to investigate numerically the stellar wind and planetary environment of this system (e.g.~\citealt{2016A&A...588A..28A, 2016A&A...594A..95A}).     

\begin{table}
\centering
\begin{tabular}{lccccr}
 \hline
 \hline 
     & \Shk\,   & I$_{\rm H\upalpha}$ & $B_{\ell}$ & $|B_{\ell}|$ & RV  \\
 All dataset:\\
\hline
\Shk\,              &   1.00  &   0.82   &   0.025  & -0.034 & -0.015 \\
I$_{\rm H\upalpha}$ &   0.82  &   1.00   &   0.046  & -0.021 & -0.056 \\
$B_{\ell}$          &   0.025 &   0.046  &   1.00   & -0.26  &  0.042 \\
$|B_{\ell}|$        &  -0.034 &   -0.021 &  -0.26   &  1.00  &  0.045 \\
RV                  &  -0.015 &   -0.056 &   0.042  &  0.045 &  1.00  \\
 \hline
 Simultaneous Obs.:\\
\Shk\,                 &   1.00  &   0.57   &  -0.095  &  0.28  & -0.26 \\
I$_{\rm H\upalpha}$ &   0.57  &   1.00   &   0.36   & -0.059 & -0.40 \\
$B_{\ell}$          &  -0.095 &   0.36   &   1.00   & -0.34  &  0.32 \\
$|B_{\ell}|$        &   0.28  &   -0.059 &  -0.34   &  1.00  &  0.184 \\
RV                  &  -0.26  &   -0.40  &   0.32   &  0.184 &  1.00  \\
 \hline
 \hline
\end{tabular}
\caption{Pearson correlation coefficients between the different activity all indicators. Top: coefficients calculated taking the entire 199 data points obtained by HARPSpol. Bottom: coefficients calculated taking the 11 simultaneous data points obtained by HARPSpol.}\label{PiersonCoeff}
\end{table}

In this study, we analysed different atmospheric layers of \ihor\ by probing activity indicators from the photosphere to the lower corona. For this analysis we have performed a detailed multi-wavelength and multi-technique characterization of the star. 

\begin{itemize}

\item We analyse the long-term variability in magnetic activity sensitive diagnostics derived using HARPS data acquired both from archival spectroscopic data and  HARPS spectropolarimetric data from our programme, completing the observations and analysis started in \textcolor{blue}{Paper\,I}. For the activity diagnostics, we use high signal-to-noise LSD profiles to derive time series measurements of photospheric activity, including the longitudinal magnetic field ($B_{\ell}$). For the chromospheric activity, we quantify variability in the \ion{Ca}{ii} $~\rm{H\,\&\,K}$ and H$\alpha$ emission profiles using activity indices \Shk\ and I$_{\rm H\upalpha}$. The \Shk\ indices from our HARPSpol data were supplemented with archival HARPS data, spanning ten years in total. 

\item A period analysis of the full S$_{\rm HK}$ dataset revealed two dominant periods, indicating the presence of two overlapping chromospheric activity cycles, with periods of 547.65\,$\pm 3$\,d (1.49\,y) and 401.01\,$\pm 7$\,d (1.09\,y), respectively. \citet{2013A&A...553L...6S,2019A&A...631A..45S} reported a 588.5\,d (1.6\,y) cycle in both X-ray and \ion{Ca}{ii} $~\rm{H\,\&\,K}$ but on including the last 1.5-year \ion{Ca}{ii} $~\rm{H\,\&\,K}$ data from our campaign, we found that the dominant periodicity in the chromospheric cycle (1.49-year) is close aligned with the 1.6-year X-ray cycle. This is relatively similar to the Sun, in which chromospheric and coronal activity cycles are aligned. The slight difference may be due to a geometrical/inclination effect given by the misalignment of the observed features in the low chromosphere and the corona, where features located near the limbo edge can be traced in the corona but less clearly in the photosphere or chromosphere. As one hemisphere of the star is obscured, if the active regions migration over the course of the activity cycle spans a wider range of latitudes than on the Sun, this could result in the double period that we find for \ihor. Alternatively, the overlapping chromospheric cycles may be due to activity on the different hemispheres evolving differently. 

\item We find a relatively strong correlation between the two chromospheric activity indicators, S$_{\rm HK}$ and I$_{\rm H\alpha}$ ($\rho_{{\rm S_{HK},I_{H\alpha}}} = 0.82$). We noted no correlation between \Shk\ and longitudinal magnetic field $B_{\ell}$, (Pearson correlation coefficient $\rho_{B_{\ell},{\rm S_{HK}}}=0.025$). This is likely a consequence of the different length scales being probed -- $B_{\ell}$ probes the large-scale photospheric magnetic field whereas \Shk\ is sensitive to disc-integrated chromospheric activity, including small-scale active regions.

\item We apply seven different techniques to evaluate \ihor's rotation period using TESS data and find a broad degree of agreement within the range 6.6 to 7.9\,d. We can refine these measurements further using Stokes V spectra when optimising stellar parameters for Zeeman Doppler imaging -- as the large-scale field does not vary as quickly as photometry (see Fig.\,\ref{Fig:Simult}) this enables a more robust measurement of the stellar rotation period.
 
\item We implemented the novel Gradient of Power Spectra analysis (GPS) over time series of brightness variations. We estimated the stellar rotation periodicity of $7.78 \pm 0.18$ from magnetic features modulation for \ihor. Furthermore, we performed a characterization of the ratio between dark spots and bright faculae features by analysing the location of gradients in the high-frequency tail of the power spectra. We analyse the location of the star in the context of the faculae to spot branching exposed in \citet[][]{PaperI,Eliana1,Eliana2,ElianaThesis} and compared it with the solar value. We derive a facular-to-spot ratio of 0.510~$\pm$~0.023 for \ihor, considerably more spot-dominated than the Sun which has $\rm S_{fac}/S_{spot} \sim 3$ throughout its activity cycle.

\item We present a multi-wavelength analysis of \ihor\  bringing together optical photometry and spectropolarimetry with FUV and NUV HST/STIS spectra. The variability in the $B_{\ell}$ over the 17-night span of the dataset is consistent with rotational modulation caused by the presence of unevenly distributed active regions that remain relatively stable over the timespan of observations. Similarly, the S$_{\rm HK}$ and I$_{H\alpha}$ indices don't show any significant change beyond rotational modulation -- at least within the noise level of these measurements. In contrast, the TESS lightcurve clearly shows variability which cannot be explained by rotational modulation alone, indicating the emergence of new flux. There is indication that the RV also shows some evolution from one rotation cycle to the next but the effect is at the 2-3$\sigma$ level so more precise measurements would be needed to confirm definitively to what extent the flux emergence that is detected in the TESS lightcurve could affect RV measurements. 

\item We analysed and compare four different HST/STIS spectra observed during cycles 25 and 26. We observed a clearly enhanced flux during the second orbit of C25 which suggests the possible association with a flaring event. In the near- and far- ultraviolet (NUV and FUV, respectively) we analysed the inter-combination lines such as \ion{C}{iii}, \ion{C}{iv}, \ion{S}{iv}, or \ion{O}{iv}, which can be good tracers for characterizing the high chromosphere and coronal transition region. Those activity indicators trace the influence of the stellar magnetic field as it emerges from the stellar interior and weaves its way through the different atmospheric layers of the star. 

The analysis here presented of \ihor\ data allowed for the recovery of the ZDI maps for the entire campaign, which is part of a forthcoming publication. In addition, those maps have been used to model the geometry of magnetic regions in the stellar surface that extends into the chromosphere, corona, and beyond \cite[e.g.,][]{2014ApJ...783...55C,2019ApJ...875L..12A,2022ApJ...941L...8G}. This has been done by applying a detailed 3D magneto-hydrodynamics (MHD) code \cite[BATS$-$R$-$US,][]{1999JCoPh.154..284P}, originally developed and validated for the solar wind and corona \cite[e.g.,][]{2013ApJ...764...23S,2014ApJ...782...81V}. This numerical approach includes all the relevant physics for calculating a coronal/wind model, having the surface magnetic field maps as a driver of a steady-state solution for the star. Coronal heating and stellar wind acceleration are self-consistently calculated via Alfven wave turbulence dissipation in addition to radiative cooling and electron heat conduction \cite[see,][]{2022ApJ...928..147A}. Models for the different phases of the magnetic cycle are generated and compared directly with the observations. The analysed observations are also used to predict the conditions experienced by the exoplanet during its orbit and through the magnetic cycle of its parent star. Moreover, the impact on habitability conditions due to the host-star magnetic field (and its evolution) is also studied in detail and will be published after the current manuscript.

\end{itemize}

\section*{Acknowledgements}
\addcontentsline{toc}{section}{Acknowledgements}
We thank the referee for careful reading of the manuscript and constructive comments which improved the original version of the manuscript. Based on observations collected at the European Organisation for Astronomical Research in the Southern Hemisphere under ESO programmes 096.D-0257, 097.D-0420, 098.D-0187, 099.D-0236, 0100.D-0535, and 0101.D-0465. Data were obtained from the ESO Science Archive Facility under request number jalvarad.212739. This work has made use of the VALD database, operated at Uppsala University, the Institute of Astronomy RAS in Moscow, and the University of Vienna. This research has made use of the SIMBAD database, operated at CDS, Strasbourg, France. This research has made use of NASA's Astrophysics Data System. E.M.A.G. and J.D.A.G. were partially supported by HST GO-15299 and GO-15512 grants. E.M.A.G and K.P.\ acknowledge support from the German \textit{Leibniz-Gemeinschaft} under project number P67/2018. F.D.S acknowledges support from a Marie Curie Action of the European Union (grant agreement 101030103). J.S.F. acknowledges support from the Agencia Estatal de Investigación of the Spanish Ministerio de Ciencia, Innovación y Universidades through grant PID2019-109522GB-C51.

\section{Data availability}\label{DA}

The data underlying this article were accessed from different large-scale facilities. The derived data from this research will be shared upon reasonable request to the corresponding author. For the analysis of the activity indicators, longitudinal magnetic field, and radial velocity evolution presented we included pipeline-processed spectroscopic HARPS~PH3,  \url{http://archive.eso.org/wdb/wdb/adp/phase3_main/form} and the HARPS-Polarimetry pipeline processed data query form \url{http://archive.eso.org/wdb/wdb/eso/repro/form}. TESS light curves and, HST/STIS data can be obtained from the High-Level Science Products (HLSP) on the Mikulski Archive for Space Telescopes (MAST, \url{http://archive.stsci.edu/}). The solar comparison data were of a quiet region on 20 April 1997, and the active region AR-NOAA8487 on 18 March 1999 \citep[see][]{Curdt2001,Curdt2004} and are available at Werner Curdt's homepage \url{https://www2.mps.mpg.de/homes/curdt/}. Optimistic and conservative estimates place the inner boundary of the habitable zone (HZ) of \ihor~at $0.94$~au and $1.20$~au, respectively available at \url{http://www.hzgallery.org} \citep{2012PASP..124..323K}.

\bibliographystyle{mnras}
\bibliography{FarbeyondtheSunII} 

\appendix

\section{Rotation period measurements - a comparison of multiple techniques}\label{App:rotation}

As noted in the main paper, we have applied seven methods to measure photometric periodicities related to the stellar rotation period: Gaussian processes (GP), quasi-periodic Gaussian Processes (QP-GP), Generalized Lomb-Scargle periodogram (GLS), Autocorrelation Function (ACF), Wavelet Power Spectra (PS), and the Gradient of Power Spectra (GPS). In Figure~\ref{fig:IotHor_GPS} we show the results from four of these methods: GLS, ACF, PS and GPS. 

Our GP approach makes use of a quasi-periodic kernel \cite[see Eq. 1 in][]{2017A&A...599A.126D} applied to sectors 2 and 3 of the TESS data for \ihor. We quantified the rotation period of the star (parameter $\Theta$) and a timescale related to the correlation time of the fluctuations ($\lambda$), which should account for the average lifetime of active regions on the surface of the star. We used a rather large prior on the rotation period $\Theta$ and $\lambda$, allowing the system to analyze values of $\Theta$ between 3 and 9 days and correlations spanning a timescale covering the whole S2 sector. Our results show that the correlation timescale should be shorter than, or of the order of, the rotation period. In the S2 analysis, we obtained a rotation $\Theta\,=\,6.64^{+0.28}_{-0.09},\,\lambda=4.7\pm0.3$, while for S3 we obtained $\Theta\,=\,7.01^{+0.16}_{-0.09},\,\lambda=7.7\pm0.7$.

For comparison, we also performed a quasi-periodic GP modelling using a simple covariance function \cite[see Eq. 3 in][]{2020A&A...634A..75B}, constructed using the Python \texttt{exoplanet} package \cite[][]{2019zndo...1998447F} with the following properties:
\begin{equation}
\label{eq_rotational_kernel}
k(\tau) = \dfrac{B}{2+C}~e^{-\tau/L}\Bigg(\cos\left(\dfrac{2\pi\tau}{P_{\rm rot}}\right)+(1+C)\Bigg) + \sigma^2\delta_{ij}, 
\end{equation}
where $P_{\rm rot}$ corresponds to the rotation period of the star, $L$ is the length-scale of exponential decay, and $B>0$, $C>0$, and $L>0$.
We chose large prior normal distributions for each hyper-parameter, specifically for the rotation period and the amplitude, letting the model test values of $\log P_{\rm rot}$ within $\mathrm{\mathcal{N}(1.97,\:0.5)}$ (corresponding to 3.2\:-\:90 days) and of $\log B$ and $\log C$ within $\mathrm{\mathcal{N}(-12.31,\:5)}$ and $\mathrm{\mathcal{N}(-4.47,\:5)}$, respectively. The central values $\mu$ of the chosen prior normal distributions were primarily set by a quick optimization test. Then, combined with a Markov chain Monte-Carlo fit, the equivalent estimate of the peak-to-peak amplitude of the oscillation is represented by $2\,B/(2+C)$ and is valued at $3.23\:10^{-3}$. By using this kernel, the GP approach gives a fair estimation of the amplitude of the activity \emph{jitter} in addition to a consistent determination of the rotation period for $P_{\rm rot}=7.23\pm0.20~\text{days}$ compared to the QP kernel described above.

We also employed a Multi-fractal approach called a Multi-fractal Temporally Weighted Detrended Fluctuation Analysis \cite[][]{Zhou+2010JSMTE, 2017AJ....153...12A}.
This technique is optimized for detecting relevant timescales and employs a model-free approach, also giving insight into the noise properties \citep[see Sec. 2 of][for a detailed explanation]{2017AJ....153...12A}.  By analyzing the S2 and S3 data, we found that the data are dominated by white noise on short timescales (up to about 1 hour), while between 1~hour to about 4~days, they show red noise.

It was possible to detect both timescales associated with these changes in noisy behaviours, and timescales associated with periodic features. In the S2 data, we observe a typical time of 4.1~days associated with the end of the red noise phase, and periodic dynamics at 7.48~days. In S3, we found timescales of 4.41~days and 7.25~days corresponding to the end of the red noise phase and clear periodic behaviour, respectively. The uncertainties on all of these measurements are approximately 0.15~days. When analysing the combined data, we also detected strong periodicity over 21.2~days.

We also applied the Lomb-Scargle periodogram~\citep[see][]{1976Ap&SS..39..447L,1982ApJ...263..835S} to the S2\,+\,S3 data for \ihor, specifically the Generalised Lomb-Scargle periodogram (GLS) version v1.03, using the formalism given by \cite{2009A&A...496..577Z}. In panel~(b) of Figure~\ref{fig:IotHor_GPS} we show the output from GLS, giving a rotation period estimate of about 7.76~days.

Panel~(c) of Figure~\ref{fig:IotHor_GPS} shows the result from the autocorrelation function applied to  S2\,+\,S3 data and resulting in a rotation period of 7.69~days. This is based on estimating the degree of self-similarity in the light-curve, see \cite{2014ApJS..211...24M}. 

The results from our application of a wavelet power spectra transform are shown in panel~(d) of Fig.~\ref{fig:IotHor_GPS}. This method is optimal for a time series with a non-stationary signal at many different frequencies. It has previously been employed for determining stellar rotation periods by~\cite{2009A&A...506...41G}. We use the power spectra transform analysis based on the 6th order Paul wavelet, \citep[see][]{1992AnRFM..24..395F,1998BAMS...79...61T} and recover a rotation period of 7.94 days.

Finally, we apply the GPS method to the normalized and stitched S2\,+\,S3 LCs. In \citet{PaperI,Eliana1,Eliana2} they found that the gradient of the wavelet power spectra transform, and in particular, the position of the high-frequency inflection point (i.e. the inflection point, IP, with the highest frequency) is connected to the stellar rotation period. The rotation period is determined from the profile of the high-frequency tail of the smoothed wavelet power spectrum (using a Paul wavelet, order 6), i.e. its part in between about a day and a quarter of the rotation period. In particular, they identify the point where the gradient of the power spectrum (GPS) plotted in log-log scale (i.e. $d \ln P( \nu ) / (d \ln \nu) $, where $P$ is the power spectral density and $\nu$ is frequency) reaches its maximum value. That point corresponds to the inflection point, i.e. the concavity of the power spectrum, plotted in log-log scale, changes sign there. It was shown that the period corresponding to the inflection point is connected to the rotation period of a star by a calibration factor which is a function of stellar effective temperature, metallicity, and activity. They validated their new method against Kepler light curves and the Total Solar Irradiance (TSI) data from the Sun. For \ihor, we recovered a rotation period of $7.78 \pm 0.18$ days using this method.

\begin{figure}
\centering 
\includegraphics[trim={0 0 0 0cm},clip,width=0.45\textwidth]{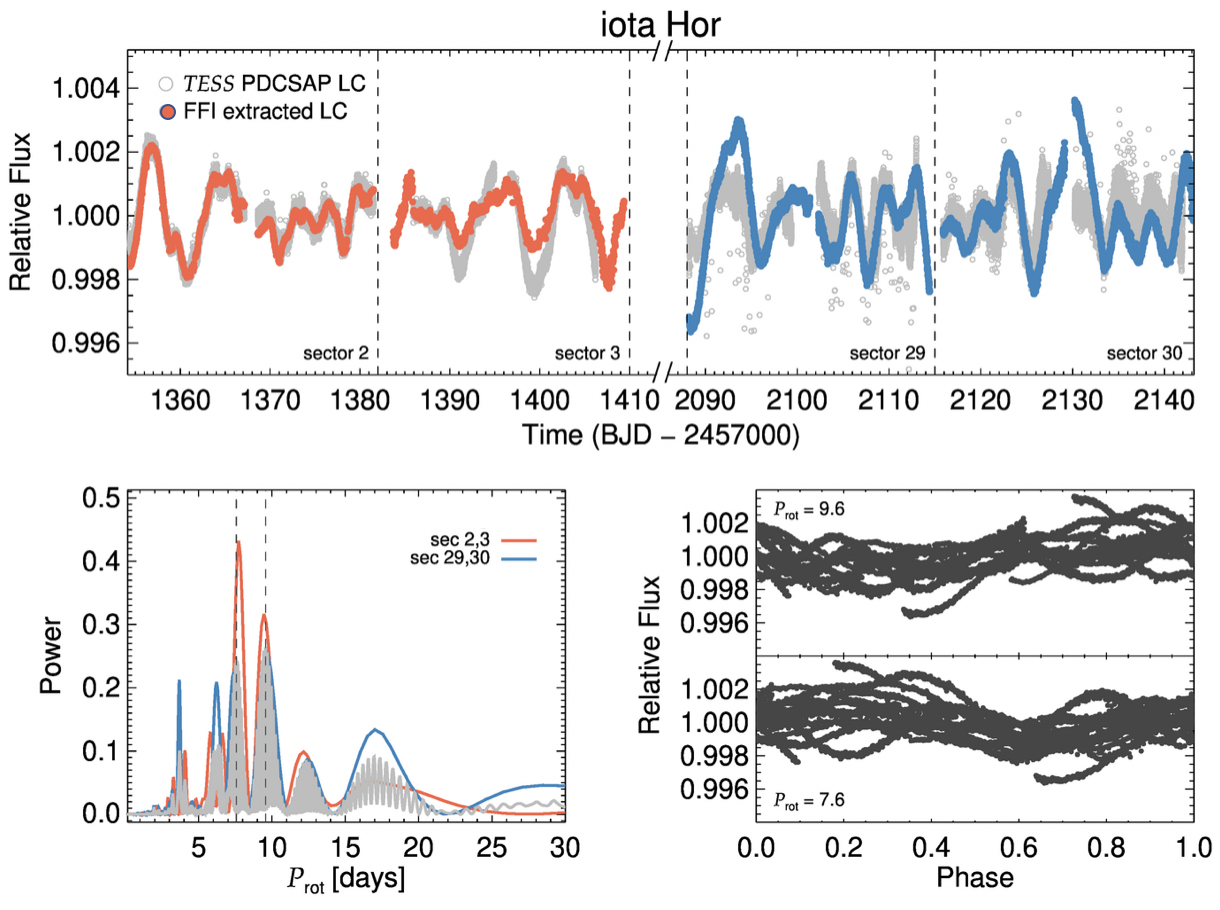}\hspace{0.0cm}
\caption{Rotation period analysis for \ihor. The upper panel shows detrended and stitched TESS light curves from sectors 2, 3, 29, and 30. The PDCSAP lightcurve (gray open symbols) clearly shows more variability than when extracting the lightcurve from full frame images, (FFI, orange and blue LC for sectors 2-3 and 29-30, respectively). Bottom panel left: Lomb-Scargle periodogram analysis. Bottom panel right: face fold FFI extracted LC for 9.6 and 7.6~days.} 
\label{fig:IotHor_DoNascimento}
\end{figure}

\begin{figure}
\centering 
\includegraphics[trim={0 0 0 0cm},clip,width=0.45\textwidth]{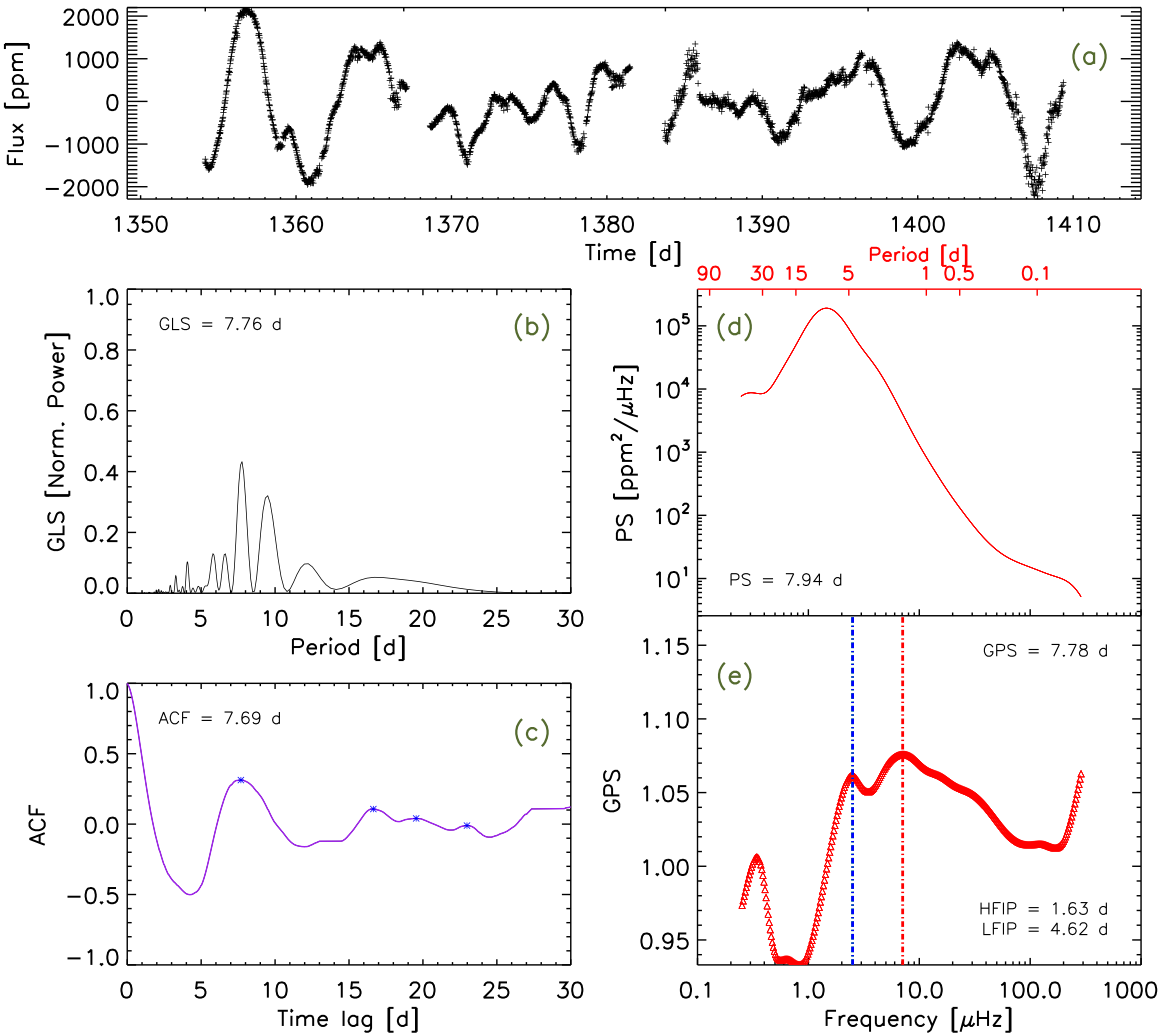}
\caption{The \ihor\ rotation period analysis for the normalized and stitched TESS light curves from Sectors~2 and 3, from 23 August 2018 to 17 October 2018 in panel~(A). Panels (B) and (C) show the corresponding Generalised Lomb-Scargle (GLS) periodogram and autocorrelation function (ACF), respectively. panel~(D) shows the global wavelet power spectrum, PS, calculated with the 6th~degree Paul wavelet. Panel~(E) shows the gradient of the power spectrum plotted in panel~(D). Red and blue dotted lines in panel~(E) indicate the location of the high and low-frequency inflection points location, respectively.}
\label{fig:IotHor_GPS}
\end{figure}

\begin{figure*} 
\includegraphics[trim=0.cm .0cm .0cm 0.cm, clip=true,scale=0.23]{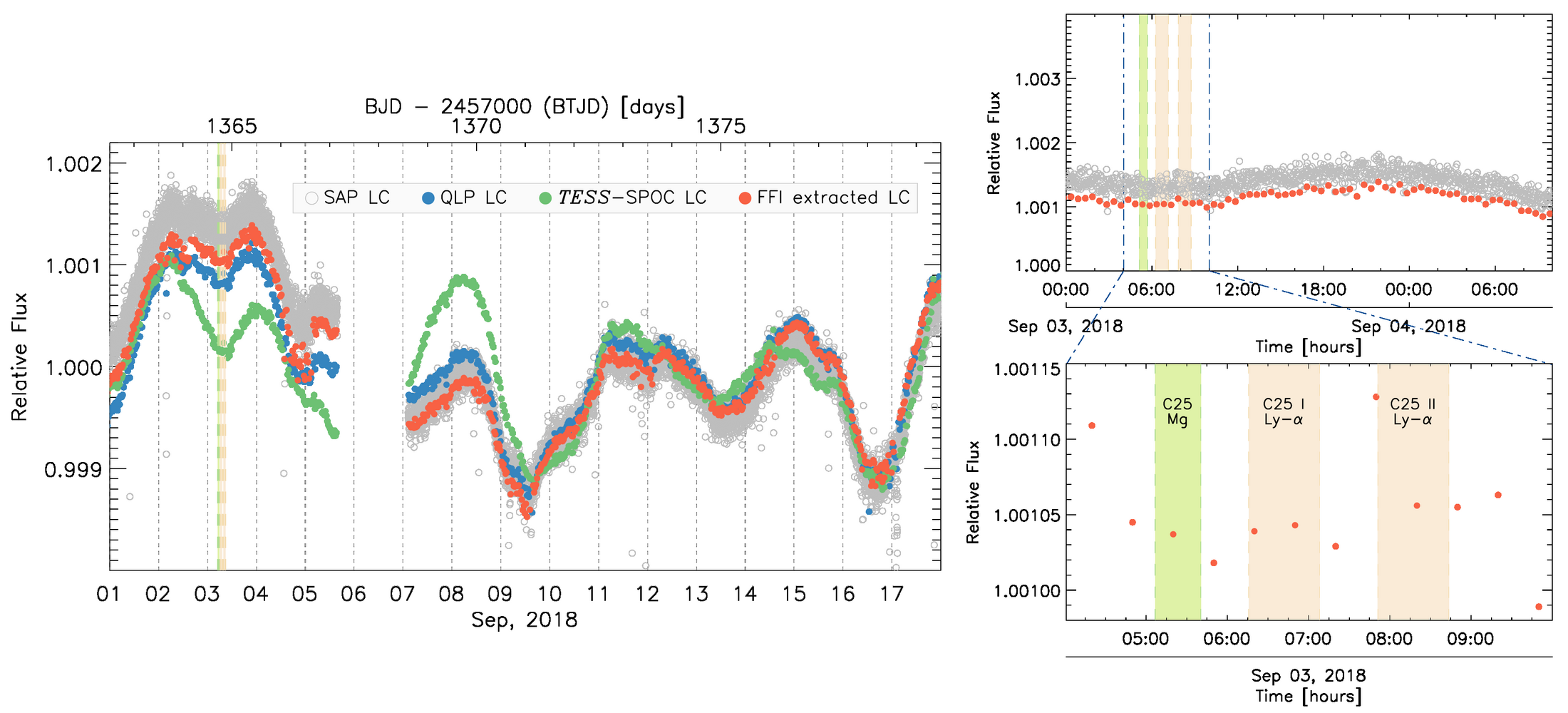}
\caption{Zoom into the S2 TESS-LC section [from 1362.5 to 1379.5 BTJD days] during 3 different STIS/HST visits for Cycle 25. The 3 HST orbits are organized for observing just ones the NUV region around the \ion{Mg}{II}$~\textrm{H\,\&\,K}$ line, and two visits for observing the FUV region nearby the Ly$\alpha$ line. \ihor.}\label{Fig15}
\end{figure*}

\section{Telluric lines in H$\alpha$}\label{appTelluric}

Removed telluric lines from H$\alpha$ are coloured in red in Figure~\ref{Fig3}. See the Rowland table of The solar spectrum book \cite{1966sst..book.....M}, where the equivalent width values are taken. 

\begin{table}\label{telha}
\centering
\begin{tabular}{c r r c}
 \hline\hline
 $\lambda$ & W                        & $\lambda$ & W \\
 $[\rm nm]$ & $[\rm m\text{\normalfont\AA}]$  &  $[\rm nm]$    & $[\rm m\text{\normalfont\AA}]$ \\
 \hline\hline
     & & & \\[-2pt]
655.7857 &	1.5  &  656.4061 &	4.5 \\ 
655.8149 &	7.0  & 656.4206  &	14.0 \\
655.8650 &	1.5  & 656.5545  &	3.0 \\
655.8955 &	1.5  &  656.7850 &	2.0 \\
656.0555 &	22.0 &  656.8806 &	3.5 \\	
656.1097 &	5.0  & 656.8900  &	5.0 \\
656.3521 &	4.5  &	         & \\
 \hline
 \end{tabular}
 \caption{Removed telluric lines from H$\alpha$ are coloured in red in Figure~\ref{Fig3}. See the Rowland table of The solar spectrum book \citet[][]{1966sst..book.....M}, where the equivalent width values are taken.}
 \end{table}

\section{Journal of observations}\label{app:journal}

\begin{table*}
\centering
\begin{threeparttable}
\begin{tabular}{c c c c c c c c c c c c }    
\hline
& & & & & & & &\\[-10pt]
BJD (+2400000.) & \textcolor{black}{S/N I }& \textcolor{black}{S/N V 
}& Airmass &\textcolor{black}{ RV   }& S$_{\rm HK}$ & eS$_{\rm HK}$ & I$_{\rm H\upalpha}$ & eI$_{\rm H\upalpha}$& \multicolumn{1}{c}{$B_{\ell}$} & $\sigma_{B_{\ell}}$  & \multicolumn{1}{c}{$N_{\ell}$}     \\[2pt]
[days] & \textcolor{black}{[@551 nm]
} & \textcolor{black}{[@551 nm]} & &\textcolor{black}{ [m s$^{-1}$] }  &   &  &  &  & [G] &  [G] & [G]   \\[2pt]

\hline
ESO EPOCH P99A\\
\hline                                                            
57934.90224  &  710  &    676  &   1.340 &  16909.2 $\pm$   2.5  &   0.2270  &   0.0060   &  0.51123   &  0.00139   &   -3.14    &   0.61   &   0.11     \\
57935.91343  &  765  &    742  &   1.281 &  16902.6 $\pm$   2.3  &   0.2254  &   0.0056   &  0.51165   &  0.00128   &   -2.67    &   0.56   &   0.26     \\
57936.90045  &  592  &    572  &   1.324 &  16892.9 $\pm$   2.7  &   0.2260  &   0.0074   &  0.51122   &  0.00164   &   -2.78    &   0.73   &  -1.09     \\
57937.90757  &  739  &    706  &   1.283 &  16904.5 $\pm$   2.4  &   0.2266  &   0.0058   &  0.51103   &  0.00134   &   -1.42    &   0.58   &   0.46     \\
57938.89985  &  264  &    260  &   1.303 &  16907.0 $\pm$   3.9  &   0.2092  &   0.0173   &  0.51171   &  0.00370   &   -2.27    &   1.63   &   1.75     \\
57942.91940  &  803  &    765  &   1.198 &  16919.2 $\pm$   2.3  &   0.2280  &   0.0052   &  0.51075   &  0.00124   &   -2.68    &   0.54   &   0.19     \\
57943.92830  &  675  &    639  &   1.168 &  16911.4 $\pm$   2.5  &   0.2246  &   0.0059   &  0.51137   &  0.00147   &   -0.92    &   0.63   &   0.17     \\
57944.89264  &  721  &    696  &   1.268 &  16894.6 $\pm$   2.4  &   0.2233  &   0.0058   &  0.51157   &  0.00135   &   -0.53    &   0.60   &   0.93     \\
57945.92100  &  855  &    826  &   1.174 &  16903.8 $\pm$   2.2  &   0.2239  &   0.0049   &  0.51153   &  0.00116   &   -1.16    &   0.50   &  -0.07     \\
57946.91205  &  654  &    631  &   1.189 &  16928.3 $\pm$   2.5  &   0.2259  &   0.0063   &  0.51178   &  0.00150   &   -2.54    &   0.66   &  -0.14     \\
57948.89648  &  660  &    642  &   1.219 &  16905.9 $\pm$   2.7  &   0.2312  &   0.0063   &  0.51168   &  0.00148   &   -1.33    &   0.66   &   0.01     \\
57949.91318  &  335  &    309  &   1.167 &  16916.8 $\pm$   3.8  &   0.2160  &   0.0121   &  0.51136   &  0.00293   &   -3.49    &   1.26   &  -0.45     \\
\hline
ESO EPOCH P99B\\
\hline
57961.89812  & 833  &    806  &   1.132 &  16934.1 $\pm$   2.3  &   0.2363  &   0.0051   &  0.51221   &  0.00119   &   -0.48    &   0.52   &  -0.33     \\
57962.90690  & 352  &    226  &   1.113 &  16951.4 $\pm$   3.5  &   0.2303  &   0.0125   &  0.51293   &  0.00283   &    2.27    &   2.07   &  -2.26     \\
57964.91010  & 964  &    932  &   1.102 &  16941.0 $\pm$   2.1  &   0.2325  &   0.0044   &  0.51163   &  0.00103   &   -1.19    &   0.45   &  -0.10     \\
57967.90436  & 350  &    335  &   1.099 &  16931.2 $\pm$   3.6  &   0.2284  &   0.0122   &  0.51144   &  0.00279   &    3.97    &   1.22   &  -0.13     \\
57968.93869  & 589  &    545  &   1.076 &  16949.8 $\pm$   2.7  &   0.2325  &   0.0073   &  0.51203   &  0.00165   &    1.31    &   0.73   &   0.69     \\
57970.92374  & 577  &    554  &   1.079 &  16959.7 $\pm$   2.8  &   0.2413  &   0.0070   &  0.51274   &  0.00171   &   -0.58    &   0.74   &   1.01     \\
57971.93274  & 517  &    500  &   1.076 &  16960.8 $\pm$   2.9  &   0.2343  &   0.0068   &  0.51196   &  0.00198   &   -1.67    &   0.79   &   0.01     \\
57972.92529  & 537  &    526  &   1.077 &  16942.7 $\pm$   2.9  &   0.2322  &   0.0078   &  0.51162   &  0.00182   &   -2.93    &   0.80   &   1.17     \\
57973.92259  & 466  &    421  &   1.077 &  16941.5 $\pm$   3.1  &   0.2326  &   0.0095   &  0.51255   &  0.00208   &   -2.53    &   0.95   &   1.36     \\
57974.93909  & 577  &    561  &   1.078 &  16942.5 $\pm$   2.7  &   0.2345  &   0.0071   &  0.51214   &  0.00171   &   -1.44    &   0.73   &   1.35     \\
\hline                      
ESO EPOCH P99C\\            
\hline                      
58013.88198  & 703  &    680  &   1.123 &  16991.1 $\pm$   2.4  &   0.2344  &   0.0060   &  0.51107   &  0.00141   &    0.47    &   0.62   &   0.91     \\
58014.88046  & 601  &    589  &   1.125 &  17005.1 $\pm$   2.6  &   0.2430  &   0.0070   &  0.51187   &  0.00163   &   -2.54    &   0.71   &   0.35     \\
58017.75162  & 638  &    603  &   1.123 &  16996.7 $\pm$   2.5  &   0.2394  &   0.0068   &  0.51277   &  0.00151   &   -0.10    &   0.67   &  -0.35     \\
58019.83871  & 836  &    809  &   1.090 &  16991.8 $\pm$   2.2  &   0.2364  &   0.0052   &  0.51156   &  0.00117   &   -0.29    &   0.52   &   1.13     \\
58020.84732  & 591  &    569  &   1.102 &  16987.1 $\pm$   2.7  &   0.2336  &   0.0065   &  0.51121   &  0.00169   &    0.48    &   0.71   &   0.37     \\
58021.87703  & 656  &    630  &   1.155 &  16985.8 $\pm$   2.5  &   0.2360  &   0.0064   &  0.51123   &  0.00149   &   -0.73    &   0.66   &  -0.22     \\
58022.79833  & 551  &    526  &   1.076 &  17001.7 $\pm$   2.9  &   0.2442  &   0.0069   &  0.51303   &  0.00182   &   -1.19    &   0.76   &  -1.38     \\
58023.82878  & 632  &    602  &   1.090 &  17000.1 $\pm$   2.6  &   0.2439  &   0.0060   &  0.51329   &  0.00159   &   -2.25    &   0.66   &  -0.39     \\
58024.76855  & 380  &    337  &   1.084 &  16999.0 $\pm$   3.4  &   0.2457  &   0.0109   &  0.51331   &  0.00258   &    1.48    &   1.21   &   2.79     \\
58025.83057  & 334  &    300  &   1.098 &  16992.1 $\pm$   3.6  &   0.2458  &   0.0149   &  0.51305   &  0.00288   &    0.88    &   1.39   &  -0.44     \\
58026.80399  & 561  &    524  &   1.079 &  17016.1 $\pm$   2.8  &   0.2437  &   0.0068   &  0.51313   &  0.00178   &    0.76    &   0.75   &  -0.67     \\
58027.80589  & 502  &    481  &   1.082 &  16994.7 $\pm$   2.9  &   0.2392  &   0.0090   &  0.51246   &  0.00190   &    0.86    &   0.87   &  -0.64     \\
58028.79881  & 575  &    545  &   1.079 &  16978.9 $\pm$   2.8  &   0.2379  &   0.0077   &  0.51152   &  0.00168   &   -0.08    &   0.75   &  -0.57     \\
\hline
ESO EPOCH P100A\\
\hline
58069.60736  &  847  &    816 &   1.125 &  16946.7 $\pm$   2.3  &   0.2488  &   0.0044   &  0.51332   &  0.00120   &   -5.46    &   0.51   &  -0.72     \\
58070.60682  &  883  &    854 &   1.122 &  16923.2 $\pm$   2.3  &   0.2420  &   0.0044   &  0.51232   &  0.00113   &    0.39    &   0.48   &   0.09     \\
58071.64337  & 1021  &    998 &   1.081 &  16929.5 $\pm$   2.1  &   0.2393  &   0.0037   &  0.51192   &  0.00101   &    0.59    &   0.42   &  -0.34     \\
58073.63701  & 1117  &   1048 &   1.082 &  16929.7 $\pm$   2.0  &   0.2393  &   0.0036   &  0.51190   &  0.00090   &   -0.67    &   0.39   &   0.43     \\
58074.66443  &  958  &    925 &   1.078 &  16939.8 $\pm$   2.2  &   0.2427  &   0.0043   &  0.51352   &  0.00104   &   -2.58    &   0.45   &   1.11     \\
58075.64065  &  989  &    948 &   1.078 &  16950.3 $\pm$   2.2  &   0.2492  &   0.0041   &  0.51411   &  0.00101   &   -1.44    &   0.44   &   0.49     \\
58076.65331  &  643  &    603 &   1.075 &  16927.5 $\pm$   2.7  &   0.2479  &   0.0065   &  0.51334   &  0.00154   &   -1.75    &   0.68   &   0.15     \\
58077.65193  & 1067  &   1010 &   1.076 &  16936.1 $\pm$   2.1  &   0.2408  &   0.0039   &  0.51213   &  0.00093   &   -0.47    &   0.41   &   0.35     \\
58078.66084  & 1111  &   1049 &   1.079 &  16918.6 $\pm$   2.0  &   0.2378  &   0.0038   &  0.51184   &  0.00091   &    1.50    &   0.39   &  -0.18     \\
58079.65272  &  873  &    848 &   1.078 &  16928.3 $\pm$   2.3  &   0.2380  &   0.0046   &  0.51227   &  0.00114   &   -1.30    &   0.49   &  -0.25     \\
58081.65863  & 1070  &   1042 &   1.083 &  16935.0 $\pm$   2.0  &   0.2476  &   0.0040   &  0.51312   &  0.00093   &   -0.42    &   0.41   &  -0.13     \\
\hline                       
ESO EPOCH P100B\\            
\hline                       
58108.56994  &  715  &    679 &   1.076 &  16863.0 $\pm$   2.7  &   0.2466  &   0.0055   &  0.51485   &  0.00142   &    2.19    &   0.60   &  -0.25     \\
58109.62286  &  828  &    785 &   1.132 &  16900.3 $\pm$   2.4  &   0.2381  &   0.0049   &  0.51406   &  0.00117   &   -0.21    &   0.51   &  -0.06     \\
58110.61173  &  792  &    753 &   1.118 &  16926.9 $\pm$   2.3  &   0.2433  &   0.0051   &  0.51472   &  0.00123   &   -2.55    &   0.54   &   0.10     \\
58111.64169  &  681  &    654 &   1.185 &  16940.7 $\pm$   2.6  &   0.2512  &   0.0058   &  0.51679   &  0.00143   &   -3.99    &   0.63   &  -0.19     \\
58112.63716  &  874  &    838 &   1.180 &  16949.6 $\pm$   2.2  &   0.2612  &   0.0050   &  0.51840   &  0.00111   &   -0.39    &   0.50   &   0.03     \\
58113.60393  &  985  &    925 &   1.119 &  16907.1 $\pm$   2.2  &   0.2657  &   0.0045   &  0.51839   &  0.00099   &   -0.51    &   0.44   &   0.28     \\
58114.58488  & 1052  &    999 &   1.098 &  16927.0 $\pm$   2.2  &   0.2612  &   0.0040   &  0.51754   &  0.00094   &   -3.63    &   0.41   &  -0.32     \\
58115.62666  &  959  &    924 &   1.175 &  16897.0 $\pm$   2.2  &   0.2505  &   0.0045   &  0.51575   &  0.00103   &    1.09    &   0.45   &  -0.13     \\
58116.58632  & 1161  &   1114 &   1.105 &  16868.6 $\pm$   2.1  &   0.2421  &   0.0040   &  0.51461   &  0.00085   &    1.98    &   0.38   &   0.20     \\
58117.61202  &  984  &    957 &   1.154 &  16895.8 $\pm$   2.2  &   0.2428  &   0.0048   &  0.51346   &  0.00098   &    1.07    &   0.45   &  -0.39     \\
58118.59907  & 1081  &   1026 &   1.134 &  16920.1 $\pm$   2.1  &   0.2475  &   0.0041   &  0.51431   &  0.00091   &   -1.12    &   0.41   &   0.34     \\
58119.54663  &  973  &    939 &   1.079 &  16949.8 $\pm$   2.1  &   0.2597  &   0.0042   &  0.51619   &  0.00103   &   -1.64    &   0.44   &   0.31     \\
58121.53995  &  700  &    679 &   1.079 &  16911.3 $\pm$   2.6  &   0.2654  &   0.0056   &  0.51698   &  0.00140   &   -2.95    &   0.61   &   0.67     \\
58122.55886  & 1116  &   1093 &   1.094 &  16922.4 $\pm$   2.1  &   0.2611  &   0.0038   &  0.51667   &  0.00089   &   -2.55    &   0.39   &  -0.18     \\
\hline                 
\end{tabular}
\begin{tablenotes}{\vspace{-2pt}} 
\caption{Journal of observations (columns $1-4$) and measurements for each night (columns $5-12$).} 
\item {\hfill \textit{Continued on page 22}}
\end{tablenotes}
\end{threeparttable}\label{table_data_B3}
\end{table*}

\begin{table*}
\label{table_continue}      
\centering                                                         \begin{threeparttable}
\begin{tabular}{c c c c c c c c c c c c c}    
\hline\hline
& & & & & & & &\\[-4pt]
BJD (+2400000.) & \textcolor{black}{S/N I }& \textcolor{black}{S/N V 
}& Airmass &\textcolor{black}{ RV   }& S$_{\rm HK}$ & eS$_{\rm HK}$ & I$_{\rm H\upalpha}$ & eI$_{\rm H\upalpha}$& \multicolumn{1}{c}{$B_{\ell}$} & $\sigma_{B_{\ell}}$  & \multicolumn{1}{c}{$N_{\ell}$}     \\[2pt]
[days] & \textcolor{black}{[@551 nm]
} & \textcolor{black}{[@551 nm]} & &\textcolor{black}{ [m s$^{-1}$] }  &   &  &  &  & [G] &  [G] & [G]   \\[2pt]  \hline
ESO EPOCH P100C\\
\hline
& & & & & & & &\\[-3pt]  
58151.56156  & 916   & 865  &  1.286 &  16914.6 $\pm$   2.3  &   0.2638  &   0.0054   &  0.51728   &  0.00102   &   -1.50    &   0.50   &   0.16     \\
58152.55135  & 905   & 874  &  1.257 &  16905.5  $\pm$  2.3  &   0.2655  &   0.0052   &  0.51696   &  0.00107   &    0.22    &   0.48   &  -0.74     \\
58153.53730  & 991   & 962  &  1.220 &  16919.4  $\pm$  2.1  &   0.2600  &   0.0050   &  0.51604   &  0.00097   &    0.84    &   0.44   &  -0.18     \\
58155.53830  & 878   & 833  &  1.241 &  16898.9  $\pm$  2.3  &   0.2583  &   0.0052   &  0.51460   &  0.00113   &   -2.17    &   0.49   &  -0.28     \\
58156.54132  & 668   & 637  &  1.262 &  16916.4  $\pm$  2.6  &   0.2608  &   0.0068   &  0.51515   &  0.00143   &   -5.35    &   0.65   &   0.03     \\
58157.52308  & 804   & 780  &  1.211 &  16914.1  $\pm$  2.3  &   0.2672  &   0.0059   &  0.51698   &  0.00118   &   -1.54    &   0.55   &   0.09     \\
58158.52441  & 891   & 862  &  1.223 &  16889.7  $\pm$  2.4  &   0.2689  &   0.0055   &  0.51670   &  0.00107   &    1.27    &   0.50   &   0.39     \\
58159.52273  & 857   & 835  &  1.226 &  16908.2  $\pm$  2.3  &   0.2642  &   0.0054   &  0.51601   &  0.00112   &    0.70    &   0.51   &  -0.30     \\
58160.52493  & 720   & 695  &  1.243 &  16911.7  $\pm$  2.5  &   0.2639  &   0.0060   &  0.51597   &  0.00133   &   -0.96    &   0.60   &   0.17     \\
58161.53424  & 757   & 740  &  1.287 &  16883.6  $\pm$  2.6  &   0.2654  &   0.0059   &  0.51629   &  0.00126   &   -0.94    &   0.59   &  -0.17     \\
58162.51359  & 994   & 952  &  1.224 &  16918.8  $\pm$  2.2  &   0.2638  &   0.0049   &  0.51632   &  0.00097   &   -1.16    &   0.44   &  -0.01     \\
\hline                                                     
ESO EPOCH P101A\\                                          
\hline                                                     
58292.90558  & 703   & 774  &  1.423 &  17031.2  $\pm$  2.5  &   0.2445  &   0.0054   &  0.52289   &  0.00137   &    0.39    &   0.62   &   0.08     \\
58293.90560  & 441   & 487  &  1.408 &  17060.5  $\pm$  3.1  &   0.2470  &   0.0085   &  0.52396   &  0.00214   &   -2.19    &   0.98   &   0.39     \\
58294.90798  & 499   & 544  &  1.381 &  17051.8  $\pm$  3.0  &   0.2499  &   0.0069   &  0.52424   &  0.00196   &   -2.48    &   0.84   &  -1.06     \\
58295.91032  & 745   & 821  &  1.355 &  17053.0  $\pm$  2.4  &   0.2542  &   0.0047   &  0.52429   &  0.00133   &   -1.29    &   0.56   &   0.79     \\
58296.90910  & 711   & 781  &  1.349 &  17077.6  $\pm$  2.5  &   0.2534  &   0.0052   &  0.52485   &  0.00136   &    0.84    &   0.60   &  -0.32     \\
58297.88170  & 756   & 839  &  1.487 &  17063.0  $\pm$  2.3  &   0.2561  &   0.0049   &  0.52527   &  0.00129   &   -0.33    &   0.56   &   0.54     \\
58298.90085  & 683   & 762  &  1.363 &  17030.8  $\pm$  2.5  &   0.2469  &   0.0054   &  0.52453   &  0.00141   &   -1.18    &   0.62   &  -0.78     \\
58306.92059  & 477   & 512  &  1.204 &  17052.7  $\pm$  3.0  &   0.2413  &   0.0072   &  0.52327   &  0.00204   &   -1.69    &   0.87   &  -0.29     \\
\hline                                                     
ESO EPOCH P101B\\                                          
\hline                                                     
58328.87595  & 699   & 663  &  1.167 &  17001.0  $\pm$  2.5  &   0.2422  &   0.0052   &  0.51407   &  0.00138   &    0.78    &   0.61   &  -0.30     \\
58329.87836  & 755   & 748  &  1.155 &  16996.3  $\pm$  2.4  &   0.2387  &   0.0048   &  0.51288   &  0.00128   &   -2.05    &   0.56   &   0.12     \\
58331.86236  & 656   & 629  &  1.181 &  16992.6  $\pm$  2.6  &   0.2375  &   0.0048   &  0.51252   &  0.00156   &   -0.52    &   0.61   &  -0.74     \\
58332.79295  & 691   & 684  &  1.454 &  16988.7  $\pm$  2.6  &   0.2365  &   0.0055   &  0.51318   &  0.00137   &   -1.75    &   0.61   &  -0.92     \\
58333.86852  & 641   & 595  &  1.153 &  16999.2  $\pm$  2.6  &   0.2389  &   0.0056   &  0.51332   &  0.00153   &   -3.71    &   0.65   &  -0.31     \\
58335.87337  & 864   & 826  &  1.134 &  16999.9  $\pm$  2.3  &   0.2433  &   0.0041   &  0.51384   &  0.00114   &    0.98    &   0.48   &  -0.44     \\
58336.85067  & 827   & 809  &  1.176 &  17010.6  $\pm$  2.2  &   0.2377  &   0.0045   &  0.51248   &  0.00117   &   -1.49    &   0.51   &   0.83     \\
58341.86137  & 687   & 653  &  1.127 &  16997.7  $\pm$  2.5  &   0.2441  &   0.0050   &  0.51385   &  0.00144   &   -1.72    &   0.60   &  -0.32     \\
\hline
ESO EPOCH P101C\\
\hline
58363.85483  &  899   &  876  &  1.077 &  16962.5  $\pm$  2.1  &   0.2374  &   0.0039   &  0.51192   &  0.00111   &   -2.61    &   0.46   &   0.48     \\
58366.78328  &  850   &  810  &  1.145 &  16962.7  $\pm$  2.3  &   0.2332  &   0.0042   &  0.51127   &  0.00118   &   -0.65    &   0.49   &  -0.15     \\
58367.76771  &  987   &  935  &  1.174 &  16956.2  $\pm$  2.0  &   0.2372  &   0.0037   &  0.51216   &  0.00101   &   -4.21    &   0.42   &  -1.09     \\
58368.79676  &  944   &  898  &  1.113 &  16947.1  $\pm$  2.1  &   0.2395  &   0.0038   &  0.51228   &  0.00106   &   -1.03    &   0.44   &   0.84     \\
58369.86552  & 1047   &  991  &  1.079 &  16961.4  $\pm$  2.0  &   0.2369  &   0.0037   &  0.51220   &  0.00094   &    3.32    &   0.40   &   0.22     \\
58370.86078  & 1065   & 1027  &  1.079 &  16966.8  $\pm$  2.0  &   0.2435  &   0.0036   &  0.51266   &  0.00093   &    2.06    &   0.39   &   0.13     \\
58371.87136  &  758   &  729  &  1.086 &  16931.6  $\pm$  2.8  &   0.2432  &   0.0071   &  0.51296   &  0.00182   &   -2.07    &   0.79   &  -0.22     \\
58372.85848  &  847   &  809  &  1.081 &  16963.7  $\pm$  2.2  &   0.2428  &   0.0045   &  0.51184   &  0.00115   &   -3.29    &   0.50   &  -0.22     \\
58375.84499  &  676   &  643  &  1.079 &  16941.4  $\pm$  2.5  &   0.2427  &   0.0057   &  0.51188   &  0.00140   &   -3.99    &   0.63   &  -0.17     \\
58376.84750  &  741   &  716  &  1.081 &  16934.3  $\pm$  2.4  &   0.2383  &   0.0051   &  0.51247   &  0.00129   &   -0.85    &   0.57   &  -0.85     \\
58377.74118  &  663   &  642  &  1.173 &  16958.0  $\pm$  2.5  &   0.2405  &   0.0055   &  0.51277   &  0.00144   &    2.26    &   0.63   &  -0.29     \\
[3pt] 
\hline                 
\end{tabular}
\caption{Journal of observations (columns $1-4$) and measurements for each night (columns $5-12$).} 
\begin{tablenotes}{\vspace{-2pt}}
\item {\hfill \textit{Continuation of table~\ref{table_data_B3}.}}
\end{tablenotes}
\end{threeparttable}
\end{table*}

 \section{Additional HST lines}\label{App:HST}

\begin{figure*}
\centering 
\includegraphics[trim={0 0 0 0cm},clip,width=1.0\textwidth]{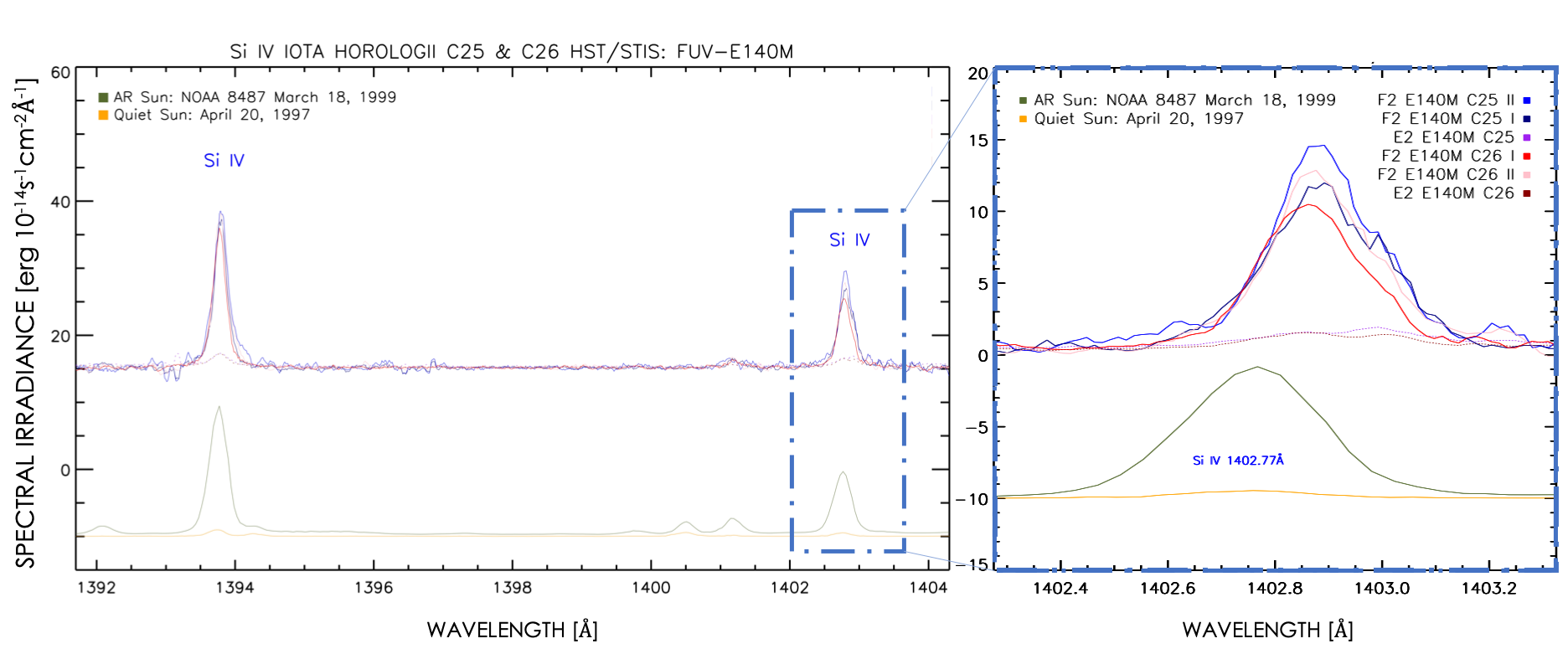}\hspace{0.0cm}
\caption{\ion{Si}{IV} spectra of \ihor\ compared with the solar active (green) and quiet regions (yellow). Enhanced emission is seen in \ihor\ during the second visit in Cycle 25 though to a lesser extent than in the \ion{C}{III} and \ion{C}{IV} lines.}
\label{Fig11}
\end{figure*}

\begin{figure*}
\centering 
\includegraphics[trim={0 0 0 0cm},clip,width=1.0\textwidth]{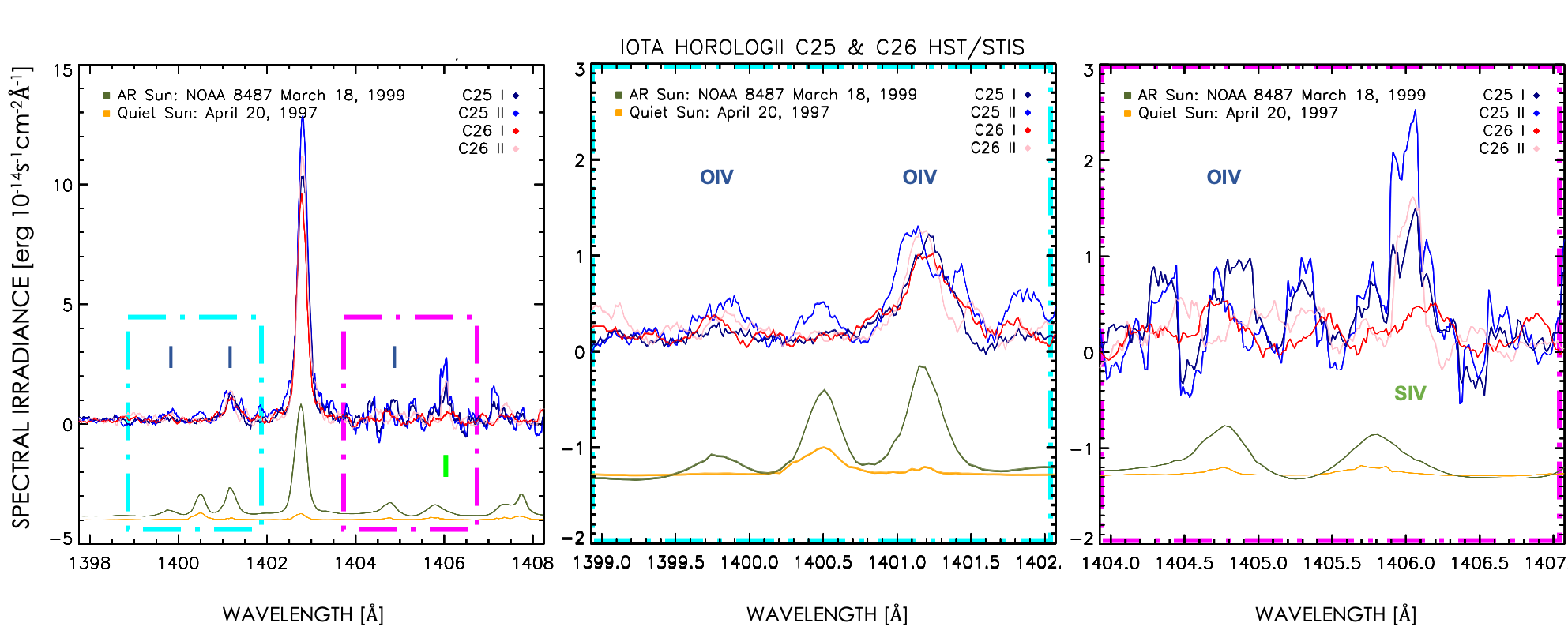}\hspace{0.0cm}
\caption{\ion{O}{IV} lines of \ihor\ at four epochs compared with solar active (green) and quiet regions (yellow) as in the previous plots of \ion{Si}{IV}, \ion{C}{III} and \ion{C}{IV}. While enhancement is clearly seen in the \ion{O}{IV} emission during the observation marked C25-II, the shape of \ion{O}{IV} 1401.0\AA\ also appears to change.}
\label{Fig12}
\end{figure*}

\label{lastpage}
\end{document}